\documentclass[lettersize,journal]{IEEEtran}
\usepackage{amsmath,amsfonts}
\usepackage{algorithm}
\usepackage{array}
\usepackage[caption=false,font=normalsize,labelfont=sf,textfont=sf]{subfig}
\usepackage{textcomp}
\usepackage{stfloats}
\usepackage{url}
\usepackage{verbatim}
\usepackage{cite}
\usepackage{amsmath,amssymb,amsfonts}
\usepackage{textcomp}
\usepackage{amsmath}
\usepackage{tikz}
\usepackage{breakurl}
\usepackage{tabularx}
\usepackage{multirow}
\usepackage{float}
\usepackage{tabularray}
\usepackage{comment}
\usepackage{amssymb}
\usepackage{bbding}
\usepackage{algpseudocode}
\usepackage{verbatim}
\usepackage{tcolorbox}
\usepackage{pifont}
\usepackage{microtype}
\usepackage{graphicx}
\usepackage{booktabs} 
\usepackage[normalem]{ulem}
\useunder{\uline}{\ul}{}
\usepackage{enumitem}
\usepackage[table,xcdraw]{xcolor}
\usepackage{dsfont}
\usepackage{algorithmicx}
\usepackage{acronym}
\usepackage{xspace}

\newcommand{\ie}{i.e.,\xspace}
\newcommand{\eg}{e.g.,\xspace}

\newcommand{\etal}{\textit{et al.}\xspace}
\newcommand{\eat}[1]{}

\acrodef{iot}[IoT]{Internet of Things}

\definecolor{dkgreen}{rgb}{0,0.6,0}

\newcommand{\tool}{{RHINO}\xspace}

\newcommand{\revise}[1]{\textcolor{black}{#1}}

\begin{document}

\title{\tool: Guided Reasoning for Mapping Network Logs to Adversarial Tactics and Techniques with Large Language Models}

\author{Fanchao Meng,
        Jiaping Gui,~\IEEEmembership{Member,~IEEE,}
        Yunbo Li,~\IEEEmembership{Student Member,~IEEE,}
        and Yue Wu,~\IEEEmembership{Senior Member,~IEEE}
\thanks{This work was supported by the National Key R\&D Program of China under Grant 2023YFB2704903. \textit{(Corresponding authors: Jiaping Gui; Yue Wu.)}}
\thanks{The authors are with the School of Computer Science, Shanghai Jiao Tong University, No. 800 Dongchuan Road, Shanghai, China (e-mail: mactavishmeng@sjtu.edu.cn, jgui@sjtu.edu.cn, 
li-yun-bo@sjtu.edu.cn, wuyue@sjtu.edu.cn).} 
\thanks{This paper has supplementary material available at http://ieeexplore.ieee.org, including prompts, detailed results, case analyses, and annotation details. Contact mactavishmeng@sjtu.edu.cn for further questions.}}

\markboth{IEEE Transactions on Information Forensics and Security}%
{Meng \MakeLowercase{\textit{et al.}}: RHINO: Guided Reasoning for Mapping Network Logs to Adversarial Tactics and Techniques}


\maketitle

\begin{abstract}

Modern Network Intrusion Detection Systems (NIDS) generate vast volumes of low-level alerts, yet these outputs remain semantically fragmented, requiring labor-intensive manual correlation with high-level adversarial behaviors. Existing solutions for automating this mapping---rule-based systems and machine learning classifiers---suffer from critical limitations: rule-based approaches fail to adapt to novel attack variations, while machine learning methods lack contextual awareness and treat tactic-technique (TT) mapping as a syntactic matching problem rather than a reasoning task. Although Large Language Models (LLMs) have shown promise in cybersecurity tasks, preliminary experiments reveal that existing LLM-based methods frequently \revise{hallucinate technique--tactic pairs} or produce decontextualized mappings due to their single-step classification approach.

To address these challenges, we introduce \tool, a framework that decomposes LLM-based attack analysis into three structured phases: (1) behavioral abstraction, where raw logs are translated into contextualized narratives; (2) partitioned inference over the ATT\&CK tactic space, generating candidate techniques by evaluating behavioral evidence against MITRE ATT\&CK knowledge; and (3) knowledge-grounded refinement, cross-referencing predictions with official MITRE definitions to reduce hallucinations. \tool bridges the semantic gap between low-level observations and adversarial intent while improving output reliability through structured reasoning.

We evaluate \tool on three benchmarks (DAPT2020, CICIDS2017, IoT23) across four backbone models (ChatGPT-4o, Gemini 2.5 Flash, Claude Sonnet 4, and DeepSeek V3). \revise{Under a scenario-documented single-label formulation covering 11 distinct techniques,} \tool achieves an average top-1 accuracy of 87.34\% under micro-averaging and 63.77\% under macro-averaging, surpassing the strongest baseline by 16.78 and 13.60 points respectively. These results demonstrate that \tool delivers accurate TT mappings across diverse attack scenarios and backbone models \revise{within this scope}, showing that structured LLM reasoning can effectively translate low-level network observations into high-level adversarial behaviors.

\end{abstract}

\begin{IEEEkeywords}
Large language models, MITRE ATT\&CK, tactic and technique mapping, threat intelligence.
\end{IEEEkeywords}

\section{Introduction}

\IEEEPARstart{M}{odern} cybersecurity defenses rely heavily on Network Intrusion Detection Systems (NIDS) to monitor and flag malicious activities within network traffic~\cite{liao2013intrusion}. These systems, ranging from signature-based tools like Snort~\cite{koziol2003intrusion} to machine learning-based anomaly detectors such as Kitsune~\cite{mirsky2018kitsune}, generate vast volumes of low-level alerts and events (\ie NIDS logs). However, these outputs remain semantically fragmented, requiring security analysts to manually correlate them with high-level adversarial behaviors—a process that is not only time-consuming but also prone to human error. Studies indicate that analysts misclassify up to 30\% of alerts when mapping them to threat frameworks like MITRE ATT\&CK~\cite{strom2018mitre}, while the sheer volume of daily alerts (often exceeding 10,000 in enterprise environments~\cite{jalalvand2024alert}) further exacerbates operational inefficiencies.

Existing solutions for automating this mapping process fall into two broad categories: rule-based systems~\cite{izzuddin2022mapping, milajerdi2019holmes, meng2026knowhow} and machine learning classifiers~\cite{hakim2024predicting, moskal2021translating, nadeem2021alert, meng2025distr}. Rule-based approaches, such as M2ASK~\cite{meng2024poster}, rely on predefined correlation rules to associate alerts with ATT\&CK techniques. While effective for known attack patterns, these methods fail to adapt to novel tactics or variations in log phrasing—for instance, treating ``FTP PASS Command detected'' and ``FTP Login Attempt'' as distinct behaviors despite their semantic equivalence. Machine learning-based methods, such as PATRL~\cite{moskal2021translating}, leverage annotated datasets to train classifiers for tactic and technique (TT) prediction. However, they often lack contextual awareness, leading to misinterpretations when behavioral cues (\eg repeated login attempts or temporal patterns) are absent from the training data. More fundamentally, both approaches treat TT mapping as a syntactic matching problem rather than a reasoning task, limiting their ability to infer adversarial intent from incomplete or noisy observations.

Recent advancements in Large Language Models (LLMs) have demonstrated potential in addressing these limitations, with applications ranging from malware analysis~\cite{yu2024maltracker} to phishing detection~\cite{li2024knowphish}. However, our preliminary experiments reveal that LLM-based methods suffer from critical shortcomings. Without structured reasoning, \revise{LLMs frequently hallucinate technique--tactic pairs (about 24\% invalid predictions, 836 of 3,493 with ChatGPT-4o)}, or produce decontextualized mappings that ignore multi-stage attack dynamics.

We observe that in real-world analyst workflows, accurate attribution requires progressive reasoning from NIDS logs to behavioral abstractions, then to adversarial intent, and finally to MITRE technique alignment. Inspired by this, to bridge this gap, we introduce \tool, a framework that decomposes LLM-based analysis into three structured phases. First, the model extracts key behaviors (\eg ``rapid sequential SSH login failures from a single IP'') from NIDS logs while preserving contextual signals like protocol semantics and temporal patterns within each traffic segment. Next, an inference stage partitions the ATT\&CK tactic space into semantically coherent groups and generates candidate techniques (\eg T1110 Brute Force) by evaluating behavioral evidence against ATT\&CK knowledge within each group. Finally, a knowledge-grounded refinement stage cross-references candidates against official MITRE definitions, \revise{reducing the share of candidates that contradict the taxonomy from 7.5\% to 2.6\%} and producing a ranked list of candidate techniques.

\tool's design addresses three core challenges in LLM-based log analysis: (1) \textbf{Complex attack reasoning.} Partitioning inference over the ATT\&CK tactic space lets the model reason over a coherent slice of attacker behavior at each step, rather than searching the full technique space at once. (2) \textbf{Semantic gaps.} Explicitly modeling network context (\eg flow asymmetry, protocol anomalies) reconstructs attacker behavior that raw statistics alone obscure. (3) \textbf{Output reliability.} The knowledge-grounded refinement mechanism enforces consistency with MITRE's taxonomy, reducing technically invalid predictions.

Our evaluation demonstrates that \tool achieves an average top-1 accuracy of 87.34\% (micro) and 63.77\% (macro) for technique prediction across advanced backbones, including ChatGPT-4o, Gemini 2.5 Flash, Claude Sonnet 4, and DeepSeek V3, outperforming the strongest baseline by 16.78 and 13.60 points respectively. \revise{These gains are achieved while preserving generalization across diverse attack datasets (DAPT2020, CICIDS2017, IoT23).} The framework's multi-stage design proves critical, with ablation studies showing the abstraction module alone accounts for a 68 percentage points accuracy drop when removed, by contextualizing low-level observations.
\revise{When integrated with a production NIDS, \tool provides ATT\&CK mappings without compromising the NIDS's detection accuracy and achieves strong performance ($>$90\% top-1 accuracy) on the evaluated datasets.} These results demonstrate \tool's effectiveness in bridging the gap between raw network data and actionable threat intelligence.

This paper makes the following contributions:

\begin{itemize}[leftmargin=*, noitemsep, topsep=0pt]
	\item \textbf{Structured Reasoning Pipeline.} We design \tool, a three-stage pipeline that reformulates TT mapping from a classification problem into a structured reasoning task, addressing the semantic gap between low-level network observations and high-level adversarial behaviors.
	
	\item \textbf{Context-Aware Abstraction:} We introduce a semantic parsing module that translates low-level network logs into high-level behavioral narratives, preserving critical contextual signals (\eg protocol semantics, temporal patterns) to bridge the gap between raw observations and adversarial intent.
	
	\item \textbf{Partitioned Inference and Knowledge-Grounded Refinement.} We partition the ATT\&CK tactic space into semantically coherent groups and conduct independent inference passes over each group to mitigate bias toward dominant techniques. We further introduce a refinement step that grounds candidate predictions against official MITRE definitions, reducing hallucinated technique identifiers.
\end{itemize}

\noindent \revise{\textbf{Scope of This Work.} \tool is designed for post-detection analysis. We assume that an upstream NIDS or correlation engine has already identified the input as malicious, and \tool maps that traffic to MITRE ATT\&CK techniques. \tool is not a detector, and none of the results in this paper characterize its behavior on benign traffic, NIDS false positives, or inputs with insufficient evidence. All claims about \tool's capability are therefore limited to post-detection mapping.}

\section{Background and Motivation}

\begin{figure*}[ht!]
    \centering
    \includegraphics[width=0.95\textwidth]{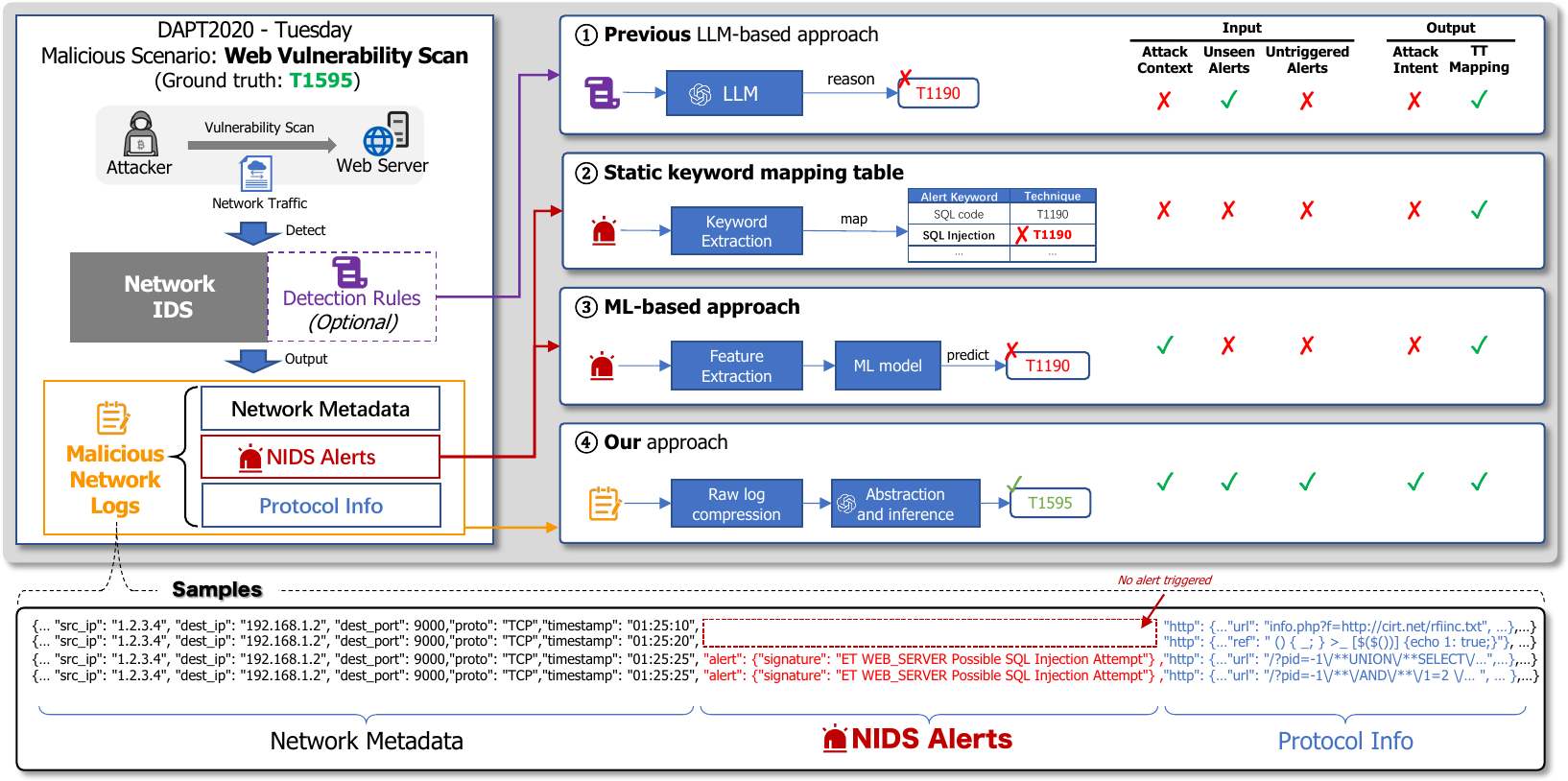}
    \caption{An example of attack-to-tactic/technique (TT) mapping for web server vulnerability scanning (in the DAPT2020-Tuesday dataset). The attacker probes for injection points, with traffic analyzed by a NIDS. Traditional methods (rule-based alerts, statistical keyword tables, or ML) and prior LLM-based approaches rely on alert text or rule parsing. In contrast, our method processes compressed attack context logs through LLM reasoning, \revise{enabling TT mapping from raw flow content rather than from alert signature text or rules. ``Untriggered alerts'' denotes records within the flagged malicious logs that carry no NIDS alert}.}
	\label{fig:example}
\end{figure*}

Accurate mapping of network logs to adversarial tactics and techniques (TTs) is a foundational capability for modern cybersecurity defenses. This process enables security teams to reconstruct attack kill chains and derive actionable threat intelligence, playing a critical role in incident response and mitigation strategies. However, operational challenges arise from the inherent limitations of Network Intrusion Detection Systems (NIDS). Enterprise environments routinely face over 10,000 daily alerts~\cite{jalalvand2024alert}, yet these alerts remain semantically fragmented—requiring manual correlation to high-level adversarial behaviors. Studies demonstrate that this manual process introduces error rates up to 30\%~\cite{strom2018mitre} and creates significant operational bottlenecks in Security Operations Centers (SOCs).

To illustrate these challenges, we analyze the DAPT2020 Tuesday dataset's web vulnerability scanning scenario (Figure~\ref{fig:example}). In this attack, adversaries employ automated tools to probe web servers for injection points (\eg SQL injection, command injection, remote file inclusion). The generated traffic undergoes NIDS analysis, producing malicious network logs comprising three key components: (1) network metadata (IP/port, timestamps, protocol types), (2) NIDS alerts (signature-based detections), and (3) protocol-specific information (\eg HTTP URLs, hostnames).

The transformation from raw logs to TTs involves two critical semantic transitions: first, converting low-level network traffic into attack intent, and second, mapping this intent to MITRE ATT\&CK techniques. Traditional approaches exhibit three key limitations:

\ding{172} \textit{LLM-based rule parsing}~\cite{daniel2023labeling,wudali2025rule} processes NIDS detection rules and alert signatures via LLM analysis. While flexible across rule sets, this method lacks contextual awareness from actual attack logs, \revise{failing to infer true intent when an alert's signature text is uninformative or absent}.

\ding{173} \textit{Static keyword mapping}~\cite{izzuddin2022mapping,meng2024poster} relies on pre-defined alert-to-TT lookup tables. This approach discards contextual signals and fails when encountering novel alert keywords.

\ding{174} \textit{Machine learning models}~\cite{hakim2024predicting, moskal2021translating} predict TTs from alert features but struggle with generalization—performance degrades for new alert signatures.

\textbf{Opportunities in Context-Aware Analysis.} Our key observation is that these approaches discard the most discriminative signal: the protocol-level attack context embedded in raw logs. For instance, an HTTP request containing \texttt{admin.php?cmd=rm+-rf+/} \revise{may raise only a generic alert whose signature text does not name the behavior,} yet clearly indicates command injection. By contrast, our method (Figure~\ref{fig:example}-\ding{175}) processes compressed log summaries—retaining critical protocol semantics while reducing data volume—and uses LLM reasoning to directly infer TTs. \revise{This reduces the mapping step's dependence on NIDS signature coverage: the mapping does not require a NIDS signature or alert text as input, but reasons over the traffic representation itself. We emphasize that this does not imply detection capability: \tool operates after detection, and traffic that does not trigger the NIDS is outside the scope of this work.}

\section{Approach Overview}

\begin{figure*}[!ht]
	\centering
	\includegraphics[scale=0.7]{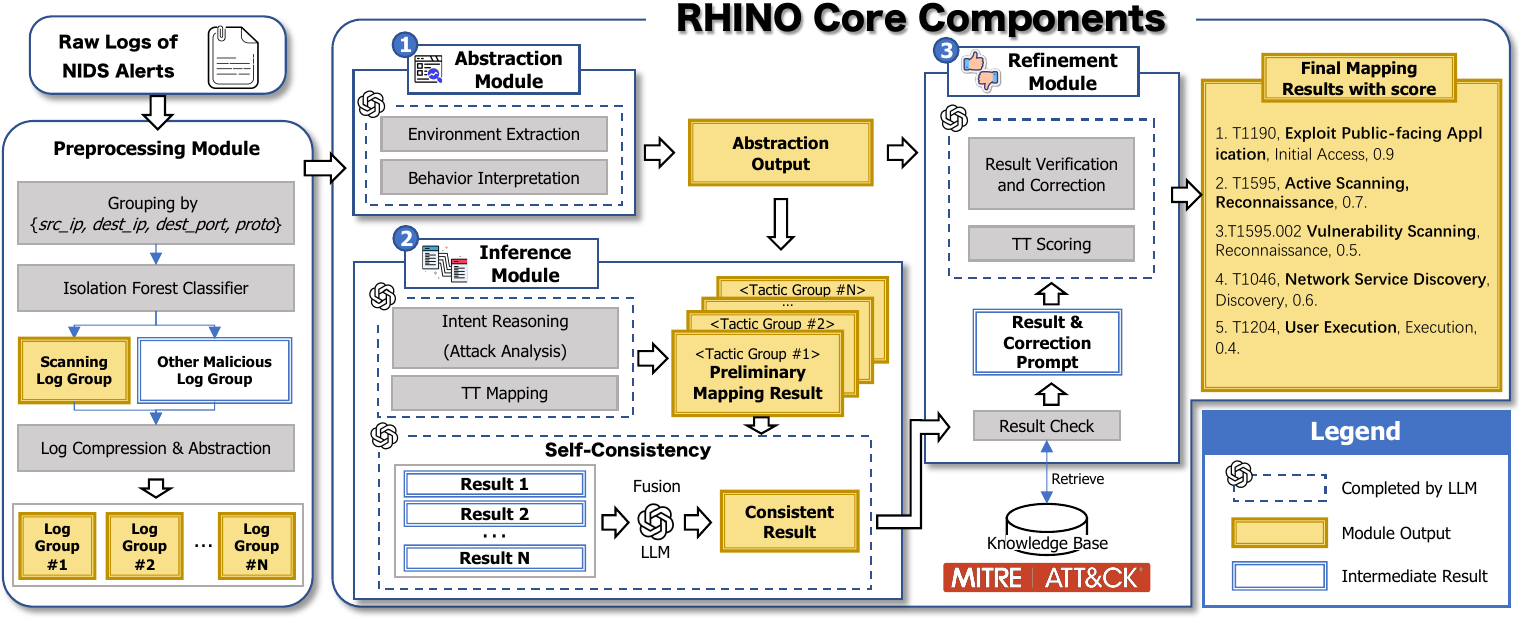}
    \caption{A detailed illustration of the \tool framework. When reasoning about attacks, we first compress raw NIDS logs to identify corresponding network behaviors. These behaviors are then mapped to attack tactics and techniques (TTs) through partitioned inference over the ATT\&CK tactic space. Finally, the mapped results are compared against standardized TT definitions in MITRE ATT\&CK to ensure accuracy and reduce any invalid TTs.}
	\label{fig:system_architecture}
\end{figure*}

Mapping raw NIDS logs to adversarial tactics and techniques (TTs) demands more than syntactic pattern matching; it requires a nuanced understanding of the behavioral intent and contextual signals embedded within fragmented, noisy observations. 
To bridge this semantic gap, \tool transforms unstructured NIDS logs into structured MITRE ATT\&CK mappings through a multi-stage reasoning pipeline. Figure~\ref{fig:system_architecture} illustrates the end-to-end reasoning process, which is divided into four modules, each addressing a critical challenge in LLM-based log analysis. \revise{\tool performs post-detection mapping. Its input is a traffic representation that an upstream NIDS or correlation engine has already identified as malicious; \tool's output is a ranked list of MITRE ATT\&CK tactic–technique candidates for that input.}

\noindent \textbf{Preprocessing and Context Preservation.} Raw NIDS logs are inherently verbose and repetitive, often exceeding LLM context windows while lacking semantic cohesion. The preprocessing module tackles this by aggregating logs into flow-level summaries, filtering low-information noise (\eg bulk port scans), and extracting protocol-specific features (\eg HTTP methods, SMB commands). This compression preserves behavioral signals—such as traffic directionality, packet asymmetries, and session timing—while reducing input volume by 83.5\% in our experiments. To mitigate information loss, application-layer fields are statistically sampled to retain representative examples without overwhelming the model's token budget.

\noindent \textbf{Behavioral Abstraction via Semantic Parsing.} The parsed flow summaries are then translated into natural language descriptions that capture adversarial behaviors (\eg ``rapid sequential SSH login failures from a single IP''). Unlike traditional classifiers that treat logs as isolated events, this stage explicitly models network context—such as source-destination relationships, protocol anomalies, and temporal patterns—to reconstruct attacker actions as coherent narratives. These descriptions serve as a bridge between low-level statistics and high-level reasoning, enabling the system to infer intent from incomplete or ambiguous observations.

\noindent \textbf{Partitioned Inference.} \tool maps abstracted behaviors to MITRE TTs through a partitioned reasoning strategy. First, the LLM hypothesizes the attacker's objective (\eg ``gaining initial access via credential brute-forcing'') from the behavioral narrative. The 14 ATT\&CK tactics are then divided into 5 semantically coherent groups, and the model performs an independent inference pass over each group to evaluate how well the observed behaviors align with the constituent techniques (\eg T1110: Brute Force). The candidate techniques produced across all groups are then consolidated into a unified mapping result.

\noindent \textbf{Knowledge-Grounded Refinement.} The initial TT candidates undergo a refinement step that grounds each candidate against MITRE ATT\&CK definitions retrieved from the official taxonomy. \revise{Immediately before this step, a code-level validation routine looks up each candidate in a local copy of the ATT\&CK taxonomy, attaches the official technique description, and annotates any detected error, which is either a non-existent technique identifier or a technique–tactic mismatch. The annotated candidates are then passed to the LLM, which is instructed to correct the flagged items and to score each candidate.} This stage reduces invalid predictions by \revise{4.9 percentage points (from 7.5\% to 2.6\%)} and produces a ranked list of candidate techniques.

\section{Design}

The core challenge in mapping NIDS logs to adversarial TTs lies in bridging the semantic gap between low-level observations (\eg FTP login attempts) and high-level attacker behaviors (\eg brute-force password guessing). Traditional approaches treat this as a classification task, either through rule-based pattern matching or supervised learning. However, these methods fail to capture the contextual and sequential reasoning that the mapping task requires. \tool addresses this limitation through a structured, multi-phase reasoning pipeline aligned with the structure of the ATT\&CK taxonomy, while leveraging the inferential capabilities of LLMs.

\subsection{Preprocessing Module}
Raw NIDS logs exhibit three properties that hinder direct LLM processing: \textit{volumetric redundancy} (thousands of similar entries per session), \textit{protocol-specific verbosity} (excessive protocol-specific metadata), and \textit{behavioral fragmentation} (attack patterns distributed across multiple flows). A typical credential brute-forcing attempt, for instance, may produce hundreds of FTP log entries documenting retransmissions, failed handshakes, and authentication attempts. Left unprocessed, this verbosity would quickly exhaust the context window of even the most capable language models while obscuring the underlying attack pattern. 

To address this, our preprocessing module implements a hierarchical compression strategy. NIDS logs are first grouped by their essential communication attributes—source and destination IP addresses, destination port, transport protocol, and application-layer service. This five-tuple grouping deliberately excludes ephemeral source ports, which typically carry no semantic value in attack scenarios. Each resulting flow group is then characterized by statistical features that capture its behavioral essence: packet and byte counts in both directions, session duration, and TCP flag distributions. This transformation reduces a 283,491-entry DDoS trace from CICIDS2017 to a 312-token representation, while retaining its characteristic pattern of numerous short-lived connections and asymmetric packet flows.

The module incorporates specialized handling for high-volume, low-information activities like port scanning. When a source IP contacts more than fifty distinct destinations—a strong indicator of scanning behavior—we employ an Isolation Forest model to separate repetitive probe traffic from potentially interesting outliers. This approach proved particularly effective in our evaluation, filtering out 13.9\% of flows in the DAPT2020 dataset without compromising attack visibility. For stateful protocols like HTTP, SMB, and FTP, we enrich the flow representations by extracting and strategically sampling application-layer fields (\eg HTTP methods, FTP commands). We further apply binomial sampling~\cite{lindley1957binomial} to preserve the semantic signal in fields like HTTP URIs while maintaining manageable input sizes.

\subsection{Abstraction Module}
\label{subsec:abstraction}
While the preprocessing module produces efficient numerical summaries, these representations remain inaccessible to the semantic reasoning capabilities of language models. The abstraction module bridges this gap by translating flow statistics into natural language narratives that capture the security-relevant aspects of network behavior.

This translation is guided by a carefully constructed prompt that instructs the language model to analyze the flow summary from a network-security perspective. Given a flow summary $f_i \in \mathcal{F}$, the language model $\mathcal{M}$ generates two complementary outputs: $a_i$ and $d_i$. The former is a structured enumeration of key attributes (IP addresses, ports, protocol features), while the latter is a free-form behavioral description (\eg \textit{``The traffic is directed from a source IP 172.16.0.1 to a destination IP within a private network (192.168.10.50)... with a high frequency of authentication attempts and consistent failures to authenticate ...''}). The description $d_i$ incorporates critical contextual cues relevant to security analysis—such as traffic directionality (\eg distinguishing client-initiated requests from server responses), temporal patterns (\eg burstiness or periodicity), and protocol-specific semantics (\eg interpreting sequences of SMB commands as potential share enumeration).

The dual-output design serves two complementary roles in the downstream pipeline. The structured metadata anchors the subsequent reasoning process in observable evidence, while the narrative description provides the rich contextual fabric needed for accurate intent inference.

\subsection{Inference Module}
\label{subsec:inference_module}
Mapping abstracted behaviors to MITRE ATT\&CK techniques presents unique challenges. The tactic-technique space is large and semantically nuanced, with many techniques exhibiting overlapping or context-dependent definitions. Our preliminary experiments revealed that conventional prompting approaches led to two persistent failure modes: hallucination of non-existent techniques (occurring in 4.9\% of candidate TTs with ChatGPT-4o) and misclassification due to over-reliance on lexical patterns rather than behavioral context.

The inference module addresses these issues through a partitioned reasoning strategy. The process begins with the language model generating the attacker's intent (\eg \textit{``The log data points to a brute force attack targeting the FTP service on the destination IP, with the attacker attempting to gain unauthorized access by guessing login credentials''}) based on the behavioral narrative.

Formally, let $\mathcal{B}_i = (a_i, d_i)$ denote a behavior description from the abstraction module. The inference module generates an attacker intent $I_i$ using the LLM $\mathcal{M}$ with an intent-focused prompt $p_{\text{intent}}$:

$$
I_i = \mathcal{M}(p_{\text{intent}}, \mathcal{B}_i)
$$

These intents then guide a structured exploration of the MITRE framework. We partition the 14 ATT\&CK tactics into five non-overlapping groups along the attack lifecycle: $G_1$, pre-compromise activity (Reconnaissance, Resource Development, Initial Access); $G_2$, post-compromise foothold (Execution, Persistence, Privilege Escalation, Defense Evasion); $G_3$, pre-lateral movement (Credential Access, Discovery); $G_4$, lateral movement and staging (Lateral Movement, Collection); and $G_5$, outbound action and impact (Command and Control, Exfiltration, Impact). This stage-based grouping reflects how tactics co-occur in practice. For instance, an adversary that has just gained execution typically pursues persistence, privilege escalation, and defense evasion in close succession, all of which fall within $G_2$. By focusing on these stages, the model can reason over a coherent slice of attacker behavior in each inference pass. The choice of five groups balances search-space reduction against token consumption, since finer partitioning increases the number of inference calls roughly linearly. The ablation in Section~\ref{sec:ablation} validates this design against alternative groupings (3 groups, 7 groups, random 5-group, and no partitioning).

For each tactic group $G_j$, the model evaluates how well the observed behaviors align with constituent techniques using prompt $p_{\text{tt}}$:

$$
\mathcal{Y}_i^{(j)} = \mathcal{M}(p_{\text{tt}}, \mathcal{B}_i, I_i, G_j) \quad \forall j \in [1,5]
$$

Here, $\mathcal{Y}_i^{(j)}$ is a list of triplets $(t, c, r)$, where $t$ denotes the technique, $c$ its corresponding tactic, and $r$ the reasoning justification. Conducting independent inference passes over each group prevents techniques from frequently referenced tactics from dominating the candidate set, surfacing less common techniques that monolithic inference tends to overlook.

Finally, the module consolidates all partial mappings using a fusion prompt $p_{\text{fusion}}$:

$$
\hat{\mathcal{Y}}_i = \mathcal{M}(p_{\text{fusion}}, \mathcal{B}_i, I_i, \{\mathcal{Y}_i^{(j)}\}_{j=1}^5)
$$

The output $\hat{\mathcal{Y}}_i = [(t_1, c_1, r_1), ..., (t_k, c_k, r_k)]$ with $k \leq 5$ is an unranked list of up to five tactic-technique pairs, each accompanied by a justification (\eg \textit{``T1110: The logs show a high number of failed authentication attempts using `USER' and `PASS' commands, which is indicative of a brute force attack.''}).

Our evaluation showed this approach improved technique prediction accuracy by at most 31.8\% compared to monolithic prompting strategies, while reducing hallucinated outputs to just 0.6\% of cases.

\subsection{Refinement Module}
The final component of our pipeline addresses a critical operational requirement: ensuring that all output mappings are not only plausible but also semantically grounded in both the observed data and official MITRE definitions. Even technically valid technique proposals may lack proper contextual fit—for instance, suggesting T1190 (Exploit Public-facing Application) for a sequence of requests that are targeting to discover potential injection points simply because the requests contain the SQL keyword.

The refinement module addresses two failure modes that we observed in the inference output: technique identifiers that do not correspond to any current MITRE ATT\&CK entry, and technique-tactic associations that violate the official taxonomy. 

Operationally, the module first retrieves the official MITRE ATT\&CK definition for each candidate technique in $\hat{\mathcal{Y}}_i$ from a local copy of the taxonomy. \revise{A code-level check attaches the retrieved description to each candidate and flags non-existent identifiers and tactics outside the technique's legitimate set. The annotated candidates, together with the original flow summary and the behavioral narrative, are then assembled into a refinement prompt $p_{\text{refine}}$ that asks the language model to correct the flagged candidates and to rank them by how well each matches the observed behavior.}

Formally, let $\mathcal{D}(\hat{\mathcal{Y}}_i)$ denote the set of retrieved official MITRE ATT\&CK definitions for each technique in $\hat{\mathcal{Y}}_i$. Using an LLM $\mathcal{M}$ with the refinement prompt $p_{\text{refine}}$, the refinement step is defined as:

$$
\mathcal{S}_i = \mathcal{M}(p_{\text{refine}}, \hat{\mathcal{Y}_i}, \mathcal{B}_i, f_i, \mathcal{D}(\hat{\mathcal{Y}}_i))
$$

The final output $\tilde{\mathcal{Y}}_i = [(t'_1, c'_1, r'_1), ..., (t'_k, c'_k, r'_k)]$ is a list of refined tactic-technique pairs with associated justifications, ranked as described above.
\section{Experiments}
\label{sec:experiments}

We present a comprehensive empirical assessment of \tool, designed to rigorously evaluate its effectiveness in mapping network behaviors to MITRE ATT\&CK techniques. The evaluation framework addresses six critical research dimensions through systematic experiments across diverse attack scenarios, comparative analyses with state-of-the-art baselines, and detailed component-level ablation studies. All experiments were conducted on a dedicated evaluation platform equipped with an Intel Core i5-13400 processor (16GB RAM), with each experimental condition repeated across three independent trials to ensure statistical reliability.

Specifically, to evaluate \tool, we focus on answering the following research questions:

\noindent \textbf{RQ1 (Performance)}: How does the performance of \tool compare with that of baseline methods?

\noindent \textbf{RQ2 (Ablation)}: How does each module within \tool contribute to the overall mapping performance?

\noindent \textbf{RQ3 (Error analysis)}: What are the common errors in this task, and how can \tool mitigate these errors to improve its overall performance?

\noindent \textbf{RQ4 (Practicality)}: Is \tool practical and effective in real-world mapping task?

\noindent \textbf{RQ5 (Contamination)}: Does \tool's accuracy depend on memorization of dataset-specific identifiers in the backbone LLMs?

\noindent \textbf{RQ6 (Sensitivity to Input Perturbations)}: Does \tool's mapping output remain stable when fields of its abstraction-module input are perturbed?

\begin{table*}[]
	\centering
	\caption{Distribution of attack categories of three datasets used in our experiment. DAPT2020~\cite{myneni2020dapt} includes simulated APT attack processes, covering multiple stages of the attack. CICIDS2017~\cite{sharafaldin2018toward} contains various types of network intrusion samples, focusing on traditional intrusion detection scenarios. IoT23~\cite{garcia2020iot23} provides various types of malicious traffic from IoT botnets.}
	\label{tab:dataset_distribution}
	\scalebox{0.77}{
		\begin{tabular}{|c|c|c|c|ccccclc|}
\hline
\textbf{Dataset} & \textbf{\begin{tabular}[c]{@{}c@{}}\# Malicious \\ Log lines\end{tabular}} & \textbf{\# IPs} & \textbf{\begin{tabular}[c]{@{}c@{}}Traffic\\ Duration (h)\end{tabular}} & \multicolumn{7}{c|}{\textbf{\# Flow Instances Corresponding to Each Attack Category}} \\ \hline
 &  &  &  & \cellcolor[HTML]{EFEFEF}Network Scan & \cellcolor[HTML]{EFEFEF}Account Discovery & \cellcolor[HTML]{EFEFEF}Directory Bruteforce & \cellcolor[HTML]{EFEFEF}Command Injection & \cellcolor[HTML]{EFEFEF}SQL Injection & \multicolumn{1}{c}{\cellcolor[HTML]{EFEFEF}Backdoor} & \cellcolor[HTML]{EFEFEF}CSRF \\
 &  &  &  & 7,653 & 133 & 9,968 & 12 & 55 & \multicolumn{1}{c}{20} & 7 \\
 &  &  &  & \cellcolor[HTML]{EFEFEF}Account Bruteforce & \cellcolor[HTML]{EFEFEF}Malware Download & \cellcolor[HTML]{EFEFEF}Web Vulnerability Scan & \cellcolor[HTML]{EFEFEF}Credential Access & \cellcolor[HTML]{EFEFEF}  & \cellcolor[HTML]{EFEFEF} & \cellcolor[HTML]{EFEFEF} \\
\multirow{-4}{*}{\textbf{DAPT2020}} & \multirow{-4}{*}{1,287,644} & \multirow{-4}{*}{28} & \multirow{-4}{*}{96} & 141 & 2 & 2,574 & 796 &  &  &  \\ \hline
 &  &  &  & \cellcolor[HTML]{EFEFEF}FTP-Patator & \cellcolor[HTML]{EFEFEF}SSH-Patator & \cellcolor[HTML]{EFEFEF}DoS/DDoS & \cellcolor[HTML]{EFEFEF}Heartbleed & \cellcolor[HTML]{EFEFEF}Web Attack & \multicolumn{1}{c}{\cellcolor[HTML]{EFEFEF}Portscan} & \cellcolor[HTML]{EFEFEF}Botnet \\
\multirow{-2}{*}{\textbf{CICIDS2017}} & \multirow{-2}{*}{1,073,698} & \multirow{-2}{*}{27} & \multirow{-2}{*}{96} & 3,972 & 2,961 & 266,778 & 11 & 104 & \multicolumn{1}{c}{159,066} & 736 \\ \hline
 &  &  &  & \cellcolor[HTML]{EFEFEF}DDoS & \cellcolor[HTML]{EFEFEF}C\&C & \cellcolor[HTML]{EFEFEF}PortScan & \cellcolor[HTML]{EFEFEF}FileDownload & \cellcolor[HTML]{EFEFEF}Attack & \cellcolor[HTML]{EFEFEF} & \cellcolor[HTML]{EFEFEF} \\
\multirow{-2}{*}{\textbf{IoT23}} & \multirow{-2}{*}{4,450,709} & \multirow{-2}{*}{1,684,627} & \multirow{-2}{*}{72} & 3,592,851 & 6,801 & 3,386,241 & 11 & 3,586 &  &  \\ \hline
\end{tabular}
	}
\end{table*}

\subsection{Evaluation Settings}
\label{sec:setup}

\noindent \textbf{Datasets.} The evaluation leverages three benchmark datasets (Table~\ref{tab:dataset_distribution}) representing distinct threat paradigms, each selected for their complementary coverage of modern attack vectors. The DAPT2020 dataset comprises 1.28 million malicious flows simulating advanced persistent threats in enterprise environments, with particular emphasis on multi-stage attack sequences including initial compromise through web vulnerability exploitation, lateral movement via SMB exploitation, and data exfiltration using FTP. This dataset's value lies in its realistic emulation of adversarial tradecraft.

For traditional network intrusion scenarios, we utilize the CICIDS2017 dataset's 1.07 million lines of malicious logs spanning seven attack categories across multiple protocols (HTTP/S, FTP, SSH). The dataset's comprehensive coverage of behavioral anomalies—ranging from volumetric DoS floods to subtle heartbleed exploits—provides a robust testbed for evaluating protocol-aware analysis capabilities. 

The IoT23 dataset's 4.45 million lines of logs from IoT botnet infections (Mirai, Gafgyt variants) were selected to evaluate performance on constrained-device attack patterns. We focused on three representative scenarios (corresponding to 34-1, 48-1, and 60-1) demonstrating complete kill-chains from device compromise through DDoS payload execution, with special consideration given to the characteristic irregular beaconing intervals and high-volume attack traffic peculiar to IoT threats.

\revise{Ground-truth labels were obtained by having two annotators with network security expertise independently map each scenario label to a primary MITRE ATT\&CK technique, with explicit constraints to ensure network-layer observability. The initial independent annotations achieved a Cohen's kappa of $\kappa = 0.822$ (almost perfect agreement~\cite{landis1977kappa}), and remaining disagreements were resolved through adjudication. Each attack label is mapped to a single primary technique, reflecting the technique that the scenario documentation most directly attributes to the observed network behavior; this yields a deterministic notion of correctness and aligns with the granularity at which existing public datasets are labeled.}

\revise{\textbf{Evaluation Formulation.} Our evaluation targets a specific question: given a network flow summary produced from a documented attack scenario, how accurately does \tool recover the attack technique that the scenario documentation attributes to that behavior? The ground-truth label for each flow is therefore derived from the official documentation of the dataset's attack scenario, not from an attempt to enumerate every technique that an isolated flow could conceivably support. In reality, a single flow may be consistent with multiple techniques when judged on evidence alone; our formulation deliberately fixes the scenario-documented technique as the evaluation target so that correctness is well defined and comparable across methods. Consequently, a prediction of an equally plausible alternative technique that omits the scenario-documented label is counted as incorrect.}

\revise{\textbf{Scope of Conclusions.} The 44 attack categories across the three datasets ultimately cover 11 distinct techniques. All quantitative conclusions in this section, including the comparison against baselines and the ablation results, are therefore limited to this technique set and to the scenario-documented single-label formulation. We provide a representative evidence-only multi-label subset analysis in Section~V of the supplementary material.}

\noindent \textbf{Baselines.} We compare \tool against five baselines. All methods process the same abstracted behavioral narratives (output of the Abstraction Module, Section~\ref{subsec:abstraction}) and produce a ranked list of up to five candidate techniques.

\textsc{Vanilla} prompting serves as our minimal baseline using minimal instructions. \textsc{Chain-of-Thought} (CoT) enforces explicit reasoning steps before prediction. \textsc{Tree-of-Thought} (ToT) follows Yao et al.~\cite{yao2023tree} with three parallel reasoning paths and majority-vote aggregation. \textsc{RAG-Chain} implements a standard retrieve-then-read pipeline~\cite{lewis2020retrieval}: it encodes the abstraction output and all ATT\&CK technique definitions using a fixed embedding model, retrieves the top-10 candidates by cosine similarity, and prompts the LLM to select the final top-5. \textsc{Few-Shot} prompting~\cite{brown2020language} provides one labeled exemplar per ground-truth technique class from each dataset (seed=42). Unlike all other baselines operating in a zero-shot setting, \textsc{Few-Shot} requires per-deployment labeled exemplars; we report its results in italics in Table~\ref{tab:main_results} and discuss the protocol distinction separately.

We additionally evaluated \textsc{SecureBERT}~\cite{aghaei2022securebert} by computing the cosine similarity between the abstraction output and each ATT\&CK technique definition. \textsc{SecureBERT} achieved 0\% top-1 accuracy across all datasets, with top-5 accuracy ranging from 0.22\% to 8.20\%. Manual inspection of the embedding space revealed that distinct technique definitions produced nearly identical representations (cosine similarity between 0.95 and 0.96), indicating insufficient discriminative capacity for technique-level retrieval. Consequently, we omit \textsc{SecureBERT} from subsequent quantitative comparisons.
	
We exclude rule- and alert-based LLM mapping methods, including LNR~\cite{daniel2023labeling} and RAM~\cite{wudali2025rule}, from our primary comparison. These methods operate on Snort rules or SIEM-derived alerts—inputs that have already undergone signature-level abstraction by the detection system. In contrast, \tool processes raw NIDS logs directly; a direct comparison would therefore confound methodological differences with input modality discrepancies.

\noindent \textbf{Metrics.} We report both \textit{micro-averaged} and \textit{macro-averaged} top-$K$ accuracy ($K=1,3,5$). Micro-averaged accuracy weights each flow equally, reflecting deployment-relevant performance; macro-averaged accuracy weights each attack category equally, exposing per-category capability.

Formally, let $T$ denote the number of attack categories in a dataset, $n_s$ the number of flow samples in category $s$, and $n = \sum_{s=1}^{T} n_s$ the total number of malicious flows. For category $s$ with samples $\{(g_i, y_i)\}_{i=1}^{n_s}$, where $g_i$ is the ground-truth label and $\{y_i^{(1)}, \ldots, y_i^{(K)}\}$ are the top-$K$ predictions, the per-category accuracy at $K$ is
$$
\text{Accuracy}_s(K) = \frac{1}{n_s} \sum_{i=1}^{n_s} \mathds{1}\!\left( g_i \in \{ y_i^{(1)}, \ldots, y_i^{(K)} \} \right).
$$
Micro-averaged and macro-averaged accuracy are then defined as
$$
\text{Accuracy}_{\text{micro}}(K) = \sum_{s=1}^{T} \frac{n_s}{n} \cdot \text{Accuracy}_s(K),
$$

$$
\text{Accuracy}_{\text{macro}}(K) = \frac{1}{T} \sum_{s=1}^{T} \text{Accuracy}_s(K).
$$

\revise{We note that top-\textit{K} accuracy measures whether the scenario-documented technique is retrieved among these candidates; it does not substitute for multi-label annotation, and a prediction of an equally plausible alternative technique that omits the scenario-documented label is counted as incorrect.}

Beyond conventional top-\textit{K} accuracy metrics, we introduce two specialized measures to address specific evaluation requirements. \textit{Tactical Consistency} evaluates whether predicted techniques properly align with ground-truth tactical phases (\eg ensuring ``T1046: Network Service Discovery'' maps to the ``Discovery'' tactic). The \textit{Class-wise F1} metric provides per-technique performance analysis to identify strengths and weaknesses across ATT\&CK's heterogeneous technique landscape.

\noindent \textbf{Implementation Details.} The Isolation Forest filter is implemented with scikit-learn, configured with n\_estimators=100, contamination=0.8. The binomial sampling step uses a 95\% confidence level with margin of error $e=0.1$ and samples occurrences uniformly at random. For LLM inference, we set temperature=0 and use default values for the remaining decoding parameters. The model snapshots are: ChatGPT-4o (gpt-4o-20240806), Gemini 2.5 Flash (gemini-2.5-flash), Claude Sonnet 4 (claude-sonnet-4-0), DeepSeek V3 (deepseek-chat), and Llama-3.3-70B-Instruct served via vLLM on a single A100 80GB GPU.

\begin{table*}[!ht]
\centering
\caption{Comparison of Micro-Averaged Accuracy across DAPT2020 (D), CICIDS2017 (C), and IoT23 (I) datasets. Accuracy (\%) is reported for Top-1, Top-3, and Top-5 predictions at both Technique and Tactic levels. Each cell shows mean\,$\pm$\,std over three independent trials. \textbf{Bold} and \underline{underlined} values denote the best and second-best results within the zero-shot methods (Ours, CoT, ToT, Vanilla, RAG) for each (backbone, dataset, Top-k) cell. \textit{Few-Shot (FS\textsuperscript{\dag})} uses in-distribution labeled exemplars and is reported in italics; it is not included in the bold/underline comparison.}
\label{tab:main_results}
\resizebox{\textwidth}{!}{%
\renewcommand{\arraystretch}{1.05}
\setlength{\tabcolsep}{3pt}
\begin{tabular}{|c|cccccccccccccccccccc|}
\hline
 & \multicolumn{1}{c|}{} & \multicolumn{1}{c|}{} & \multicolumn{6}{c|}{Top-1} & \multicolumn{6}{c|}{Top-3} & \multicolumn{6}{c|}{Top-5} \\
\multirow{-2}{*}{} & \multicolumn{1}{c|}{\multirow{-2}{*}{Model}} & \multicolumn{1}{c|}{\multirow{-2}{*}{DS}} & Ours & CoT & ToT & Vanilla & RAG & \multicolumn{1}{c|}{\textit{FS\textsuperscript{\dag}}} & Ours & CoT & ToT & Vanilla & RAG & \multicolumn{1}{c|}{\textit{FS\textsuperscript{\dag}}} & Ours & CoT & ToT & Vanilla & RAG & \textit{FS\textsuperscript{\dag}} \\ \hline
 & \multicolumn{20}{c|}{\textbf{Technique-level}} \\ \cline{2-21} 
 & \multicolumn{1}{c|}{} & \multicolumn{1}{c|}{D} & \textbf{68.00}$_{\pm2.52}$ & \underline{60.14}$_{\pm21.92}$ & 29.10$_{\pm6.05}$ & 59.91$_{\pm0.24}$ & \multicolumn{1}{c|}{11.97$_{\pm0.05}$} & \multicolumn{1}{c|}{\textit{59.72}$_{\pm0.01}$} & \textbf{88.33}$_{\pm8.47}$ & 62.96$_{\pm22.83}$ & \underline{73.81}$_{\pm6.08}$ & 63.68$_{\pm2.42}$ & \multicolumn{1}{c|}{12.16$_{\pm0.00}$} & \multicolumn{1}{c|}{\textit{82.04}$_{\pm0.34}$} & \textbf{97.57}$_{\pm0.55}$ & 63.08$_{\pm22.92}$ & \underline{73.94}$_{\pm6.11}$ & 64.22$_{\pm3.12}$ & \multicolumn{1}{c|}{12.17$_{\pm0.00}$} & \textit{84.45}$_{\pm0.59}$ \\
 & \multicolumn{1}{c|}{} & \multicolumn{1}{c|}{C} & \underline{98.54}$_{\pm0.69}$ & 2.61$_{\pm0.60}$ & 46.11$_{\pm0.75}$ & 2.91$_{\pm0.07}$ & \multicolumn{1}{c|}{\textbf{99.41}$_{\pm0.00}$} & \multicolumn{1}{c|}{\textit{99.49}$_{\pm0.01}$} & \underline{99.56}$_{\pm0.06}$ & 14.18$_{\pm19.90}$ & 46.17$_{\pm0.75}$ & 14.36$_{\pm19.79}$ & \multicolumn{1}{c|}{\textbf{99.60}$_{\pm0.14}$} & \multicolumn{1}{c|}{\textit{99.70}$_{\pm0.03}$} & \textbf{99.62}$_{\pm0.00}$ & 25.94$_{\pm19.79}$ & 69.62$_{\pm26.33}$ & 14.49$_{\pm19.67}$ & \multicolumn{1}{c|}{\underline{99.60}$_{\pm0.14}$} & \textit{99.82}$_{\pm0.08}$ \\
 & \multicolumn{1}{c|}{} & \multicolumn{1}{c|}{I} & 97.90$_{\pm3.14}$ & 99.35$_{\pm0.16}$ & \textbf{99.55}$_{\pm0.10}$ & 5.98$_{\pm0.09}$ & \multicolumn{1}{c|}{\underline{99.49}$_{\pm0.00}$} & \multicolumn{1}{c|}{\textit{99.87}$_{\pm0.00}$} & \underline{99.76}$_{\pm0.08}$ & \textbf{99.96}$_{\pm0.01}$ & 99.71$_{\pm0.10}$ & 5.98$_{\pm0.09}$ & \multicolumn{1}{c|}{99.49$_{\pm0.00}$} & \multicolumn{1}{c|}{\textit{99.87}$_{\pm0.00}$} & 99.81$_{\pm0.15}$ & \textbf{99.97}$_{\pm0.00}$ & \underline{99.91}$_{\pm0.10}$ & 6.09$_{\pm0.02}$ & \multicolumn{1}{c|}{99.61$_{\pm0.00}$} & \textit{99.99}$_{\pm0.00}$ \\
 & \multicolumn{1}{c|}{\multirow{-4}{*}{GPT}} & \multicolumn{1}{c|}{\cellcolor[HTML]{EFEFEF}Avg} & \cellcolor[HTML]{EFEFEF}\textbf{88.15} & \cellcolor[HTML]{EFEFEF}54.03 & \cellcolor[HTML]{EFEFEF}58.25 & \cellcolor[HTML]{EFEFEF}22.93 & \multicolumn{1}{c|}{\cellcolor[HTML]{EFEFEF}{\ul 70.29}} & \multicolumn{1}{c|}{\cellcolor[HTML]{EFEFEF}\textit{86.36}} & \cellcolor[HTML]{EFEFEF}\textbf{95.88} & \cellcolor[HTML]{EFEFEF}59.03 & \cellcolor[HTML]{EFEFEF}{\ul 73.23} & \cellcolor[HTML]{EFEFEF}28.01 & \multicolumn{1}{c|}{\cellcolor[HTML]{EFEFEF}70.42} & \multicolumn{1}{c|}{\cellcolor[HTML]{EFEFEF}\textit{93.87}} & \cellcolor[HTML]{EFEFEF}\textbf{99.00} & \cellcolor[HTML]{EFEFEF}63.00 & \cellcolor[HTML]{EFEFEF}{\ul 81.16} & \cellcolor[HTML]{EFEFEF}28.27 & \multicolumn{1}{c|}{\cellcolor[HTML]{EFEFEF}70.46} & \cellcolor[HTML]{EFEFEF}\textit{94.75} \\ \cline{2-21} 
 & \multicolumn{1}{c|}{} & \multicolumn{1}{c|}{D} & \textbf{81.66}$_{\pm12.25}$ & 60.60$_{\pm0.37}$ & 44.94$_{\pm19.51}$ & \underline{66.75}$_{\pm0.05}$ & \multicolumn{1}{c|}{11.92$_{\pm0.00}$} & \multicolumn{1}{c|}{\textit{59.56}$_{\pm0.02}$} & \textbf{94.47}$_{\pm4.92}$ & 78.73$_{\pm4.81}$ & 79.89$_{\pm5.06}$ & \underline{81.83}$_{\pm0.59}$ & \multicolumn{1}{c|}{11.97$_{\pm0.00}$} & \multicolumn{1}{c|}{\textit{96.35}$_{\pm3.03}$} & \textbf{94.68}$_{\pm5.04}$ & 79.24$_{\pm4.75}$ & \underline{84.79}$_{\pm2.93}$ & 81.94$_{\pm0.61}$ & \multicolumn{1}{c|}{11.97$_{\pm0.00}$} & \textit{98.74}$_{\pm0.05}$ \\
 & \multicolumn{1}{c|}{} & \multicolumn{1}{c|}{C} & \underline{81.43}$_{\pm30.25}$ & 72.55$_{\pm22.30}$ & 63.90$_{\pm0.23}$ & 2.93$_{\pm0.03}$ & \multicolumn{1}{c|}{\textbf{98.34}$_{\pm0.00}$} & \multicolumn{1}{c|}{\textit{99.52}$_{\pm0.00}$} & \textbf{99.23}$_{\pm0.75}$ & 95.46$_{\pm4.49}$ & 64.27$_{\pm0.01}$ & 45.28$_{\pm0.26}$ & \multicolumn{1}{c|}{\underline{98.45}$_{\pm0.14}$} & \multicolumn{1}{c|}{\textit{99.86}$_{\pm0.01}$} & \textbf{99.29}$_{\pm0.73}$ & 96.30$_{\pm5.20}$ & 64.69$_{\pm0.72}$ & \underline{98.90}$_{\pm0.71}$ & \multicolumn{1}{c|}{98.61$_{\pm0.00}$} & \textit{99.93}$_{\pm0.00}$ \\
 & \multicolumn{1}{c|}{} & \multicolumn{1}{c|}{I} & 96.10$_{\pm3.14}$ & 37.22$_{\pm54.15}$ & \textbf{99.79}$_{\pm0.01}$ & \underline{99.60}$_{\pm0.08}$ & \multicolumn{1}{c|}{99.49$_{\pm0.00}$} & \multicolumn{1}{c|}{\textit{99.87}$_{\pm0.00}$} & 98.06$_{\pm3.14}$ & \underline{99.89}$_{\pm0.09}$ & \textbf{99.94}$_{\pm0.00}$ & 99.79$_{\pm0.08}$ & \multicolumn{1}{c|}{99.61$_{\pm0.00}$} & \multicolumn{1}{c|}{\textit{99.87}$_{\pm0.00}$} & 98.14$_{\pm3.20}$ & \underline{99.90}$_{\pm0.09}$ & \textbf{99.95}$_{\pm0.00}$ & 99.84$_{\pm0.01}$ & \multicolumn{1}{c|}{99.61$_{\pm0.00}$} & \textit{99.87}$_{\pm0.00}$ \\
 & \multicolumn{1}{c|}{\multirow{-4}{*}{Claude}} & \multicolumn{1}{c|}{\cellcolor[HTML]{EFEFEF}Avg} & \cellcolor[HTML]{EFEFEF}\textbf{86.40} & \cellcolor[HTML]{EFEFEF}56.79 & \cellcolor[HTML]{EFEFEF}69.54 & \cellcolor[HTML]{EFEFEF}56.43 & \multicolumn{1}{c|}{\cellcolor[HTML]{EFEFEF}{\ul 69.92}} & \multicolumn{1}{c|}{\cellcolor[HTML]{EFEFEF}\textit{86.32}} & \cellcolor[HTML]{EFEFEF}\textbf{97.25} & \cellcolor[HTML]{EFEFEF}{\ul 91.36} & \cellcolor[HTML]{EFEFEF}81.37 & \cellcolor[HTML]{EFEFEF}75.63 & \multicolumn{1}{c|}{\cellcolor[HTML]{EFEFEF}70.01} & \multicolumn{1}{c|}{\cellcolor[HTML]{EFEFEF}\textit{98.69}} & \cellcolor[HTML]{EFEFEF}\textbf{97.37} & \cellcolor[HTML]{EFEFEF}91.81 & \cellcolor[HTML]{EFEFEF}83.14 & \cellcolor[HTML]{EFEFEF}{\ul 93.56} & \multicolumn{1}{c|}{\cellcolor[HTML]{EFEFEF}70.06} & \cellcolor[HTML]{EFEFEF}\textit{99.51} \\ \cline{2-21} 
 & \multicolumn{1}{c|}{} & \multicolumn{1}{c|}{D} & 69.04$_{\pm17.98}$ & \textbf{82.47}$_{\pm4.70}$ & \underline{77.44}$_{\pm13.49}$ & 71.91$_{\pm0.37}$ & \multicolumn{1}{c|}{11.92$_{\pm0.00}$} & \multicolumn{1}{c|}{\textit{65.87}$_{\pm0.01}$} & \textbf{95.17}$_{\pm3.31}$ & 85.94$_{\pm8.20}$ & \underline{93.15}$_{\pm4.07}$ & 72.48$_{\pm0.37}$ & \multicolumn{1}{c|}{11.97$_{\pm0.00}$} & \multicolumn{1}{c|}{\textit{97.97}$_{\pm0.00}$} & \textbf{95.66}$_{\pm3.03}$ & 88.92$_{\pm8.03}$ & \underline{93.94}$_{\pm3.93}$ & 75.64$_{\pm4.61}$ & \multicolumn{1}{c|}{11.97$_{\pm0.00}$} & \textit{98.47}$_{\pm0.00}$ \\
 & \multicolumn{1}{c|}{} & \multicolumn{1}{c|}{C} & \underline{96.52}$_{\pm5.13}$ & 25.72$_{\pm19.71}$ & 67.37$_{\pm19.73}$ & 2.93$_{\pm0.01}$ & \multicolumn{1}{c|}{\textbf{98.27}$_{\pm0.00}$} & \multicolumn{1}{c|}{\textit{99.52}$_{\pm0.00}$} & \textbf{99.25}$_{\pm0.73}$ & 55.73$_{\pm29.90}$ & 67.49$_{\pm19.73}$ & 20.71$_{\pm30.63}$ & \multicolumn{1}{c|}{\underline{98.37}$_{\pm0.00}$} & \multicolumn{1}{c|}{\textit{99.74}$_{\pm0.00}$} & \textbf{99.26}$_{\pm0.73}$ & 55.75$_{\pm29.91}$ & 90.27$_{\pm0.01}$ & 56.08$_{\pm0.01}$ & \multicolumn{1}{c|}{\underline{98.62}$_{\pm0.00}$} & \textit{99.79}$_{\pm0.00}$ \\
 & \multicolumn{1}{c|}{} & \multicolumn{1}{c|}{I} & \textbf{99.79}$_{\pm0.07}$ & 5.63$_{\pm0.16}$ & 5.93$_{\pm0.00}$ & 5.49$_{\pm0.08}$ & \multicolumn{1}{c|}{\underline{99.49}$_{\pm0.00}$} & \multicolumn{1}{c|}{\textit{99.87}$_{\pm0.00}$} & \textbf{99.79}$_{\pm0.07}$ & 5.82$_{\pm0.08}$ & 5.94$_{\pm0.00}$ & 68.16$_{\pm54.01}$ & \multicolumn{1}{c|}{\underline{99.49}$_{\pm0.00}$} & \multicolumn{1}{c|}{\textit{99.99}$_{\pm0.00}$} & \textbf{99.83}$_{\pm0.00}$ & 5.82$_{\pm0.08}$ & 5.96$_{\pm0.00}$ & 99.50$_{\pm0.17}$ & \multicolumn{1}{c|}{\underline{99.61}$_{\pm0.00}$} & \textit{99.99}$_{\pm0.00}$ \\
 & \multicolumn{1}{c|}{\multirow{-4}{*}{Gemini}} & \multicolumn{1}{c|}{\cellcolor[HTML]{EFEFEF}Avg} & \cellcolor[HTML]{EFEFEF}\textbf{88.45} & \cellcolor[HTML]{EFEFEF}37.94 & \cellcolor[HTML]{EFEFEF}50.25 & \cellcolor[HTML]{EFEFEF}26.78 & \multicolumn{1}{c|}{\cellcolor[HTML]{EFEFEF}{\ul 69.89}} & \multicolumn{1}{c|}{\cellcolor[HTML]{EFEFEF}\textit{88.42}} & \cellcolor[HTML]{EFEFEF}\textbf{98.07} & \cellcolor[HTML]{EFEFEF}49.16 & \cellcolor[HTML]{EFEFEF}55.53 & \cellcolor[HTML]{EFEFEF}53.78 & \multicolumn{1}{c|}{\cellcolor[HTML]{EFEFEF}{\ul 69.94}} & \multicolumn{1}{c|}{\cellcolor[HTML]{EFEFEF}\textit{99.23}} & \cellcolor[HTML]{EFEFEF}\textbf{98.25} & \cellcolor[HTML]{EFEFEF}50.16 & \cellcolor[HTML]{EFEFEF}63.39 & \cellcolor[HTML]{EFEFEF}{\ul 77.07} & \multicolumn{1}{c|}{\cellcolor[HTML]{EFEFEF}70.07} & \cellcolor[HTML]{EFEFEF}\textit{99.42} \\ \cline{2-21} 
 & \multicolumn{1}{c|}{} & \multicolumn{1}{c|}{D} & \textbf{65.17}$_{\pm5.21}$ & \underline{52.56}$_{\pm23.21}$ & 40.34$_{\pm26.69}$ & 52.16$_{\pm2.42}$ & \multicolumn{1}{c|}{19.41$_{\pm6.61}$} & \multicolumn{1}{c|}{\textit{67.79}$_{\pm0.05}$} & \underline{79.77}$_{\pm4.90}$ & 76.42$_{\pm8.43}$ & \textbf{80.93}$_{\pm3.16}$ & 57.67$_{\pm0.39}$ & \multicolumn{1}{c|}{21.84$_{\pm8.77}$} & \multicolumn{1}{c|}{\textit{94.90}$_{\pm4.74}$} & \underline{80.18}$_{\pm5.05}$ & 79.76$_{\pm7.53}$ & \textbf{84.89}$_{\pm3.60}$ & 57.87$_{\pm0.40}$ & \multicolumn{1}{c|}{22.02$_{\pm9.04}$} & \textit{98.75}$_{\pm0.03}$ \\
 & \multicolumn{1}{c|}{} & \multicolumn{1}{c|}{C} & \underline{94.21}$_{\pm4.21}$ & 39.29$_{\pm31.07}$ & 39.13$_{\pm31.36}$ & 5.12$_{\pm4.83}$ & \multicolumn{1}{c|}{\textbf{99.41}$_{\pm0.00}$} & \multicolumn{1}{c|}{\textit{99.48}$_{\pm0.00}$} & \textbf{99.63}$_{\pm0.02}$ & 55.94$_{\pm44.05}$ & 47.03$_{\pm31.35}$ & 5.69$_{\pm4.37}$ & \multicolumn{1}{c|}{\underline{99.60}$_{\pm0.14}$} & \multicolumn{1}{c|}{\textit{99.72}$_{\pm0.00}$} & \textbf{99.73}$_{\pm0.17}$ & 56.09$_{\pm44.24}$ & 58.41$_{\pm11.62}$ & 17.16$_{\pm18.05}$ & \multicolumn{1}{c|}{\underline{99.69}$_{\pm0.00}$} & \textit{99.92}$_{\pm0.01}$ \\
 & \multicolumn{1}{c|}{} & \multicolumn{1}{c|}{I} & \textbf{99.76}$_{\pm0.07}$ & 5.68$_{\pm0.09}$ & 68.34$_{\pm54.22}$ & 5.62$_{\pm0.08}$ & \multicolumn{1}{c|}{\underline{97.69}$_{\pm3.14}$} & \multicolumn{1}{c|}{\textit{99.87}$_{\pm0.00}$} & \textbf{99.90}$_{\pm0.15}$ & 5.91$_{\pm0.02}$ & \underline{99.77}$_{\pm0.07}$ & 5.63$_{\pm0.07}$ & \multicolumn{1}{c|}{99.50$_{\pm0.01}$} & \multicolumn{1}{c|}{\textit{99.87}$_{\pm0.00}$} & \textbf{99.99}$_{\pm0.01}$ & 5.96$_{\pm0.08}$ & \underline{99.82}$_{\pm0.14}$ & 37.02$_{\pm53.98}$ & \multicolumn{1}{c|}{99.55$_{\pm0.07}$} & \textit{99.87}$_{\pm0.00}$ \\
 & \multicolumn{1}{c|}{\multirow{-4}{*}{DeepSeek}} & \multicolumn{1}{c|}{\cellcolor[HTML]{EFEFEF}Avg} & \cellcolor[HTML]{EFEFEF}\textbf{86.38} & \cellcolor[HTML]{EFEFEF}32.51 & \cellcolor[HTML]{EFEFEF}49.27 & \cellcolor[HTML]{EFEFEF}20.97 & \multicolumn{1}{c|}{\cellcolor[HTML]{EFEFEF}{\ul 72.17}} & \multicolumn{1}{c|}{\cellcolor[HTML]{EFEFEF}\textit{89.05}} & \cellcolor[HTML]{EFEFEF}\textbf{93.10} & \cellcolor[HTML]{EFEFEF}46.09 & \cellcolor[HTML]{EFEFEF}{\ul 75.91} & \cellcolor[HTML]{EFEFEF}23.00 & \multicolumn{1}{c|}{\cellcolor[HTML]{EFEFEF}73.65} & \multicolumn{1}{c|}{\cellcolor[HTML]{EFEFEF}\textit{98.16}} & \cellcolor[HTML]{EFEFEF}\textbf{93.30} & \cellcolor[HTML]{EFEFEF}47.27 & \cellcolor[HTML]{EFEFEF}{\ul 81.04} & \cellcolor[HTML]{EFEFEF}37.35 & \multicolumn{1}{c|}{\cellcolor[HTML]{EFEFEF}73.75} & \cellcolor[HTML]{EFEFEF}\textit{99.51} \\ \cline{2-21} 
 & \multicolumn{20}{c|}{\textbf{Tactic-level}} \\ \cline{2-21} 
 & \multicolumn{1}{c|}{} & \multicolumn{1}{c|}{D} & \textbf{68.00}$_{\pm2.52}$ & \underline{60.42}$_{\pm21.69}$ & 32.59$_{\pm7.17}$ & 47.17$_{\pm22.70}$ & \multicolumn{1}{c|}{13.78$_{\pm0.05}$} & \multicolumn{1}{c|}{\textit{59.73}$_{\pm0.02}$} & \textbf{88.40}$_{\pm8.49}$ & 63.27$_{\pm22.44}$ & \underline{75.27}$_{\pm5.70}$ & 61.94$_{\pm1.28}$ & \multicolumn{1}{c|}{14.12$_{\pm0.01}$} & \multicolumn{1}{c|}{\textit{82.06}$_{\pm0.34}$} & \textbf{97.80}$_{\pm0.38}$ & 63.65$_{\pm22.73}$ & \underline{75.44}$_{\pm5.72}$ & 62.13$_{\pm1.31}$ & \multicolumn{1}{c|}{14.14$_{\pm0.01}$} & \textit{84.53}$_{\pm0.66}$ \\
 & \multicolumn{1}{c|}{} & \multicolumn{1}{c|}{C} & \underline{99.37}$_{\pm0.05}$ & 2.61$_{\pm0.60}$ & 46.14$_{\pm0.72}$ & 2.91$_{\pm0.07}$ & \multicolumn{1}{c|}{\textbf{99.47}$_{\pm0.01}$} & \multicolumn{1}{c|}{\textit{99.49}$_{\pm0.01}$} & \underline{99.56}$_{\pm0.06}$ & 22.09$_{\pm20.14}$ & 46.18$_{\pm0.70}$ & 14.49$_{\pm20.03}$ & \multicolumn{1}{c|}{\textbf{99.66}$_{\pm0.14}$} & \multicolumn{1}{c|}{\textit{99.73}$_{\pm0.03}$} & \underline{99.62}$_{\pm0.01}$ & 33.85$_{\pm19.91}$ & 63.87$_{\pm29.91}$ & 22.40$_{\pm19.91}$ & \multicolumn{1}{c|}{\textbf{99.68}$_{\pm0.14}$} & \textit{99.84}$_{\pm0.08}$ \\
 & \multicolumn{1}{c|}{} & \multicolumn{1}{c|}{I} & 97.90$_{\pm3.14}$ & 99.35$_{\pm0.15}$ & \textbf{99.56}$_{\pm0.11}$ & 5.98$_{\pm0.09}$ & \multicolumn{1}{c|}{\underline{99.49}$_{\pm0.00}$} & \multicolumn{1}{c|}{\textit{99.87}$_{\pm0.00}$} & 99.77$_{\pm0.08}$ & \textbf{99.98}$_{\pm0.01}$ & \underline{99.80}$_{\pm0.07}$ & 6.00$_{\pm0.08}$ & \multicolumn{1}{c|}{99.51$_{\pm0.00}$} & \multicolumn{1}{c|}{\textit{99.87}$_{\pm0.00}$} & 99.77$_{\pm0.08}$ & \textbf{99.98}$_{\pm0.01}$ & \underline{99.94}$_{\pm0.06}$ & 6.10$_{\pm0.01}$ & \multicolumn{1}{c|}{99.63$_{\pm0.00}$} & \textit{99.99}$_{\pm0.00}$ \\
 & \multicolumn{1}{c|}{\multirow{-4}{*}{GPT}} & \multicolumn{1}{c|}{\cellcolor[HTML]{EFEFEF}Avg} & \cellcolor[HTML]{EFEFEF}\textbf{88.42} & \cellcolor[HTML]{EFEFEF}54.13 & \cellcolor[HTML]{EFEFEF}59.43 & \cellcolor[HTML]{EFEFEF}18.69 & \multicolumn{1}{c|}{\cellcolor[HTML]{EFEFEF}{\ul 70.91}} & \multicolumn{1}{c|}{\cellcolor[HTML]{EFEFEF}\textit{86.36}} & \cellcolor[HTML]{EFEFEF}\textbf{95.91} & \cellcolor[HTML]{EFEFEF}61.78 & \cellcolor[HTML]{EFEFEF}{\ul 73.75} & \cellcolor[HTML]{EFEFEF}27.48 & \multicolumn{1}{c|}{\cellcolor[HTML]{EFEFEF}71.10} & \multicolumn{1}{c|}{\cellcolor[HTML]{EFEFEF}\textit{93.89}} & \cellcolor[HTML]{EFEFEF}\textbf{99.06} & \cellcolor[HTML]{EFEFEF}65.83 & \cellcolor[HTML]{EFEFEF}{\ul 79.75} & \cellcolor[HTML]{EFEFEF}30.21 & \multicolumn{1}{c|}{\cellcolor[HTML]{EFEFEF}71.15} & \cellcolor[HTML]{EFEFEF}\textit{94.79} \\ \cline{2-21} 
 & \multicolumn{1}{c|}{} & \multicolumn{1}{c|}{D} & \textbf{81.82}$_{\pm12.40}$ & 64.04$_{\pm4.63}$ & 45.23$_{\pm19.70}$ & \underline{66.80}$_{\pm0.05}$ & \multicolumn{1}{c|}{11.92$_{\pm0.00}$} & \multicolumn{1}{c|}{\textit{59.56}$_{\pm0.02}$} & \textbf{94.54}$_{\pm4.92}$ & 79.27$_{\pm4.78}$ & 80.21$_{\pm5.26}$ & \underline{81.95}$_{\pm0.55}$ & \multicolumn{1}{c|}{13.87$_{\pm0.00}$} & \multicolumn{1}{c|}{\textit{96.40}$_{\pm3.03}$} & \textbf{94.81}$_{\pm5.03}$ & 79.44$_{\pm4.77}$ & \underline{85.14}$_{\pm2.96}$ & 82.10$_{\pm0.61}$ & \multicolumn{1}{c|}{13.95$_{\pm0.01}$} & \textit{98.85}$_{\pm0.05}$ \\
 & \multicolumn{1}{c|}{} & \multicolumn{1}{c|}{C} & \textbf{99.53}$_{\pm0.05}$ & 72.55$_{\pm22.30}$ & 64.03$_{\pm0.01}$ & 2.93$_{\pm0.03}$ & \multicolumn{1}{c|}{\underline{98.39}$_{\pm0.00}$} & \multicolumn{1}{c|}{\textit{99.52}$_{\pm0.00}$} & \textbf{99.65}$_{\pm0.04}$ & 98.05$_{\pm0.01}$ & 64.24$_{\pm0.04}$ & 45.71$_{\pm0.89}$ & \multicolumn{1}{c|}{\underline{98.52}$_{\pm0.14}$} & \multicolumn{1}{c|}{\textit{99.86}$_{\pm0.01}$} & \textbf{99.71}$_{\pm0.03}$ & 98.89$_{\pm0.72}$ & 64.67$_{\pm0.74}$ & \underline{99.33}$_{\pm0.72}$ & \multicolumn{1}{c|}{98.70$_{\pm0.00}$} & \textit{99.93}$_{\pm0.00}$ \\
 & \multicolumn{1}{c|}{} & \multicolumn{1}{c|}{I} & 98.04$_{\pm3.14}$ & 37.22$_{\pm54.15}$ & \textbf{99.80}$_{\pm0.01}$ & \underline{99.66}$_{\pm0.00}$ & \multicolumn{1}{c|}{99.49$_{\pm0.00}$} & \multicolumn{1}{c|}{\textit{99.87}$_{\pm0.00}$} & \textbf{99.99}$_{\pm0.00}$ & 99.90$_{\pm0.09}$ & \underline{99.95}$_{\pm0.01}$ & 99.79$_{\pm0.09}$ & \multicolumn{1}{c|}{99.63$_{\pm0.00}$} & \multicolumn{1}{c|}{\textit{99.95}$_{\pm0.07}$} & \textbf{99.99}$_{\pm0.00}$ & 99.92$_{\pm0.09}$ & \underline{99.97}$_{\pm0.00}$ & 99.84$_{\pm0.01}$ & \multicolumn{1}{c|}{99.63$_{\pm0.00}$} & \textit{99.99}$_{\pm0.00}$ \\
 & \multicolumn{1}{c|}{\multirow{-4}{*}{Claude}} & \multicolumn{1}{c|}{\cellcolor[HTML]{EFEFEF}Avg} & \cellcolor[HTML]{EFEFEF}\textbf{93.13} & \cellcolor[HTML]{EFEFEF}57.94 & \cellcolor[HTML]{EFEFEF}69.69 & \cellcolor[HTML]{EFEFEF}56.46 & \multicolumn{1}{c|}{\cellcolor[HTML]{EFEFEF}{\ul 69.93}} & \multicolumn{1}{c|}{\cellcolor[HTML]{EFEFEF}\textit{86.32}} & \cellcolor[HTML]{EFEFEF}\textbf{98.06} & \cellcolor[HTML]{EFEFEF}{\ul 92.41} & \cellcolor[HTML]{EFEFEF}81.47 & \cellcolor[HTML]{EFEFEF}75.82 & \multicolumn{1}{c|}{\cellcolor[HTML]{EFEFEF}70.67} & \multicolumn{1}{c|}{\cellcolor[HTML]{EFEFEF}\textit{98.74}} & \cellcolor[HTML]{EFEFEF}\textbf{98.17} & \cellcolor[HTML]{EFEFEF}92.75 & \cellcolor[HTML]{EFEFEF}83.26 & \cellcolor[HTML]{EFEFEF}{\ul 93.76} & \multicolumn{1}{c|}{\cellcolor[HTML]{EFEFEF}70.76} & \cellcolor[HTML]{EFEFEF}\textit{99.59} \\ \cline{2-21} 
 & \multicolumn{1}{c|}{} & \multicolumn{1}{c|}{D} & 70.67$_{\pm16.51}$ & \textbf{82.09}$_{\pm4.13}$ & 77.49$_{\pm13.59}$ & \underline{78.92}$_{\pm4.76}$ & \multicolumn{1}{c|}{13.72$_{\pm0.00}$} & \multicolumn{1}{c|}{\textit{65.87}$_{\pm0.01}$} & \textbf{95.62}$_{\pm3.53}$ & 88.09$_{\pm6.41}$ & \underline{89.08}$_{\pm6.40}$ & 79.88$_{\pm5.07}$ & \multicolumn{1}{c|}{13.88$_{\pm0.00}$} & \multicolumn{1}{c|}{\textit{98.02}$_{\pm0.00}$} & \textbf{96.00}$_{\pm3.11}$ & \underline{90.34}$_{\pm6.46}$ & 89.92$_{\pm5.93}$ & 83.20$_{\pm0.29}$ & \multicolumn{1}{c|}{13.93$_{\pm0.00}$} & \textit{98.62}$_{\pm0.00}$ \\
 & \multicolumn{1}{c|}{} & \multicolumn{1}{c|}{C} & \textbf{99.18}$_{\pm0.54}$ & 50.22$_{\pm43.84}$ & 67.41$_{\pm20.06}$ & 2.97$_{\pm0.01}$ & \multicolumn{1}{c|}{\underline{98.32}$_{\pm0.00}$} & \multicolumn{1}{c|}{\textit{99.52}$_{\pm0.00}$} & \textbf{99.66}$_{\pm0.02}$ & \underline{99.00}$_{\pm0.86}$ & 67.51$_{\pm20.08}$ & 9.07$_{\pm4.17}$ & \multicolumn{1}{c|}{98.43$_{\pm0.00}$} & \multicolumn{1}{c|}{\textit{99.74}$_{\pm0.00}$} & \textbf{99.68}$_{\pm0.01}$ & \underline{99.02}$_{\pm0.86}$ & 91.15$_{\pm1.37}$ & 44.44$_{\pm29.37}$ & \multicolumn{1}{c|}{98.69$_{\pm0.00}$} & \textit{99.79}$_{\pm0.00}$ \\
 & \multicolumn{1}{c|}{} & \multicolumn{1}{c|}{I} & \textbf{99.79}$_{\pm0.07}$ & 5.82$_{\pm0.17}$ & 5.94$_{\pm0.01}$ & 5.77$_{\pm0.16}$ & \multicolumn{1}{c|}{\underline{99.49}$_{\pm0.00}$} & \multicolumn{1}{c|}{\textit{99.87}$_{\pm0.00}$} & \textbf{99.79}$_{\pm0.07}$ & 6.03$_{\pm0.08}$ & 5.95$_{\pm0.01}$ & 68.44$_{\pm54.12}$ & \multicolumn{1}{c|}{\underline{99.51}$_{\pm0.00}$} & \multicolumn{1}{c|}{\textit{99.99}$_{\pm0.00}$} & \textbf{99.83}$_{\pm0.01}$ & 6.06$_{\pm0.08}$ & 6.00$_{\pm0.01}$ & \underline{99.79}$_{\pm0.09}$ & \multicolumn{1}{c|}{99.63$_{\pm0.00}$} & \textit{99.99}$_{\pm0.00}$ \\
 & \multicolumn{1}{c|}{\multirow{-4}{*}{Gemini}} & \multicolumn{1}{c|}{\cellcolor[HTML]{EFEFEF}Avg} & \cellcolor[HTML]{EFEFEF}\textbf{89.88} & \cellcolor[HTML]{EFEFEF}46.04 & \cellcolor[HTML]{EFEFEF}50.28 & \cellcolor[HTML]{EFEFEF}29.22 & \multicolumn{1}{c|}{\cellcolor[HTML]{EFEFEF}{\ul 70.51}} & \multicolumn{1}{c|}{\cellcolor[HTML]{EFEFEF}\textit{88.42}} & \cellcolor[HTML]{EFEFEF}\textbf{98.36} & \cellcolor[HTML]{EFEFEF}64.37 & \cellcolor[HTML]{EFEFEF}54.18 & \cellcolor[HTML]{EFEFEF}52.46 & \multicolumn{1}{c|}{\cellcolor[HTML]{EFEFEF}{\ul 70.61}} & \multicolumn{1}{c|}{\cellcolor[HTML]{EFEFEF}\textit{99.25}} & \cellcolor[HTML]{EFEFEF}\textbf{98.50} & \cellcolor[HTML]{EFEFEF}65.14 & \cellcolor[HTML]{EFEFEF}62.36 & \cellcolor[HTML]{EFEFEF}{\ul 75.81} & \multicolumn{1}{c|}{\cellcolor[HTML]{EFEFEF}70.75} & \cellcolor[HTML]{EFEFEF}\textit{99.47} \\ \cline{2-21} 
 & \multicolumn{1}{c|}{} & \multicolumn{1}{c|}{D} & \textbf{65.20}$_{\pm5.23}$ & \underline{52.51}$_{\pm23.20}$ & 40.90$_{\pm26.66}$ & 22.32$_{\pm19.34}$ & \multicolumn{1}{c|}{21.22$_{\pm6.60}$} & \multicolumn{1}{c|}{\textit{67.79}$_{\pm0.05}$} & \underline{79.82}$_{\pm4.92}$ & 72.72$_{\pm6.93}$ & \textbf{79.84}$_{\pm3.33}$ & 30.75$_{\pm16.73}$ & \multicolumn{1}{c|}{23.76$_{\pm8.77}$} & \multicolumn{1}{c|}{\textit{94.95}$_{\pm4.74}$} & \underline{80.29}$_{\pm5.09}$ & 73.16$_{\pm7.16}$ & \textbf{81.79}$_{\pm0.35}$ & 57.27$_{\pm0.61}$ & \multicolumn{1}{c|}{24.00$_{\pm9.04}$} & \textit{98.82}$_{\pm0.04}$ \\
 & \multicolumn{1}{c|}{} & \multicolumn{1}{c|}{C} & \textbf{99.54}$_{\pm0.05}$ & 41.88$_{\pm26.58}$ & 62.01$_{\pm4.13}$ & 7.70$_{\pm4.25}$ & \multicolumn{1}{c|}{\underline{99.47}$_{\pm0.00}$} & \multicolumn{1}{c|}{\textit{99.48}$_{\pm0.00}$} & \underline{99.64}$_{\pm0.02}$ & 55.89$_{\pm44.00}$ & 64.63$_{\pm0.74}$ & 10.86$_{\pm0.24}$ & \multicolumn{1}{c|}{\textbf{99.67}$_{\pm0.14}$} & \multicolumn{1}{c|}{\textit{99.72}$_{\pm0.00}$} & \underline{99.74}$_{\pm0.17}$ & 58.62$_{\pm44.54}$ & 76.05$_{\pm18.96}$ & 40.44$_{\pm26.92}$ & \multicolumn{1}{c|}{\textbf{99.77}$_{\pm0.01}$} & \textit{99.92}$_{\pm0.01}$ \\
 & \multicolumn{1}{c|}{} & \multicolumn{1}{c|}{I} & \textbf{99.76}$_{\pm0.07}$ & 5.77$_{\pm0.16}$ & 68.38$_{\pm54.25}$ & 5.72$_{\pm0.00}$ & \multicolumn{1}{c|}{\underline{97.69}$_{\pm3.14}$} & \multicolumn{1}{c|}{\textit{99.87}$_{\pm0.00}$} & \textbf{99.90}$_{\pm0.15}$ & 6.07$_{\pm0.14}$ & \underline{99.83}$_{\pm0.01}$ & 5.73$_{\pm0.00}$ & \multicolumn{1}{c|}{99.51$_{\pm0.01}$} & \multicolumn{1}{c|}{\textit{99.87}$_{\pm0.00}$} & \textbf{99.99}$_{\pm0.01}$ & 6.20$_{\pm0.09}$ & \underline{99.89}$_{\pm0.08}$ & 37.20$_{\pm54.12}$ & \multicolumn{1}{c|}{99.56$_{\pm0.07}$} & \textit{99.99}$_{\pm0.00}$ \\
\multirow{-34}{*}{\rotatebox{90}{\textbf{Micro-Average(\%)}}} & \multicolumn{1}{c|}{\multirow{-4}{*}{DeepSeek}} & \multicolumn{1}{c|}{\cellcolor[HTML]{EFEFEF}Avg} & \cellcolor[HTML]{EFEFEF}\textbf{88.17} & \cellcolor[HTML]{EFEFEF}33.39 & \cellcolor[HTML]{EFEFEF}57.10 & \cellcolor[HTML]{EFEFEF}11.91 & \multicolumn{1}{c|}{\cellcolor[HTML]{EFEFEF}{\ul 72.79}} & \multicolumn{1}{c|}{\cellcolor[HTML]{EFEFEF}\textit{89.05}} & \cellcolor[HTML]{EFEFEF}\textbf{93.12} & \cellcolor[HTML]{EFEFEF}44.89 & \cellcolor[HTML]{EFEFEF}{\ul 81.43} & \cellcolor[HTML]{EFEFEF}15.78 & \multicolumn{1}{c|}{\cellcolor[HTML]{EFEFEF}74.31} & \multicolumn{1}{c|}{\cellcolor[HTML]{EFEFEF}\textit{98.18}} & \cellcolor[HTML]{EFEFEF}\textbf{93.34} & \cellcolor[HTML]{EFEFEF}45.99 & \cellcolor[HTML]{EFEFEF}{\ul 85.91} & \cellcolor[HTML]{EFEFEF}44.97 & \multicolumn{1}{c|}{\cellcolor[HTML]{EFEFEF}74.44} & \cellcolor[HTML]{EFEFEF}\textit{99.58} \\ \hline
 & \multicolumn{20}{c|}{\textbf{Technique-level}} \\ \cline{2-21} 
 & \multicolumn{1}{c|}{} & \multicolumn{1}{c|}{D} & \textbf{44.79}$_{\pm1.49}$ & \underline{35.06}$_{\pm6.18}$ & 33.34$_{\pm5.50}$ & 31.27$_{\pm1.65}$ & \multicolumn{1}{c|}{12.78$_{\pm1.32}$} & \multicolumn{1}{c|}{\textit{34.52}$_{\pm0.66}$} & \textbf{64.64}$_{\pm5.05}$ & 44.10$_{\pm10.10}$ & \underline{50.60}$_{\pm6.04}$ & 40.54$_{\pm4.30}$ & \multicolumn{1}{c|}{19.60$_{\pm0.00}$} & \multicolumn{1}{c|}{\textit{55.95}$_{\pm0.88}$} & \textbf{70.18}$_{\pm0.61}$ & 45.98$_{\pm9.80}$ & \underline{54.60}$_{\pm5.65}$ & 44.78$_{\pm3.83}$ & \multicolumn{1}{c|}{24.36$_{\pm0.00}$} & \textit{67.77}$_{\pm0.17}$ \\
 & \multicolumn{1}{c|}{} & \multicolumn{1}{c|}{C} & \textbf{76.79}$_{\pm0.49}$ & 29.48$_{\pm3.71}$ & 57.44$_{\pm1.61}$ & 33.32$_{\pm3.01}$ & \multicolumn{1}{c|}{\underline{71.52}$_{\pm0.00}$} & \multicolumn{1}{c|}{\textit{79.02}$_{\pm0.11}$} & \textbf{81.76}$_{\pm0.85}$ & 54.32$_{\pm8.99}$ & 65.15$_{\pm1.48}$ & 45.76$_{\pm10.40}$ & \multicolumn{1}{c|}{\underline{72.10}$_{\pm0.03}$} & \multicolumn{1}{c|}{\textit{87.66}$_{\pm0.88}$} & \textbf{83.16}$_{\pm0.57}$ & 61.60$_{\pm6.12}$ & \underline{73.39}$_{\pm12.61}$ & 58.05$_{\pm9.50}$ & \multicolumn{1}{c|}{72.10$_{\pm0.03}$} & \textit{91.56}$_{\pm0.93}$ \\
 & \multicolumn{1}{c|}{} & \multicolumn{1}{c|}{I} & \underline{68.27}$_{\pm1.85}$ & 65.55$_{\pm14.41}$ & 50.08$_{\pm1.15}$ & \textbf{73.49}$_{\pm1.54}$ & \multicolumn{1}{c|}{48.60$_{\pm0.00}$} & \multicolumn{1}{c|}{\textit{64.64}$_{\pm0.44}$} & \underline{80.68}$_{\pm13.11}$ & \textbf{90.71}$_{\pm2.60}$ & 59.53$_{\pm1.57}$ & 73.98$_{\pm1.29}$ & \multicolumn{1}{c|}{48.60$_{\pm0.00}$} & \multicolumn{1}{c|}{\textit{89.64}$_{\pm0.44}$} & 81.37$_{\pm13.63}$ & \textbf{92.92}$_{\pm1.87}$ & \underline{89.62}$_{\pm1.93}$ & 74.98$_{\pm0.01}$ & \multicolumn{1}{c|}{49.14$_{\pm0.00}$} & \textit{99.99}$_{\pm0.00}$ \\
 & \multicolumn{1}{c|}{\multirow{-4}{*}{GPT}} & \multicolumn{1}{c|}{\cellcolor[HTML]{EFEFEF}Avg} & \cellcolor[HTML]{EFEFEF}\textbf{63.28} & \cellcolor[HTML]{EFEFEF}43.36 & \cellcolor[HTML]{EFEFEF}{\ul 46.95} & \cellcolor[HTML]{EFEFEF}46.03 & \multicolumn{1}{c|}{\cellcolor[HTML]{EFEFEF}44.30} & \multicolumn{1}{c|}{\cellcolor[HTML]{EFEFEF}\textit{59.39}} & \cellcolor[HTML]{EFEFEF}\textbf{75.69} & \cellcolor[HTML]{EFEFEF}{\ul 63.04} & \cellcolor[HTML]{EFEFEF}58.43 & \cellcolor[HTML]{EFEFEF}53.43 & \multicolumn{1}{c|}{\cellcolor[HTML]{EFEFEF}46.77} & \multicolumn{1}{c|}{\cellcolor[HTML]{EFEFEF}\textit{77.75}} & \cellcolor[HTML]{EFEFEF}\textbf{78.24} & \cellcolor[HTML]{EFEFEF}66.83 & \cellcolor[HTML]{EFEFEF}{\ul 72.54} & \cellcolor[HTML]{EFEFEF}59.27 & \multicolumn{1}{c|}{\cellcolor[HTML]{EFEFEF}48.53} & \cellcolor[HTML]{EFEFEF}\textit{86.44} \\ \cline{2-21} 
 & \multicolumn{1}{c|}{} & \multicolumn{1}{c|}{D} & \textbf{52.16}$_{\pm6.15}$ & \underline{39.26}$_{\pm3.50}$ & 35.05$_{\pm0.90}$ & 35.97$_{\pm0.19}$ & \multicolumn{1}{c|}{14.31$_{\pm0.00}$} & \multicolumn{1}{c|}{\textit{29.00}$_{\pm1.18}$} & \textbf{64.48}$_{\pm1.32}$ & 54.39$_{\pm1.82}$ & 54.18$_{\pm3.46}$ & \underline{60.38}$_{\pm0.52}$ & \multicolumn{1}{c|}{22.24$_{\pm0.00}$} & \multicolumn{1}{c|}{\textit{63.72}$_{\pm0.40}$} & \textbf{64.53}$_{\pm1.35}$ & 55.14$_{\pm2.03}$ & 62.33$_{\pm0.58}$ & \underline{62.71}$_{\pm0.34}$ & \multicolumn{1}{c|}{22.24$_{\pm0.00}$} & \textit{70.01}$_{\pm1.45}$ \\
 & \multicolumn{1}{c|}{} & \multicolumn{1}{c|}{C} & \textbf{75.58}$_{\pm3.60}$ & 66.25$_{\pm9.39}$ & \underline{74.36}$_{\pm0.26}$ & 34.40$_{\pm2.52}$ & \multicolumn{1}{c|}{62.44$_{\pm0.00}$} & \multicolumn{1}{c|}{\textit{81.44}$_{\pm0.00}$} & \textbf{85.68$_{\pm0.75}$} & \underline{82.87}$_{\pm0.13}$ & 77.91$_{\pm0.12}$ & 64.44$_{\pm2.63}$ & \multicolumn{1}{c|}{62.99$_{\pm0.03}$} & \multicolumn{1}{c|}{\textit{93.68}$_{\pm0.20}$} & \underline{86.79}$_{\pm0.49}$ & 83.00$_{\pm0.12}$ & 78.22$_{\pm0.01}$ & \textbf{88.60}$_{\pm1.56}$ & \multicolumn{1}{c|}{63.03$_{\pm0.00}$} & \textit{94.74}$_{\pm0.00}$ \\
 & \multicolumn{1}{c|}{} & \multicolumn{1}{c|}{I} & \underline{57.93}$_{\pm13.08}$ & 57.75$_{\pm14.05}$ & 49.98$_{\pm0.04}$ & \textbf{78.37}$_{\pm11.75}$ & \multicolumn{1}{c|}{48.60$_{\pm0.00}$} & \multicolumn{1}{c|}{\textit{64.89}$_{\pm0.00}$} & \underline{83.03}$_{\pm7.18}$ & 81.77$_{\pm2.36}$ & 76.23$_{\pm0.74}$ & \textbf{99.92}$_{\pm0.04}$ & \multicolumn{1}{c|}{49.14$_{\pm0.00}$} & \multicolumn{1}{c|}{\textit{89.89}$_{\pm0.00}$} & \textbf{99.99}$_{\pm0.00}$ & 86.50$_{\pm2.95}$ & 82.11$_{\pm1.47}$ & \underline{99.96}$_{\pm0.00}$ & \multicolumn{1}{c|}{49.14$_{\pm0.00}$} & \textit{89.89}$_{\pm0.00}$ \\
 & \multicolumn{1}{c|}{\multirow{-4}{*}{Claude}} & \multicolumn{1}{c|}{\cellcolor[HTML]{EFEFEF}Avg} & \cellcolor[HTML]{EFEFEF}\textbf{61.89} & \cellcolor[HTML]{EFEFEF}{\ul 54.42} & \cellcolor[HTML]{EFEFEF}53.13 & \cellcolor[HTML]{EFEFEF}49.58 & \multicolumn{1}{c|}{\cellcolor[HTML]{EFEFEF}41.78} & \multicolumn{1}{c|}{\cellcolor[HTML]{EFEFEF}\textit{58.44}} & \cellcolor[HTML]{EFEFEF}\textbf{77.73} & \cellcolor[HTML]{EFEFEF}73.01 & \cellcolor[HTML]{EFEFEF}69.44 & \cellcolor[HTML]{EFEFEF}{\ul 74.91} & \multicolumn{1}{c|}{\cellcolor[HTML]{EFEFEF}44.79} & \multicolumn{1}{c|}{\cellcolor[HTML]{EFEFEF}\textit{82.43}} & \cellcolor[HTML]{EFEFEF}\textbf{83.77} & \cellcolor[HTML]{EFEFEF}74.88 & \cellcolor[HTML]{EFEFEF}74.22 & \cellcolor[HTML]{EFEFEF}{\ul 83.76} & \multicolumn{1}{c|}{\cellcolor[HTML]{EFEFEF}44.80} & \cellcolor[HTML]{EFEFEF}\textit{84.88} \\ \cline{2-21} 
 & \multicolumn{1}{c|}{} & \multicolumn{1}{c|}{D} & \textbf{56.08}$_{\pm3.61}$ & 47.09$_{\pm5.51}$ & \underline{51.74}$_{\pm7.64}$ & 35.24$_{\pm0.72}$ & \multicolumn{1}{c|}{14.31$_{\pm0.00}$} & \multicolumn{1}{c|}{\textit{31.75}$_{\pm0.66}$} & \underline{67.26}$_{\pm3.50}$ & 55.48$_{\pm5.43}$ & \textbf{76.10}$_{\pm16.64}$ & 51.96$_{\pm1.92}$ & \multicolumn{1}{c|}{22.24$_{\pm0.00}$} & \multicolumn{1}{c|}{\textit{59.05}$_{\pm4.18}$} & \underline{72.30}$_{\pm4.16}$ & 65.79$_{\pm4.77}$ & \textbf{78.59}$_{\pm14.47}$ & 55.54$_{\pm1.81}$ & \multicolumn{1}{c|}{22.24$_{\pm0.00}$} & \textit{59.41}$_{\pm7.03}$ \\
 & \multicolumn{1}{c|}{} & \multicolumn{1}{c|}{C} & \textbf{76.50}$_{\pm0.85}$ & 54.67$_{\pm15.28}$ & \underline{63.72}$_{\pm3.02}$ & 34.83$_{\pm0.33}$ & \multicolumn{1}{c|}{61.35$_{\pm0.00}$} & \multicolumn{1}{c|}{\textit{81.44}$_{\pm0.00}$} & \textbf{84.97}$_{\pm0.86}$ & \underline{84.01}$_{\pm2.51}$ & 69.57$_{\pm3.46}$ & 55.56$_{\pm16.02}$ & \multicolumn{1}{c|}{63.43$_{\pm0.00}$} & \multicolumn{1}{c|}{\textit{91.82}$_{\pm3.97}$} & \textbf{85.75}$_{\pm0.77}$ & \underline{84.71}$_{\pm2.54}$ & 77.35$_{\pm0.21}$ & 70.50$_{\pm5.89}$ & \multicolumn{1}{c|}{63.49$_{\pm0.00}$} & \textit{92.48}$_{\pm0.82}$ \\
 & \multicolumn{1}{c|}{} & \multicolumn{1}{c|}{I} & \textbf{65.69}$_{\pm1.95}$ & 40.30$_{\pm14.46}$ & \underline{49.44}$_{\pm1.32}$ & 33.10$_{\pm13.34}$ & \multicolumn{1}{c|}{48.60$_{\pm0.00}$} & \multicolumn{1}{c|}{\textit{64.13}$_{\pm0.00}$} & \textbf{66.67}$_{\pm1.76}$ & 40.69$_{\pm14.30}$ & 53.12$_{\pm0.74}$ & \underline{53.19}$_{\pm21.51}$ & \multicolumn{1}{c|}{48.60$_{\pm0.00}$} & \multicolumn{1}{c|}{\textit{99.24}$_{\pm0.00}$} & \underline{69.05}$_{\pm0.00}$ & 40.69$_{\pm14.30}$ & 61.70$_{\pm1.70}$ & \textbf{78.74}$_{\pm3.53}$ & \multicolumn{1}{c|}{49.14$_{\pm0.00}$} & \textit{99.99}$_{\pm0.00}$ \\
 & \multicolumn{1}{c|}{\multirow{-4}{*}{Gemini}} & \multicolumn{1}{c|}{\cellcolor[HTML]{EFEFEF}Avg} & \cellcolor[HTML]{EFEFEF}\textbf{66.09} & \cellcolor[HTML]{EFEFEF}47.35 & \cellcolor[HTML]{EFEFEF}{\ul 54.97} & \cellcolor[HTML]{EFEFEF}34.39 & \multicolumn{1}{c|}{\cellcolor[HTML]{EFEFEF}41.42} & \multicolumn{1}{c|}{\cellcolor[HTML]{EFEFEF}\textit{59.11}} & \cellcolor[HTML]{EFEFEF}\textbf{72.97} & \cellcolor[HTML]{EFEFEF}60.06 & \cellcolor[HTML]{EFEFEF}{\ul 66.26} & \cellcolor[HTML]{EFEFEF}53.57 & \multicolumn{1}{c|}{\cellcolor[HTML]{EFEFEF}44.76} & \multicolumn{1}{c|}{\cellcolor[HTML]{EFEFEF}\textit{83.37}} & \cellcolor[HTML]{EFEFEF}\textbf{75.70} & \cellcolor[HTML]{EFEFEF}63.73 & \cellcolor[HTML]{EFEFEF}{\ul 72.55} & \cellcolor[HTML]{EFEFEF}68.26 & \multicolumn{1}{c|}{\cellcolor[HTML]{EFEFEF}44.96} & \cellcolor[HTML]{EFEFEF}\textit{83.96} \\ \cline{2-21} 
 & \multicolumn{1}{c|}{} & \multicolumn{1}{c|}{D} & \textbf{43.27}$_{\pm3.56}$ & \underline{32.72}$_{\pm4.37}$ & 29.86$_{\pm2.77}$ & 31.18$_{\pm0.11}$ & \multicolumn{1}{c|}{18.73$_{\pm1.79}$} & \multicolumn{1}{c|}{\textit{27.87}$_{\pm0.80}$} & \textbf{62.68}$_{\pm3.10}$ & \underline{52.10}$_{\pm0.02}$ & 49.31$_{\pm1.68}$ & 42.58$_{\pm2.92}$ & \multicolumn{1}{c|}{28.07$_{\pm5.01}$} & \multicolumn{1}{c|}{\textit{63.12}$_{\pm1.43}$} & \textbf{67.51}$_{\pm2.58}$ & \underline{58.60}$_{\pm0.08}$ & 57.33$_{\pm0.39}$ & 48.45$_{\pm0.99}$ & \multicolumn{1}{c|}{29.68$_{\pm2.36}$} & \textit{65.18}$_{\pm1.14}$ \\
 & \multicolumn{1}{c|}{} & \multicolumn{1}{c|}{C} & \textbf{75.23}$_{\pm1.31}$ & 55.36$_{\pm2.54}$ & 60.93$_{\pm5.26}$ & 42.33$_{\pm6.17}$ & \multicolumn{1}{c|}{\underline{71.99}$_{\pm0.09}$} & \multicolumn{1}{c|}{\textit{77.66}$_{\pm0.10}$} & \textbf{83.24}$_{\pm1.05}$ & 67.29$_{\pm12.28}$ & 70.06$_{\pm2.49}$ & 51.74$_{\pm2.00}$ & \multicolumn{1}{c|}{\underline{72.46}$_{\pm0.13}$} & \multicolumn{1}{c|}{\textit{90.58}$_{\pm0.27}$} & \textbf{83.67}$_{\pm1.30}$ & 72.88$_{\pm14.13}$ & \underline{75.63}$_{\pm5.11}$ & 62.06$_{\pm10.82}$ & \multicolumn{1}{c|}{72.59$_{\pm0.09}$} & \textit{94.51}$_{\pm0.20}$ \\
 & \multicolumn{1}{c|}{} & \multicolumn{1}{c|}{I} & \textbf{72.94}$_{\pm0.22}$ & 46.93$_{\pm15.71}$ & 46.04$_{\pm14.40}$ & 48.78$_{\pm0.31}$ & \multicolumn{1}{c|}{\underline{49.07}$_{\pm13.96}$} & \multicolumn{1}{c|}{\textit{64.89}$_{\pm0.00}$} & \textbf{91.34}$_{\pm14.68}$ & 58.44$_{\pm5.42}$ & \underline{61.75}$_{\pm0.49}$ & 49.02$_{\pm0.38}$ & \multicolumn{1}{c|}{56.93$_{\pm1.85}$} & \multicolumn{1}{c|}{\textit{89.89}$_{\pm0.00}$} & \textbf{99.80}$_{\pm0.12}$ & 60.16$_{\pm3.53}$ & \underline{76.94}$_{\pm1.84}$ & 67.91$_{\pm14.47}$ & \multicolumn{1}{c|}{57.11$_{\pm1.84}$} & \textit{89.89}$_{\pm0.00}$ \\
 & \multicolumn{1}{c|}{\multirow{-4}{*}{DeepSeek}} & \multicolumn{1}{c|}{\cellcolor[HTML]{EFEFEF}Avg} & \cellcolor[HTML]{EFEFEF}\textbf{63.81} & \cellcolor[HTML]{EFEFEF}45.00 & \cellcolor[HTML]{EFEFEF}45.61 & \cellcolor[HTML]{EFEFEF}40.76 & \multicolumn{1}{c|}{\cellcolor[HTML]{EFEFEF}{\ul 46.60}} & \multicolumn{1}{c|}{\cellcolor[HTML]{EFEFEF}\textit{56.81}} & \cellcolor[HTML]{EFEFEF}\textbf{79.09} & \cellcolor[HTML]{EFEFEF}59.28 & \cellcolor[HTML]{EFEFEF}{\ul 60.37} & \cellcolor[HTML]{EFEFEF}47.78 & \multicolumn{1}{c|}{\cellcolor[HTML]{EFEFEF}52.49} & \multicolumn{1}{c|}{\cellcolor[HTML]{EFEFEF}\textit{81.20}} & \cellcolor[HTML]{EFEFEF}\textbf{83.66} & \cellcolor[HTML]{EFEFEF}63.88 & \cellcolor[HTML]{EFEFEF}{\ul 69.97} & \cellcolor[HTML]{EFEFEF}59.47 & \multicolumn{1}{c|}{\cellcolor[HTML]{EFEFEF}53.13} & \cellcolor[HTML]{EFEFEF}\textit{83.19} \\ \cline{2-21} 
 & \multicolumn{20}{c|}{\textbf{Tactic-level}} \\ \cline{2-21} 
 & \multicolumn{1}{c|}{} & \multicolumn{1}{c|}{D} & \textbf{46.38}$_{\pm3.16}$ & \underline{40.41}$_{\pm6.77}$ & 36.10$_{\pm4.80}$ & 37.39$_{\pm2.14}$ & \multicolumn{1}{c|}{13.24$_{\pm1.32}$} & \multicolumn{1}{c|}{\textit{34.65}$_{\pm0.58}$} & \textbf{63.82}$_{\pm6.97}$ & 53.96$_{\pm5.98}$ & \underline{56.46}$_{\pm8.11}$ & 51.63$_{\pm5.00}$ & \multicolumn{1}{c|}{30.74$_{\pm0.13}$} & \multicolumn{1}{c|}{\textit{56.15}$_{\pm0.74}$} & \textbf{72.12}$_{\pm2.86}$ & 59.08$_{\pm6.29}$ & \underline{62.86}$_{\pm8.38}$ & 58.13$_{\pm4.53}$ & \multicolumn{1}{c|}{30.89$_{\pm0.01}$} & \textit{69.08}$_{\pm1.95}$ \\
 & \multicolumn{1}{c|}{} & \multicolumn{1}{c|}{C} & \underline{76.79}$_{\pm0.49}$ & 29.48$_{\pm3.71}$ & 59.37$_{\pm0.47}$ & 33.36$_{\pm3.01}$ & \multicolumn{1}{c|}{\textbf{77.56}$_{\pm0.41}$} & \multicolumn{1}{c|}{\textit{79.02}$_{\pm0.11}$} & \textbf{81.77}$_{\pm0.93}$ & 54.44$_{\pm9.15}$ & 65.55$_{\pm2.02}$ & 45.70$_{\pm10.50}$ & \multicolumn{1}{c|}{\underline{78.57}$_{\pm0.03}$} & \multicolumn{1}{c|}{\textit{89.42}$_{\pm0.20}$} & \textbf{83.34}$_{\pm0.37}$ & 61.58$_{\pm6.14}$ & 73.79$_{\pm12.61}$ & 58.09$_{\pm9.50}$ & \multicolumn{1}{c|}{\underline{80.13}$_{\pm0.06}$} & \textit{92.87}$_{\pm1.67}$ \\
 & \multicolumn{1}{c|}{} & \multicolumn{1}{c|}{I} & 68.76$_{\pm1.53}$ & \underline{68.98}$_{\pm13.91}$ & 53.02$_{\pm1.85}$ & \textbf{73.49}$_{\pm1.54}$ & \multicolumn{1}{c|}{48.60$_{\pm0.00}$} & \multicolumn{1}{c|}{\textit{64.64}$_{\pm0.44}$} & \underline{81.41}$_{\pm13.71}$ & \textbf{99.80}$_{\pm0.12}$ & 74.50$_{\pm0.11}$ & 74.75$_{\pm0.01}$ & \multicolumn{1}{c|}{48.90$_{\pm0.00}$} & \multicolumn{1}{c|}{\textit{89.89}$_{\pm0.00}$} & 81.66$_{\pm14.13}$ & \underline{99.80}$_{\pm0.12}$ & \textbf{99.94}$_{\pm0.00}$ & 75.02$_{\pm0.03}$ & \multicolumn{1}{c|}{49.44$_{\pm0.00}$} & \textit{99.99}$_{\pm0.00}$ \\
 & \multicolumn{1}{c|}{\multirow{-4}{*}{GPT}} & \multicolumn{1}{c|}{\cellcolor[HTML]{EFEFEF}Avg} & \cellcolor[HTML]{EFEFEF}\textbf{63.98} & \cellcolor[HTML]{EFEFEF}46.29 & \cellcolor[HTML]{EFEFEF}{\ul 49.50} & \cellcolor[HTML]{EFEFEF}48.08 & \multicolumn{1}{c|}{\cellcolor[HTML]{EFEFEF}46.47} & \multicolumn{1}{c|}{\cellcolor[HTML]{EFEFEF}\textit{59.44}} & \cellcolor[HTML]{EFEFEF}\textbf{75.67} & \cellcolor[HTML]{EFEFEF}{\ul 69.40} & \cellcolor[HTML]{EFEFEF}65.50 & \cellcolor[HTML]{EFEFEF}57.36 & \multicolumn{1}{c|}{\cellcolor[HTML]{EFEFEF}52.74} & \multicolumn{1}{c|}{\cellcolor[HTML]{EFEFEF}\textit{78.49}} & \cellcolor[HTML]{EFEFEF}\textbf{79.04} & \cellcolor[HTML]{EFEFEF}73.49 & \cellcolor[HTML]{EFEFEF}{\ul 78.86} & \cellcolor[HTML]{EFEFEF}63.75 & \multicolumn{1}{c|}{\cellcolor[HTML]{EFEFEF}53.49} & \cellcolor[HTML]{EFEFEF}\textit{87.31} \\ \cline{2-21} 
 & \multicolumn{1}{c|}{} & \multicolumn{1}{c|}{D} & \textbf{55.88}$_{\pm4.48}$ & \underline{45.84}$_{\pm4.82}$ & 37.74$_{\pm2.72}$ & 37.48$_{\pm0.26}$ & \multicolumn{1}{c|}{14.31$_{\pm0.00}$} & \multicolumn{1}{c|}{\textit{29.00}$_{\pm1.18}$} & \textbf{68.18}$_{\pm1.25}$ & 61.28$_{\pm3.49}$ & 60.53$_{\pm3.48}$ & \underline{67.22}$_{\pm1.21}$ & \multicolumn{1}{c|}{27.18$_{\pm0.00}$} & \multicolumn{1}{c|}{\textit{68.72}$_{\pm0.40}$} & 71.60$_{\pm1.60}$ & 65.35$_{\pm3.40}$ & \underline{72.12}$_{\pm0.49}$ & \textbf{72.46}$_{\pm1.08}$ & \multicolumn{1}{c|}{29.04$_{\pm0.46}$} & \textit{72.54}$_{\pm1.38}$ \\
 & \multicolumn{1}{c|}{} & \multicolumn{1}{c|}{C} & \textbf{75.58}$_{\pm3.60}$ & 65.84$_{\pm9.22}$ & \underline{74.36}$_{\pm0.26}$ & 34.59$_{\pm2.52}$ & \multicolumn{1}{c|}{67.54$_{\pm0.00}$} & \multicolumn{1}{c|}{\textit{81.44}$_{\pm0.00}$} & \textbf{85.69}$_{\pm0.96}$ & \underline{82.87}$_{\pm0.13}$ & 77.36$_{\pm0.91}$ & 66.57$_{\pm4.31}$ & \multicolumn{1}{c|}{70.63$_{\pm0.26}$} & \multicolumn{1}{c|}{\textit{93.68}$_{\pm0.20}$} & \underline{88.73}$_{\pm2.06}$ & 83.31$_{\pm0.12}$ & 77.68$_{\pm0.98}$ & \textbf{90.74}$_{\pm0.11}$ & \multicolumn{1}{c|}{72.58$_{\pm0.09}$} & \textit{94.74}$_{\pm0.00}$ \\
 & \multicolumn{1}{c|}{} & \multicolumn{1}{c|}{I} & \underline{58.53}$_{\pm12.99}$ & 57.99$_{\pm13.83}$ & 50.00$_{\pm0.00}$ & \textbf{92.01}$_{\pm2.37}$ & \multicolumn{1}{c|}{48.60$_{\pm0.00}$} & \multicolumn{1}{c|}{\textit{64.89}$_{\pm0.00}$} & \underline{91.25}$_{\pm13.87}$ & 84.95$_{\pm1.24}$ & 80.40$_{\pm2.36}$ & \textbf{99.95}$_{\pm0.02}$ & \multicolumn{1}{c|}{49.44$_{\pm0.00}$} & \multicolumn{1}{c|}{\textit{96.62}$_{\pm5.83}$} & \textbf{99.99}$_{\pm0.00}$ & 99.00$_{\pm0.83}$ & 89.96$_{\pm0.85}$ & \underline{99.96}$_{\pm0.00}$ & \multicolumn{1}{c|}{50.17$_{\pm0.00}$} & \textit{99.99}$_{\pm0.00}$ \\
 & \multicolumn{1}{c|}{\multirow{-4}{*}{Claude}} & \multicolumn{1}{c|}{\cellcolor[HTML]{EFEFEF}Avg} & \cellcolor[HTML]{EFEFEF}\textbf{63.33} & \cellcolor[HTML]{EFEFEF}{\ul 56.56} & \cellcolor[HTML]{EFEFEF}54.03 & \cellcolor[HTML]{EFEFEF}54.69 & \multicolumn{1}{c|}{\cellcolor[HTML]{EFEFEF}43.48} & \multicolumn{1}{c|}{\cellcolor[HTML]{EFEFEF}\textit{58.44}} & \cellcolor[HTML]{EFEFEF}\textbf{81.71} & \cellcolor[HTML]{EFEFEF}76.37 & \cellcolor[HTML]{EFEFEF}72.76 & \cellcolor[HTML]{EFEFEF}{\ul 77.91} & \multicolumn{1}{c|}{\cellcolor[HTML]{EFEFEF}49.08} & \multicolumn{1}{c|}{\cellcolor[HTML]{EFEFEF}\textit{86.34}} & \cellcolor[HTML]{EFEFEF}{\ul 86.77} & \cellcolor[HTML]{EFEFEF}82.55 & \cellcolor[HTML]{EFEFEF}79.92 & \cellcolor[HTML]{EFEFEF}\textbf{87.72} & \multicolumn{1}{c|}{\cellcolor[HTML]{EFEFEF}50.60} & \cellcolor[HTML]{EFEFEF}\textit{89.09} \\ \cline{2-21} 
 & \multicolumn{1}{c|}{} & \multicolumn{1}{c|}{D} & \textbf{60.73}$_{\pm4.02}$ & 49.18$_{\pm8.71}$ & \underline{52.95}$_{\pm6.78}$ & 47.87$_{\pm3.24}$ & \multicolumn{1}{c|}{14.54$_{\pm0.00}$} & \multicolumn{1}{c|}{\textit{31.75}$_{\pm0.66}$} & \underline{76.12}$_{\pm1.45}$ & 60.95$_{\pm6.67}$ & \textbf{77.29}$_{\pm15.77}$ & 69.21$_{\pm3.53}$ & \multicolumn{1}{c|}{27.41$_{\pm0.00}$} & \multicolumn{1}{c|}{\textit{64.05}$_{\pm0.00}$} & \underline{76.94}$_{\pm3.89}$ & 61.82$_{\pm4.34}$ & \textbf{77.49}$_{\pm16.09}$ & 70.21$_{\pm2.65}$ & \multicolumn{1}{c|}{36.21$_{\pm2.36}$} & \textit{70.40}$_{\pm0.92}$ \\
 & \multicolumn{1}{c|}{} & \multicolumn{1}{c|}{C} & \textbf{73.47}$_{\pm4.40}$ & 47.72$_{\pm14.90}$ & 60.44$_{\pm7.01}$ & 32.22$_{\pm0.15}$ & \multicolumn{1}{c|}{\underline{67.02}$_{\pm0.00}$} & \multicolumn{1}{c|}{\textit{81.44}$_{\pm0.00}$} & \textbf{85.01}$_{\pm0.86}$ & \underline{83.52}$_{\pm2.69}$ & 65.83$_{\pm7.52}$ & 48.40$_{\pm5.57}$ & \multicolumn{1}{c|}{69.71$_{\pm0.00}$} & \multicolumn{1}{c|}{\textit{91.82}$_{\pm0.00}$} & \textbf{85.05}$_{\pm1.29}$ & \underline{83.88}$_{\pm2.35}$ & 75.86$_{\pm4.75}$ & 63.56$_{\pm11.16}$ & \multicolumn{1}{c|}{80.97$_{\pm0.41}$} & \textit{94.51}$_{\pm0.20}$ \\
 & \multicolumn{1}{c|}{} & \multicolumn{1}{c|}{I} & \textbf{66.42}$_{\pm1.61}$ & 43.80$_{\pm15.55}$ & 50.95$_{\pm0.74}$ & \underline{52.06}$_{\pm14.51}$ & \multicolumn{1}{c|}{48.60$_{\pm0.00}$} & \multicolumn{1}{c|}{\textit{64.13}$_{\pm0.00}$} & \underline{67.40}$_{\pm1.75}$ & 59.13$_{\pm15.15}$ & 58.31$_{\pm1.98}$ & \textbf{74.36}$_{\pm24.83}$ & \multicolumn{1}{c|}{48.90$_{\pm0.00}$} & \multicolumn{1}{c|}{\textit{99.24}$_{\pm0.00}$} & \textbf{99.82}$_{\pm0.15}$ & 69.92$_{\pm1.94}$ & \underline{81.28}$_{\pm14.94}$ & 78.41$_{\pm13.62}$ & \multicolumn{1}{c|}{57.41$_{\pm1.84}$} & \textit{99.99}$_{\pm0.00}$ \\
 & \multicolumn{1}{c|}{\multirow{-4}{*}{Gemini}} & \multicolumn{1}{c|}{\cellcolor[HTML]{EFEFEF}Avg} & \cellcolor[HTML]{EFEFEF}\textbf{66.87} & \cellcolor[HTML]{EFEFEF}46.90 & \cellcolor[HTML]{EFEFEF}{\ul 54.78} & \cellcolor[HTML]{EFEFEF}44.05 & \multicolumn{1}{c|}{\cellcolor[HTML]{EFEFEF}43.39} & \multicolumn{1}{c|}{\cellcolor[HTML]{EFEFEF}\textit{59.11}} & \cellcolor[HTML]{EFEFEF}\textbf{76.18} & \cellcolor[HTML]{EFEFEF}{\ul 67.87} & \cellcolor[HTML]{EFEFEF}67.14 & \cellcolor[HTML]{EFEFEF}63.99 & \multicolumn{1}{c|}{\cellcolor[HTML]{EFEFEF}48.67} & \multicolumn{1}{c|}{\cellcolor[HTML]{EFEFEF}\textit{85.04}} & \cellcolor[HTML]{EFEFEF}\textbf{87.27} & \cellcolor[HTML]{EFEFEF}71.87 & \cellcolor[HTML]{EFEFEF}{\ul 78.21} & \cellcolor[HTML]{EFEFEF}70.73 & \multicolumn{1}{c|}{\cellcolor[HTML]{EFEFEF}58.20} & \cellcolor[HTML]{EFEFEF}\textit{88.30} \\ \cline{2-21} 
 & \multicolumn{1}{c|}{} & \multicolumn{1}{c|}{D} & \textbf{45.91}$_{\pm3.61}$ & 33.80$_{\pm8.06}$ & \underline{35.28}$_{\pm1.84}$ & 27.32$_{\pm2.00}$ & \multicolumn{1}{c|}{19.12$_{\pm2.03}$} & \multicolumn{1}{c|}{\textit{27.87}$_{\pm0.80}$} & \textbf{65.09}$_{\pm2.39}$ & 57.05$_{\pm1.08}$ & \underline{58.08}$_{\pm1.05}$ & 46.64$_{\pm1.60}$ & \multicolumn{1}{c|}{33.02$_{\pm5.01}$} & \multicolumn{1}{c|}{\textit{68.12}$_{\pm1.43}$} & \underline{79.41}$_{\pm1.80}$ & 72.87$_{\pm4.97}$ & \textbf{80.76}$_{\pm12.58}$ & 74.69$_{\pm2.79}$ & \multicolumn{1}{c|}{34.23$_{\pm4.04}$} & \textit{68.30}$_{\pm1.53}$ \\
 & \multicolumn{1}{c|}{} & \multicolumn{1}{c|}{C} & \underline{75.34}$_{\pm1.40}$ & 55.59$_{\pm2.55}$ & 61.51$_{\pm5.10}$ & 43.63$_{\pm4.79}$ & \multicolumn{1}{c|}{\textbf{77.98}$_{\pm0.36}$} & \multicolumn{1}{c|}{\textit{77.66}$_{\pm0.10}$} & \textbf{83.71}$_{\pm0.66}$ & 65.93$_{\pm13.04}$ & 69.29$_{\pm3.03}$ & 53.25$_{\pm4.19}$ & \multicolumn{1}{c|}{\underline{79.65}$_{\pm0.13}$} & \multicolumn{1}{c|}{\textit{90.58}$_{\pm0.27}$} & \textbf{85.74}$_{\pm0.83}$ & \underline{84.54}$_{\pm2.62}$ & 74.06$_{\pm5.42}$ & 63.33$_{\pm11.20}$ & \multicolumn{1}{c|}{80.20$_{\pm1.07}$} & \textit{92.48}$_{\pm0.80}$ \\
 & \multicolumn{1}{c|}{} & \multicolumn{1}{c|}{I} & \textbf{72.94}$_{\pm0.22}$ & 48.05$_{\pm15.34}$ & 46.71$_{\pm14.34}$ & \underline{49.20}$_{\pm0.00}$ & \multicolumn{1}{c|}{49.07$_{\pm13.96}$} & \multicolumn{1}{c|}{\textit{64.89}$_{\pm0.00}$} & \textbf{91.34}$_{\pm14.68}$ & 62.79$_{\pm3.64}$ & \underline{66.62}$_{\pm0.46}$ & 51.41$_{\pm1.47}$ & \multicolumn{1}{c|}{57.17$_{\pm1.85}$} & \multicolumn{1}{c|}{\textit{89.89}$_{\pm0.00}$} & \underline{92.77}$_{\pm11.52}$ & 67.23$_{\pm6.39}$ & 74.24$_{\pm0.41}$ & \textbf{99.67}$_{\pm0.05}$ & \multicolumn{1}{c|}{58.02$_{\pm1.63}$} & \textit{99.99}$_{\pm0.00}$ \\
\multirow{-34}{*}{\rotatebox{90}{\textbf{Macro-Average(\%)}}} & \multicolumn{1}{c|}{\multirow{-4}{*}{DeepSeek}} & \multicolumn{1}{c|}{\cellcolor[HTML]{EFEFEF}Avg} & \cellcolor[HTML]{EFEFEF}\textbf{64.73} & \cellcolor[HTML]{EFEFEF}45.81 & \cellcolor[HTML]{EFEFEF}47.83 & \cellcolor[HTML]{EFEFEF}40.05 & \multicolumn{1}{c|}{\cellcolor[HTML]{EFEFEF}{\ul 48.72}} & \multicolumn{1}{c|}{\cellcolor[HTML]{EFEFEF}\textit{56.81}} & \cellcolor[HTML]{EFEFEF}\textbf{80.05} & \cellcolor[HTML]{EFEFEF}61.92 & \cellcolor[HTML]{EFEFEF}{\ul 64.66} & \cellcolor[HTML]{EFEFEF}50.43 & \multicolumn{1}{c|}{\cellcolor[HTML]{EFEFEF}56.61} & \multicolumn{1}{c|}{\cellcolor[HTML]{EFEFEF}\textit{82.86}} & \cellcolor[HTML]{EFEFEF}{\textbf{85.97}} & \cellcolor[HTML]{EFEFEF}74.88 & \cellcolor[HTML]{EFEFEF}76.35 & \cellcolor[HTML]{EFEFEF}\underline{79.23} & \multicolumn{1}{c|}{\cellcolor[HTML]{EFEFEF}57.48} & \cellcolor[HTML]{EFEFEF}\textit{86.92} \\ \hline
\end{tabular}}
\end{table*}

\subsection{RQ1: Performance Evaluation}
\label{subsec:rq1_performance}

We evaluate \tool against four zero-shot baselines (Vanilla, CoT, ToT, RAG-Chain) and report Few-Shot prompting separately as a labeled-data reference (italicized in Table~\ref{tab:main_results}).

\noindent \textbf{Key Findings.} As shown in Table~\ref{tab:main_results}, \tool achieves an overall top-1 micro-averaged accuracy of 87.34\%. This is 16.78~points above the strongest baseline (RAG-Chain, 70.57\%); ToT, CoT, and Vanilla all fall below 60\%. Individual baselines can match or exceed \tool in narrow conditions. For instance, RAG-Chain reaches 98.34\% on CICIDS2017 with Claude against \tool's 81.43\%. However, these advantages do not generalize. RAG-Chain drops to 11.92-19.41\% on DAPT2020, and CoT and ToT fall below 10\% on IoT23 in several settings. By contrast, \tool's lowest result across all twelve conditions remains 65.17\%. This indicates that \tool \revise{maintains usable accuracy across heterogeneous scenarios}, while every baseline has at least one regime where its accuracy collapses below 20\%.

The same ordering holds under macro-averaging (Table~\ref{tab:main_results}). \tool reaches an average macro top-1 accuracy of 63.77\%, 13.60~points above the strongest baseline (ToT, 50.16\%). The gap against RAG-Chain widens from 16.78~points under micro-averaging to 20.24~points under macro-averaging, indicating that RAG-Chain's micro-level competitiveness relies on majority-class performance and degrades when each class is weighted equally. Tactic-level results follow the same pattern. \tool is therefore the only zero-shot method that maintains a usable accuracy range across every backbone-dataset combination under both averaging schemes and both granularities.

\revise{The CICIDS2017 micro-averaged top-1 results for Claude Sonnet~4 carry large standard deviations
(\(81.43 \pm 30.25\%\)).
This is a sample-counting effect rather than evidence of instability
across the dataset: roughly \(53\%\) of CICIDS2017 flows belong to a
single DoS Hulk flow group that is summarized and mapped as one unit,
so under micro-averaging a rank change on this one group can move
top-1 accuracy by roughly \(53\) points, while macro-averaged accuracy
and top-3/top-5 accuracy remain far more stable. We regard this
concentration of flows in a few large groups as a limitation of the
CICIDS2017 evaluation. Per-run results are given in Section~III
(Tables~V--X) of the supplemental material.}

Vanilla exceeds \tool on several IoT23 macro conditions (\eg 92.01\% vs. 58.53\% at tactic-level Top-1 with Claude). However, on DAPT2020 and CICIDS2017, Vanilla's tactic-level macro Top-1 stays between 27.32\% and 47.87\%, indicating that this is a dataset-specific observation rather than a general advantage of vanilla prompting.

\noindent \textbf{Comparison with Few-Shot Prompting.} Unlike the four zero-shot baselines, Few-Shot receives one labeled exemplar per ground-truth class, so we report it separately as a reference.

Few-Shot reaches an overall top-1 micro accuracy of 87.54\%, close to \tool's 87.34\%, and exceeds \tool on nine of twelve conditions. The ranking reverses under macro-averaging: \tool reaches 63.77\% against Few-Shot's 58.44\%, and outperforms Few-Shot on seven of twelve conditions. This reversal reflects a known limitation of in-context learning: Few-Shot tends to match samples against the provided exemplars, and samples outside the exemplar coverage fall back to surface-level guesses with little reasoning support. Few-Shot also requires per-deployment labeled exemplars that must be re-acquired when the attack distribution shifts, whereas \tool operates without any labeled data.

\noindent \textbf{Statistical Significance.} We use the Wilcoxon signed-rank test~\cite{wilcoxon1945individual} to assess the significance of \tool's improvements over each baseline. Each (backbone, dataset) combination provides one paired observation, with the mean accuracy across three trials taken as the representative value; this yields twelve pairs per baseline. We perform one-sided tests under the alternative hypothesis that \tool achieves higher accuracy than the baseline, and report per-baseline results without family-wise correction since each comparison is directional and specified a priori. As shown in Table~\ref{tab:wilcoxon}, \tool's advantage over CoT, ToT, and Vanilla is significant at $p<0.011$ under both micro- and macro-averaging at both granularities. The micro-technique comparison against RAG-Chain is not significant ($p=0.259$), reflecting their parity when majority classes dominate accuracy. Under macro-averaging, however, \tool outperforms RAG-Chain on twelve of twelve pairs at the technique level ($p<0.001$) and ten of twelve at the tactic level ($p=0.001$). \revise{The statistical evidence supports \tool's gains}, while baselines reach parity only when class imbalance favors them.

\begin{table}[!ht]
\centering
\caption{Wilcoxon signed-rank test results (one-sided, $H_1$: \tool $>$ baseline) for top-1 accuracy.  W/L indicates the number of backbone-dataset combinations where \tool wins or loses. Tech and Tac denote the technique and tactic levels. Significance levels: $^{*}p<0.05$, $^{**}p<0.01$, $^{***}p<0.001$.}
\label{tab:wilcoxon}
\footnotesize
\setlength{\tabcolsep}{2.5pt}
\scalebox{0.9}{
\begin{tabular}{l|cc|cc|cc|cc}
\hline
\multirow{2}{*}{\textbf{Baseline}} & \multicolumn{2}{c|}{\textbf{Tech Micro}} & \multicolumn{2}{c|}{\textbf{Tech Macro}} & \multicolumn{2}{c|}{\textbf{Tac Micro}} & \multicolumn{2}{c}{\textbf{Tac Macro}} \\
\cline{2-9}
 & W/L & $p$ & W/L & $p$ & W/L & $p$ & W/L & $p$ \\
\hline
CoT     & 10/2 & 0.003$^{**}$ & 12/0 & $<$0.001$^{***}$ & 10/2 & 0.002$^{**}$ & 11/1 & $<$0.001$^{***}$ \\
ToT     & 9/3  & 0.003$^{**}$ & 12/0 & $<$0.001$^{***}$ & 9/3  & 0.003$^{**}$ & 12/0 & $<$0.001$^{***}$ \\
Vanilla & 10/2 & 0.001$^{**}$ & 10/2 & 0.003$^{**}$ & 10/2 & 0.001$^{**}$ & 10/2 & 0.010$^{*}$ \\
RAG     & 6/6  & 0.259        & 12/0 & $<$0.001$^{***}$ & 9/3  & 0.032$^{*}$  & 10/2 & 0.001$^{**}$ \\
\hline
\end{tabular}}
\end{table}

\noindent \textbf{Class-wise F1 Analysis.} To further dissect performance across the diverse set of techniques in MITRE ATT\&CK, we evaluate class-wise F1 scores using the One-vs-Rest (OvR) approach~\cite{chmielnicki2016using}. In this setup, each technique is treated as an independent binary classification task, where the target technique is considered the positive class (1) and all others are grouped as negatives (0). For top-1 predictions, the highest-scoring candidate is selected; for top-5, the correct technique is chosen if present among candidates, otherwise the top candidate is retained. Predictions and ground truth are aggregated across all datasets, computing precision, recall, and F1 per technique.

\begin{figure}[!ht]
	\centering
	\includegraphics[width=\linewidth]{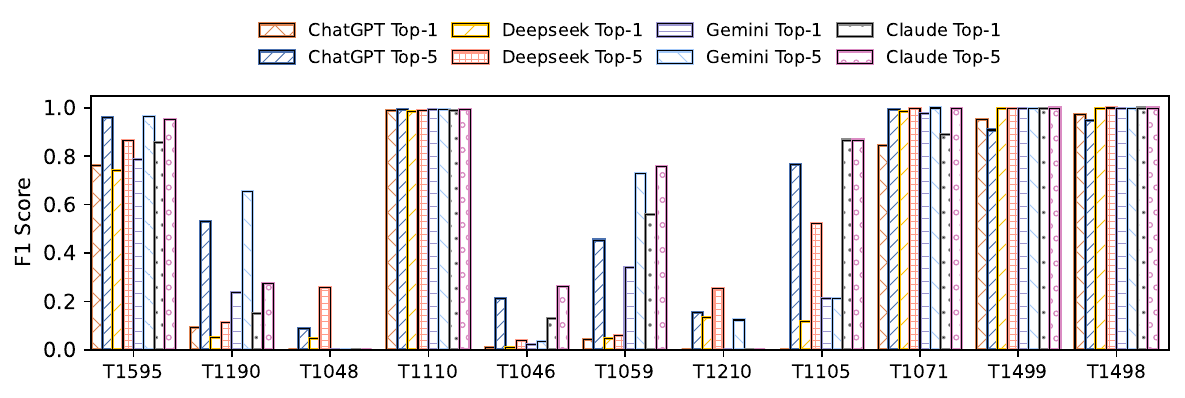}
	\caption{Class-wise F1 scores for top-1 and top-5 predictions across four representative backbone models.}
	\label{fig:class_wise_f1}
\end{figure}

Figure~\ref{fig:class_wise_f1} presents the resulting class-wise F1 scores for top-1 and top-5 predictions for four models. \revise{The evaluation covers 11 techniques in total.} Among these, techniques such as \textit{T1595 (Active Scanning)}, \textit{T1110 (Brute Force)}, \textit{T1071 (Application Layer Protocol)}, \textit{T1499 (Endpoint Denial of Service)}, and \textit{T1498 (Network Denial of Service)} achieve F1 scores above 0.7. These techniques benefit from high-volume, pattern-driven attacks (\eg repeated login attempts for \textit{T1110}), enabling reliable intent inference from traffic metadata. Conversely, techniques like \textit{T1048 (Exfiltration Over Alternative Protocol)} and \textit{T1046 (Network Service Discovery)} yield considerably lower F1 scores. This performance drop can be attributed to their more subtle and context-dependent behavioral signatures. For instance, accurately identifying \textit{T1048} often requires inspection of file content or payload, which is not accessible from metadata alone. Finally, we observe that top-5 F1 scores consistently surpass top-1 scores across most techniques, indicating that although the model may not always rank the correct technique first, it frequently includes the true technique among its top-five predictions.

\noindent \textbf{Feasibility on an Open-Weight Backbone.} To examine whether \tool's advantages persist on locally deployable models, we re-run the same evaluation using Llama-3.3-70B-Instruct; pipeline, prompts, and temperature are identical to the commercial-backbone runs.

\begin{table}[ht]
\centering
\caption{Top-1 micro-accuracy (\%) of \tool and the three zero-shot baselines on Llama-3.3-70B-Instruct across three datasets.}
\label{tab:llama_evaluation}
\footnotesize
\scalebox{0.9}{\begin{tabular}{c|l|ccc|c}
\hline
& \textbf{Method} & \textbf{DAPT2020} & \textbf{CICIDS2017} & \textbf{IoT23} & \textbf{Avg} \\ \hline
\multirow{4}{*}{\rotatebox{90}{\textbf{Micro}}}
& \tool   & 60.28 $\pm$ 5.75  & 81.68 $\pm$ 30.47 & 96.07 $\pm$ 3.14  & \textbf{79.34} \\
& CoT     & 58.07 $\pm$ 19.39 &  6.94 $\pm$ 4.49  & 66.41 $\pm$ 57.26 & 43.81 \\
& ToT     & 66.19 $\pm$ 8.53  &  2.46 $\pm$ 0.54  & 36.98 $\pm$ 54.06 & 35.21 \\
& Vanilla & 34.52 $\pm$ 16.62 &  1.27 $\pm$ 0.85  &  5.60 $\pm$ 0.00  & 13.80 \\ \hline
\end{tabular}}
\end{table}

On Llama-3.3-70B (Table~\ref{tab:llama_evaluation}), \tool achieves an average top-1 micro accuracy of 79.34\%, exceeding the strongest baseline by 35.54 points. The same pattern observed on the commercial backbones also holds here: every baseline falls below 7\% on CICIDS2017 against \tool's 81.68\%\revise{($\pm30.47$; see Section~VII of the supplemental material)}, while \tool is the only method that stays within a usable range across all three datasets. Relative to the commercial average, the Llama configuration trails by 8.00 points under micro-averaging.

\noindent \textbf{Discussion.} In summary, the results validate the effectiveness of \tool in processing complex threats. Its multi-stage design mitigates the oversimplification risks inherent in baseline methods, which often rely on heuristic mappings or truncated context. Baselines, despite excelling in narrow scenarios (\eg tactic-level predictions in IoT23), exhibit inconsistency across datasets, underscoring \tool's accuracy on complex multi-stage scenarios.

\begin{table*}[]
	\centering
	\caption{Ablation study using ChatGPT-4o on the impact of key components in \tool: the Abstraction Module, Inference Module, and Refinement Module. Three variants are tested: (1) \tool-No-Abstraction (w/o Abs.) (2) \tool-Alt-Consistency, which replaces structured inference with standard consistency decoding methods such as \textit{multi-sampling} and \textit{self-debate}; and (3) \tool-No-Refinement (w/o Ref.). The performance metrics for top-1, top-3, and top-5 accuracy are reported for each configuration. Results are color-coded, with {\color[HTML]{FF0000}red} indicating performance degradation and {\color[HTML]{00B050}green} indicating performance improvement compared to the full method of \tool.}
	\label{tab:ablation}
    \resizebox{\textwidth}{!}{%
    \renewcommand{\arraystretch}{1.05}
    \setlength{\tabcolsep}{3pt}
		\begin{tabular}{|c|c|ccccccccccccccc|}
\hline
 &  & \multicolumn{15}{c|}{\textbf{Micro-Averaged Accuracy (\%)}} \\ \cline{3-17} 
 &  & \multicolumn{5}{c|}{\textbf{Top 1}} & \multicolumn{5}{c|}{\textbf{Top 3}} & \multicolumn{5}{c|}{\textbf{Top 5}} \\
\multirow{-3}{*}{} & \multirow{-3}{*}{\textbf{Dataset}} & Ours & \begin{tabular}[c]{@{}c@{}}w/o\\ Abs.\end{tabular} & \begin{tabular}[c]{@{}c@{}}Multi\\ Sampling\end{tabular} & \begin{tabular}[c]{@{}c@{}}Self\\ Debate\end{tabular} & \multicolumn{1}{c|}{\begin{tabular}[c]{@{}c@{}}w/o\\ Ref.\end{tabular}} & Ours & \begin{tabular}[c]{@{}c@{}}w/o\\ Abs.\end{tabular} & \begin{tabular}[c]{@{}c@{}}Multi\\ Sampling\end{tabular} & \begin{tabular}[c]{@{}c@{}}Self\\ Debate\end{tabular} & \multicolumn{1}{c|}{\begin{tabular}[c]{@{}c@{}}w/o\\ Ref.\end{tabular}} & Ours & \begin{tabular}[c]{@{}c@{}}w/o\\ Abs.\end{tabular} & \begin{tabular}[c]{@{}c@{}}Multi\\ Sampling\end{tabular} & \begin{tabular}[c]{@{}c@{}}Self\\ Debate\end{tabular} & \begin{tabular}[c]{@{}c@{}}w/o\\ Ref.\end{tabular} \\ \hline
 &  &  & 50.22$_{\pm22.91}$ & 64.45$_{\pm4.97}$ & 65.11$_{\pm7.53}$ & \multicolumn{1}{c|}{59.99$_{\pm14.15}$} &  & 90.47$_{\pm2.03}$ & 66.78$_{\pm3.18}$ & 66.65$_{\pm7.28}$ & \multicolumn{1}{c|}{79.24$_{\pm11.40}$} &  & 91.33$_{\pm1.73}$ & 66.81$_{\pm3.22}$ & 69.77$_{\pm2.24}$ & 82.49$_{\pm13.16}$ \\
 & \multirow{-2}{*}{DAPT2020} & \multirow{-2}{*}{68.00$_{\pm2.52}$} & {\color[HTML]{FF0000} -17.78} & {\color[HTML]{FF0000} -3.55} & {\color[HTML]{FF0000} -2.89} & \multicolumn{1}{c|}{{\color[HTML]{FF0000} -8.00}} & \multirow{-2}{*}{88.33$_{\pm8.47}$} & {\color[HTML]{00B050} +2.13} & {\color[HTML]{FF0000} -21.56} & {\color[HTML]{FF0000} -21.68} & \multicolumn{1}{c|}{{\color[HTML]{FF0000} -9.09}} & \multirow{-2}{*}{97.57$_{\pm0.55}$} & {\color[HTML]{FF0000} -6.24} & {\color[HTML]{FF0000} -30.76} & {\color[HTML]{FF0000} -27.80} & {\color[HTML]{FF0000} -15.08} \\
 &  &  & 3.00$_{\pm0.03}$ & 79.20$_{\pm34.93}$ & 78.99$_{\pm35.27}$ & \multicolumn{1}{c|}{99.53$_{\pm0.03}$} &  & 3.05$_{\pm0.02}$ & 99.52$_{\pm0.04}$ & 79.09$_{\pm35.24}$ & \multicolumn{1}{c|}{99.59$_{\pm0.02}$} &  & 25.92$_{\pm33.10}$ & 99.55$_{\pm0.04}$ & 79.09$_{\pm35.35}$ & 99.66$_{\pm0.03}$ \\
 & \multirow{-2}{*}{CICIDS2017} & \multirow{-2}{*}{98.54$_{\pm0.69}$} & {\color[HTML]{FF0000} -95.55} & {\color[HTML]{FF0000} -19.34} & {\color[HTML]{FF0000} -19.55} & \multicolumn{1}{c|}{{\color[HTML]{00B050} +0.98}} & \multirow{-2}{*}{99.56$_{\pm0.06}$} & {\color[HTML]{FF0000} -96.51} & {\color[HTML]{FF0000} -0.04} & {\color[HTML]{FF0000} -20.47} & \multicolumn{1}{c|}{{\color[HTML]{00B050} +0.03}} & \multirow{-2}{*}{99.62$_{\pm0.00}$} & {\color[HTML]{FF0000} -73.70} & {\color[HTML]{FF0000} -0.07} & {\color[HTML]{FF0000} -20.53} & {\color[HTML]{00B050} +0.04} \\
 &  &  & 5.85$_{\pm0.08}$ & 68.45$_{\pm54.12}$ & 97.88$_{\pm3.14}$ & \multicolumn{1}{c|}{99.70$_{\pm0.00}$} &  & 6.00$_{\pm0.09}$ & 99.84$_{\pm0.01}$ & 99.71$_{\pm0.00}$ & \multicolumn{1}{c|}{99.75$_{\pm0.08}$} &  & 6.02$_{\pm0.09}$ & 99.88$_{\pm0.08}$ & 99.75$_{\pm0.07}$ & 99.76$_{\pm0.08}$ \\
 & \multirow{-2}{*}{IoT23} & \multirow{-2}{*}{97.90$_{\pm3.14}$} & {\color[HTML]{FF0000} -92.05} & {\color[HTML]{FF0000} -29.45} & {\color[HTML]{FF0000} -0.02} & \multicolumn{1}{c|}{{\color[HTML]{00B050} +1.80}} & \multirow{-2}{*}{99.76$_{\pm0.08}$} & {\color[HTML]{FF0000} -93.76} & {\color[HTML]{00B050} +0.08} & {\color[HTML]{FF0000} -0.05} & \multicolumn{1}{c|}{{\color[HTML]{FF0000} -0.01}} & \multirow{-2}{*}{99.81$_{\pm0.15}$} & {\color[HTML]{FF0000} -93.80} & {\color[HTML]{00B050} +0.07} & {\color[HTML]{FF0000} -0.06} & {\color[HTML]{FF0000} -0.05} \\
 & \cellcolor[HTML]{EFEFEF} & \cellcolor[HTML]{EFEFEF} & \cellcolor[HTML]{EFEFEF}19.69$_{\pm25.63}$ & \cellcolor[HTML]{EFEFEF}70.70$_{\pm32.97}$ & \cellcolor[HTML]{EFEFEF}80.66$_{\pm23.03}$ & \multicolumn{1}{c|}{\cellcolor[HTML]{EFEFEF}86.41$_{\pm21.03}$} & \cellcolor[HTML]{EFEFEF} & \cellcolor[HTML]{EFEFEF}33.17$_{\pm43.00}$ & \cellcolor[HTML]{EFEFEF}88.71$_{\pm16.53}$ & \cellcolor[HTML]{EFEFEF}81.82$_{\pm23.08}$ & \multicolumn{1}{c|}{\cellcolor[HTML]{EFEFEF}92.86$_{\pm11.70}$} & \cellcolor[HTML]{EFEFEF} & \cellcolor[HTML]{EFEFEF}41.09$_{\pm42.06}$ & \cellcolor[HTML]{EFEFEF}88.75$_{\pm16.53}$ & \cellcolor[HTML]{EFEFEF}82.87$_{\pm22.14}$ & \cellcolor[HTML]{EFEFEF}93.97$_{\pm10.84}$ \\
\multirow{-8}{*}{\rotatebox{90}{\textbf{Technique-level}}} & \multirow{-2}{*}{\cellcolor[HTML]{EFEFEF}Average} & \multirow{-2}{*}{\cellcolor[HTML]{EFEFEF}88.15$_{\pm15.25}$} & \cellcolor[HTML]{EFEFEF}{\color[HTML]{FF0000} -68.46} & \cellcolor[HTML]{EFEFEF}{\color[HTML]{FF0000} -17.45} & \cellcolor[HTML]{EFEFEF}{\color[HTML]{FF0000} -7.49} & \multicolumn{1}{c|}{\cellcolor[HTML]{EFEFEF}{\color[HTML]{FF0000} -1.74}} & \multirow{-2}{*}{\cellcolor[HTML]{EFEFEF}95.88$_{\pm7.07}$} & \cellcolor[HTML]{EFEFEF}{\color[HTML]{FF0000} -62.71} & \cellcolor[HTML]{EFEFEF}{\color[HTML]{FF0000} -7.17} & \cellcolor[HTML]{EFEFEF}{\color[HTML]{FF0000} -14.07} & \multicolumn{1}{c|}{\cellcolor[HTML]{EFEFEF}{\color[HTML]{FF0000} -3.02}} & \multirow{-2}{*}{\cellcolor[HTML]{EFEFEF}99.00$_{\pm1.11}$} & \cellcolor[HTML]{EFEFEF}{\color[HTML]{FF0000} -57.91} & \cellcolor[HTML]{EFEFEF}{\color[HTML]{FF0000} -10.25} & \cellcolor[HTML]{EFEFEF}{\color[HTML]{FF0000} -16.13} & \cellcolor[HTML]{EFEFEF}{\color[HTML]{FF0000} -5.03} \\ \hline
 &  &  & 48.75$_{\pm24.60}$ & 64.47$_{\pm4.96}$ & 65.11$_{\pm7.53}$ & \multicolumn{1}{c|}{60.00$_{\pm14.13}$} &  & 89.08$_{\pm3.37}$ & 66.82$_{\pm3.17}$ & 66.67$_{\pm7.27}$ & \multicolumn{1}{c|}{79.27$_{\pm11.43}$} &  & 89.96$_{\pm3.40}$ & 66.94$_{\pm3.22}$ & 69.81$_{\pm2.23}$ & 82.56$_{\pm13.22}$ \\
 & \multirow{-2}{*}{DAPT2020} & \multirow{-2}{*}{68.00$_{\pm2.52}$} & {\color[HTML]{FF0000} -19.25} & {\color[HTML]{FF0000} -3.53} & {\color[HTML]{FF0000} -2.89} & \multicolumn{1}{c|}{{\color[HTML]{FF0000} -8.01}} & \multirow{-2}{*}{88.40$_{\pm8.49}$} & {\color[HTML]{00B050} +0.68} & {\color[HTML]{FF0000} -21.58} & {\color[HTML]{FF0000} -21.73} & \multicolumn{1}{c|}{{\color[HTML]{FF0000} -9.13}} & \multirow{-2}{*}{97.80$_{\pm0.38}$} & {\color[HTML]{FF0000} -7.84} & {\color[HTML]{FF0000} -30.85} & {\color[HTML]{FF0000} -27.99} & {\color[HTML]{FF0000} -15.24} \\
 &  &  & 3.01$_{\pm0.02}$ & 81.70$_{\pm30.61}$ & 99.36$_{\pm0.02}$ & \multicolumn{1}{c|}{99.05$_{\pm0.82}$} &  & 3.05$_{\pm0.02}$ & 99.52$_{\pm0.03}$ & 99.49$_{\pm0.02}$ & \multicolumn{1}{c|}{99.59$_{\pm0.02}$} &  & 25.92$_{\pm33.10}$ & 99.59$_{\pm0.01}$ & 99.49$_{\pm0.02}$ & 99.67$_{\pm0.03}$ \\
 & \multirow{-2}{*}{CICIDS2017} & \multirow{-2}{*}{99.37$_{\pm0.05}$} & {\color[HTML]{FF0000} -96.36} & {\color[HTML]{FF0000} -17.67} & {\color[HTML]{FF0000} -0.01} & \multicolumn{1}{c|}{{\color[HTML]{FF0000} -0.33}} & \multirow{-2}{*}{99.56$_{\pm0.06}$} & {\color[HTML]{FF0000} -96.51} & {\color[HTML]{FF0000} -0.04} & {\color[HTML]{FF0000} -0.07} & \multicolumn{1}{c|}{{\color[HTML]{00B050} +0.03}} & \multirow{-2}{*}{99.62$_{\pm0.01}$} & {\color[HTML]{FF0000} -73.70} & {\color[HTML]{FF0000} -0.03} & {\color[HTML]{FF0000} -0.13} & {\color[HTML]{00B050} +0.05} \\
 &  &  & 5.86$_{\pm0.08}$ & 99.69$_{\pm0.00}$ & 97.88$_{\pm3.14}$ & \multicolumn{1}{c|}{99.70$_{\pm0.00}$} &  & 6.02$_{\pm0.09}$ & 99.84$_{\pm0.01}$ & 99.71$_{\pm0.00}$ & \multicolumn{1}{c|}{99.76$_{\pm0.08}$} &  & 6.04$_{\pm0.09}$ & 99.90$_{\pm0.08}$ & 99.75$_{\pm0.07}$ & 99.77$_{\pm0.08}$ \\
 & \multirow{-2}{*}{IoT23} & \multirow{-2}{*}{97.90$_{\pm3.14}$} & {\color[HTML]{FF0000} -92.03} & {\color[HTML]{00B050} +1.80} & {\color[HTML]{FF0000} -0.02} & \multicolumn{1}{c|}{{\color[HTML]{00B050} +1.80}} & \multirow{-2}{*}{99.77$_{\pm0.08}$} & {\color[HTML]{FF0000} -93.75} & {\color[HTML]{00B050} +0.08} & {\color[HTML]{FF0000} -0.06} & \multicolumn{1}{c|}{{\color[HTML]{FF0000} -0.01}} & \multirow{-2}{*}{99.77$_{\pm0.08}$} & {\color[HTML]{FF0000} -93.73} & {\color[HTML]{00B050} +0.14} & {\color[HTML]{FF0000} -0.01} & {\color[HTML]{00B050} +0.00} \\
 & \cellcolor[HTML]{EFEFEF} & \cellcolor[HTML]{EFEFEF} & \cellcolor[HTML]{EFEFEF}19.21$_{\pm25.37}$ & \cellcolor[HTML]{EFEFEF}81.96$_{\pm21.75}$ & \cellcolor[HTML]{EFEFEF}87.45$_{\pm17.26}$ & \multicolumn{1}{c|}{\cellcolor[HTML]{EFEFEF}86.25$_{\pm20.92}$} & \cellcolor[HTML]{EFEFEF} & \cellcolor[HTML]{EFEFEF}32.72$_{\pm42.32}$ & \cellcolor[HTML]{EFEFEF}88.73$_{\pm16.51}$ & \cellcolor[HTML]{EFEFEF}88.62$_{\pm16.86}$ & \multicolumn{1}{c|}{\cellcolor[HTML]{EFEFEF}92.87$_{\pm11.70}$} & \cellcolor[HTML]{EFEFEF} & \cellcolor[HTML]{EFEFEF}40.64$_{\pm41.46}$ & \cellcolor[HTML]{EFEFEF}88.81$_{\pm16.48}$ & \cellcolor[HTML]{EFEFEF}89.68$_{\pm14.95}$ & \cellcolor[HTML]{EFEFEF}94.00$_{\pm10.83}$ \\
\multirow{-8}{*}{\rotatebox{90}{\textbf{Tactic-level}}} & \multirow{-2}{*}{\cellcolor[HTML]{EFEFEF}Average} & \multirow{-2}{*}{\cellcolor[HTML]{EFEFEF}88.42$_{\pm15.46}$} & \cellcolor[HTML]{EFEFEF}{\color[HTML]{FF0000} -69.21} & \cellcolor[HTML]{EFEFEF}{\color[HTML]{FF0000} -6.47} & \cellcolor[HTML]{EFEFEF}{\color[HTML]{FF0000} -0.97} & \multicolumn{1}{c|}{\cellcolor[HTML]{EFEFEF}{\color[HTML]{FF0000} -2.18}} & \multirow{-2}{*}{\cellcolor[HTML]{EFEFEF}95.91$_{\pm7.05}$} & \cellcolor[HTML]{EFEFEF}{\color[HTML]{FF0000} -63.19} & \cellcolor[HTML]{EFEFEF}{\color[HTML]{FF0000} -7.18} & \cellcolor[HTML]{EFEFEF}{\color[HTML]{FF0000} -7.29} & \multicolumn{1}{c|}{\cellcolor[HTML]{EFEFEF}{\color[HTML]{FF0000} -3.04}} & \multirow{-2}{*}{\cellcolor[HTML]{EFEFEF}99.06$_{\pm0.97}$} & \cellcolor[HTML]{EFEFEF}{\color[HTML]{FF0000} -58.42} & \cellcolor[HTML]{EFEFEF}{\color[HTML]{FF0000} -10.25} & \cellcolor[HTML]{EFEFEF}{\color[HTML]{FF0000} -9.38} & \cellcolor[HTML]{EFEFEF}{\color[HTML]{FF0000} -5.06} \\ \hline
\end{tabular}
	}
\end{table*}

\subsection{RQ2: Ablation Study}
\label{sec:ablation}

To evaluate the contributions of \tool's core components—the Abstraction Module, Inference Module, and Refinement Module—we conduct an ablation study with three variants:

\begin{itemize}
	\item \textbf{\tool-No-Abstraction}, which bypasses the Abstraction Module and feeds compressed flow summaries directly to the reasoning stage;
	\item \textbf{\tool-Alt-Consistency}, replacing our structured inference strategy with standard consistency decoding methods (\eg multi-sampling and self-debate); and	
	\item \textbf{\tool-No-Refinement}, \revise{which omits the refinement stage and returns the fusion output directly.}
\end{itemize}

As summarized in Table~\ref{tab:ablation}, we report the performance of these variants using ChatGPT-4o across three datasets—DAPT2020, CICIDS2017, and IoT23—in terms of top-1, top-3, and top-5 accuracy. Detailed results for other LLMs are provided in the supplemental material. Below, we discuss the key findings from this study.

\noindent \textbf{Impact of the Abstraction Module.} Disabling the Abstraction Module leads to the most severe performance degradation, with an average top-1 accuracy drop of 68.46 percentage points at the technique level and 69.21 at the tactic level, \revise{with drops of 95.55 and 92.05 points on CICIDS2017 and IoT23}. This underscores the module's critical role in bridging low-level flow data with high-level intent. For instance, attacks like DoS lack explicit textual patterns and rely on subtle indicators (\eg flow duration and volume). Without abstraction, the model struggles to contextualize such signals, often misclassifying them.

\noindent \textbf{Effectiveness of Partitioned Reasoning.} Our structured inference strategy outperforms standard consistency methods (multi-sampling and self-debate) by about 1-17\% in average accuracy. While baseline methods improve output stability, they fail to guide the model through complex, multi-step reasoning—particularly evident in DAPT2020, where top-5 accuracy drops by $\sim$30\% compared to \tool. By partitioning the 14 tactics into 5 semantically coherent groups, our approach narrows the search space, enabling the model to focus on specific subtasks. This not only improves accuracy but also ensures better semantic alignment in predictions.

\noindent \textbf{Role of the Refinement Module.} While disabling refinement yields a modest average accuracy decline (1.74\% at technique level of top-1 accuracy), its impact varies by dataset. For DAPT2020, the absence of refinement causes an 8\% drop. In contrast, for IoT23 and CICIDS2017—where earlier modules already achieve high accuracy—the module's contribution is less pronounced. Beyond accuracy, refinement produces a ranked list of candidate techniques together with natural-language justifications that document the basis for each mapping decision.

\noindent \textbf{Partitioning Strategy Ablation}. To assess whether the partitioning choice in Section~\ref{subsec:inference_module} affects performance, we compare the default 5-group configuration against four alternatives. \textbf{V1 (3 groups)} merges the five default groups into three coarser ones along the same kill-chain ordering. \textbf{V2 (7 groups)} subdivides the kill chain into seven groups of two tactics each. \textbf{V3 (random 5 groups)} preserves the default group-size distribution but assigns tactics to groups randomly (seed=42), breaking the kill-chain ordering. \textbf{V4 (no partitioning)} performs a single inference pass over all 14 tactics. We evaluate all variants on Gemini~2.5~Flash and DeepSeek~V3 across the three datasets, reporting top-1 technique accuracy in Table~\ref{tab:partition_ablation}.

\begin{table}[]
\centering
\caption{Top-1 technique accuracy of \tool with alternative tactic partitioning strategies on Gemini~2.5~Flash and DeepSeek~V3 across three datasets: DAPT2020 (D), CICIDS2017 (C) and IoT23 (I). \textbf{Best} result is in bold; \underline{second best} is underlined. Each value is the mean across three runs.}
\label{tab:partition_ablation}
\scalebox{0.74}{
\begin{tabular}{lccccccc}
\hline
 &  & \textbf{} & \textbf{\tool} & \textbf{V1} & \textbf{V2} & \textbf{V3} & \textbf{V4} \\ \hline
\multicolumn{1}{l|}{} & \multicolumn{1}{c|}{} & \multicolumn{1}{c|}{D} & \textbf{69.04$_{\pm17.98}$} & 62.50$_{\pm20.12}$ & {\ul 67.72$_{\pm25.51}$} & 54.96$_{\pm14.93}$ & 48.91$_{\pm8.11}$ \\
\multicolumn{1}{l|}{} & \multicolumn{1}{c|}{} & \multicolumn{1}{c|}{C} & 96.52$_{\pm5.13}$ & \textbf{99.57$_{\pm0.03}$} & {\ul 99.56$_{\pm0.04}$} & 99.54$_{\pm0.08}$ & 99.52$_{\pm0.05}$ \\
\multicolumn{1}{l|}{} & \multicolumn{1}{c|}{} & \multicolumn{1}{c|}{I} & \textbf{99.79$_{\pm0.07}$} & 99.74$_{\pm0.07}$ & {\ul 99.75$_{\pm0.08}$} & 99.70$_{\pm0.01}$ & 97.92$_{\pm3.18}$ \\
\multicolumn{1}{l|}{} & \multicolumn{1}{c|}{\multirow{-4}{*}{\rotatebox{90}{\textbf{Gemini}}}} & \multicolumn{1}{c|}{\cellcolor[HTML]{EFEFEF}Ave.} & \cellcolor[HTML]{EFEFEF}{\ul 88.45$_{\pm17.36}$} & \cellcolor[HTML]{EFEFEF}87.27$_{\pm18.97}$ & \cellcolor[HTML]{EFEFEF}\textbf{89.01$_{\pm20.44}$} & \cellcolor[HTML]{EFEFEF}84.73$_{\pm23.54}$ & \cellcolor[HTML]{EFEFEF}82.12$_{\pm25.29}$ \\ \cline{2-8} 
\multicolumn{1}{l|}{} & \multicolumn{1}{c|}{} & \multicolumn{1}{c|}{D} & \textbf{65.17$_{\pm5.21}$} & {\ul 62.14$_{\pm25.36}$} & 56.54$_{\pm12.99}$ & 56.54$_{\pm18.91}$ & 53.14$_{\pm16.84}$ \\
\multicolumn{1}{l|}{} & \multicolumn{1}{c|}{} & \multicolumn{1}{c|}{C} & 94.21$_{\pm4.21}$ & {\ul 98.62$_{\pm1.55}$} & \textbf{99.21$_{\pm0.47}$} & 88.10$_{\pm19.70}$ & 98.62$_{\pm1.57}$ \\
\multicolumn{1}{l|}{} & \multicolumn{1}{c|}{} & \multicolumn{1}{c|}{I} & {\ul 99.76$_{\pm0.07}$} & 96.08$_{\pm2.99}$ & \textbf{99.88$_{\pm0.09}$} & 99.72$_{\pm0.01}$ & 97.94$_{\pm3.18}$ \\
\multicolumn{1}{l|}{\multirow{-8}{*}{\rotatebox{90}{\textbf{Micro-Average(\%)}}}} & \multicolumn{1}{c|}{\multirow{-4}{*}{\rotatebox{90}{\textbf{DeepSeek}}}} & \multicolumn{1}{c|}{\cellcolor[HTML]{EFEFEF}Ave.} & \cellcolor[HTML]{EFEFEF}\textbf{86.38$_{\pm16.43}$} & \cellcolor[HTML]{EFEFEF}{\ul 85.61$_{\pm21.79}$} & \cellcolor[HTML]{EFEFEF}85.21$_{\pm22.46}$ & \cellcolor[HTML]{EFEFEF}81.45$_{\pm23.68}$ & \cellcolor[HTML]{EFEFEF}83.23$_{\pm24.16}$ \\ \hline
\multicolumn{1}{l|}{} & \multicolumn{1}{c|}{} & \multicolumn{1}{c|}{D} & \textbf{56.08$_{\pm3.61}$} & {\ul 55.93$_{\pm5.91}$} & 47.44$_{\pm10.18}$ & 51.63$_{\pm5.01}$ & 48.71$_{\pm4.66}$ \\
\multicolumn{1}{l|}{} & \multicolumn{1}{c|}{} & \multicolumn{1}{c|}{C} & {\ul 76.50$_{\pm0.85}$} & 76.17$_{\pm0.54}$ & 76.16$_{\pm1.95}$ & \textbf{79.15$_{\pm7.30}$} & 73.42$_{\pm1.39}$ \\
\multicolumn{1}{l|}{} & \multicolumn{1}{c|}{} & \multicolumn{1}{c|}{I} & {\ul 65.69$_{\pm1.95}$} & 63.06$_{\pm0.91}$ & \textbf{72.85$_{\pm14.55}$} & 62.88$_{\pm1.85}$ & 52.99$_{\pm15.28}$ \\
\multicolumn{1}{l|}{} & \multicolumn{1}{c|}{\multirow{-4}{*}{\rotatebox{90}{\textbf{Gemini}}}} & \multicolumn{1}{c|}{\cellcolor[HTML]{EFEFEF}Ave.} & \cellcolor[HTML]{EFEFEF}\textbf{66.09$_{\pm9.09}$} & \cellcolor[HTML]{EFEFEF}65.05$_{\pm9.38}$ & \cellcolor[HTML]{EFEFEF}{\ul 65.48$_{\pm16.28}$} & \cellcolor[HTML]{EFEFEF}64.55$_{\pm12.81}$ & \cellcolor[HTML]{EFEFEF}58.37$_{\pm13.97}$ \\ \cline{2-8} 
\multicolumn{1}{l|}{} & \multicolumn{1}{c|}{} & \multicolumn{1}{c|}{D} & 43.27$_{\pm3.56}$ & 46.17$_{\pm3.90}$ & 44.49$_{\pm2.90}$ & {\ul 48.28$_{\pm4.73}$} & \textbf{51.20$_{\pm4.32}$} \\
\multicolumn{1}{l|}{} & \multicolumn{1}{c|}{} & \multicolumn{1}{c|}{C} & \textbf{75.23$_{\pm1.31}$} & 70.86$_{\pm5.46}$ & {\ul 73.22$_{\pm5.23}$} & 72.20$_{\pm3.42}$ & 70.10$_{\pm6.07}$ \\
\multicolumn{1}{l|}{} & \multicolumn{1}{c|}{} & \multicolumn{1}{c|}{I} & {\ul 72.94$_{\pm0.22}$} & 62.35$_{\pm13.74}$ & \textbf{76.24$_{\pm18.14}$} & 72.02$_{\pm1.15}$ & 63.60$_{\pm16.34}$ \\
\multicolumn{1}{l|}{\multirow{-8}{*}{\rotatebox{90}{\textbf{Macro-Average(\%)}}}} & \multicolumn{1}{c|}{\multirow{-4}{*}{\rotatebox{90}{\textbf{DeepSeek}}}} & \multicolumn{1}{c|}{\cellcolor[HTML]{EFEFEF}Ave.} & \cellcolor[HTML]{EFEFEF}{\ul 63.81$_{\pm15.56}$} & \cellcolor[HTML]{EFEFEF}59.80$_{\pm13.28}$ & \cellcolor[HTML]{EFEFEF}\textbf{64.65$_{\pm17.93}$} & \cellcolor[HTML]{EFEFEF}64.17$_{\pm12.28}$ & \cellcolor[HTML]{EFEFEF}61.63$_{\pm12.24}$ \\ \hline
\end{tabular}
}
\end{table}

Three patterns emerge under micro-averaging. First, all three kill-chain partitionings (V1, default, V2) outperform no partitioning (V4) on both backbones: even the weakest kill-chain configuration exceeds V4 by 5.15 points on Gemini and 1.98 points on DeepSeek on average across the three datasets. Second, random tactic assignment with the same group sizes (V3) does not deliver this gain: it sits below V4 on DeepSeek and below every kill-chain configuration on both backbones, indicating that the gain comes from the kill-chain ordering rather than from partitioning the search space alone. Third, within the kill-chain family the differences are small, with the spread between 3, 5, and 7 groups within 1.74 points on Gemini and 1.17 points on DeepSeek. \revise{Under macro-averaging the ordering is less stable: on DeepSeek V3, V1 (59.80) falls below V4 (61.63) and V3 (64.17) is on par with the kill-chain configurations, so the advantage of kill-chain ordering holds on average rather than in every setting.} We adopt the 5-group configuration as it balances search-space reduction against token consumption.

\subsection{RQ3: Error Analysis}
\label{sec:error_analysis}

This section presents a comprehensive analysis of error cases in our method, focusing on two main aspects: (1) mapping errors in technique identification and (2) tactical consistency in technique-to-tactic associations. These analyses aim to uncover systemic weaknesses and identify opportunities for improvement in the model's reasoning and refinement processes.

\begin{table}[]
	\centering
	\caption{Error rates (in percentage) of different error types for \tool (Ours) and the baseline method (ToT) using ChatGPT-4o across three datasets DAPT2020, CICIDS2017, IoT23 and in total.}
	\label{tab:error_analysis}
    \scalebox{0.85}{
	\begin{tabular}{|c|c|ccc|}
\hline
 &  & \multicolumn{3}{c|}{\textbf{Error Rate (\%)}} \\ \cline{3-5} 
\multirow{-2}{*}{\textbf{Dataset}} & \multirow{-2}{*}{\textbf{Method}} & \begin{tabular}[c]{@{}c@{}}Inference\\ Error\end{tabular} & \begin{tabular}[c]{@{}c@{}}Technique\\ Confusion\end{tabular} & \begin{tabular}[c]{@{}c@{}}Over-\\ inference\end{tabular} \\ \hline
 & Ours & 5.29 & 1.80 & 5.35 \\
\multirow{-2}{*}{DAPT2020} & ToT & 14.02 & 2.72 & 3.84 \\ \hline
 & Ours & 2.76 & 0 & 0 \\
\multirow{-2}{*}{CICIDS2017} & ToT & 10.28 & 0 & 0 \\ \hline
 & Ours & 1.26 & 0.05 & 0 \\
\multirow{-2}{*}{IoT23} & ToT & 6.89 & 0 & 0 \\ \hline
\rowcolor[HTML]{EFEFEF} 
\cellcolor[HTML]{EFEFEF} & Ours & 9.31 & 1.85 & 5.35 \\
\rowcolor[HTML]{EFEFEF} 
\multirow{-2}{*}{\cellcolor[HTML]{EFEFEF}\textbf{Total}} & ToT & 31.19 & 2.72 & 3.84 \\ \hline
\end{tabular}
}
\end{table}

\subsubsection{Mapping Error Analysis}

To quantify the effectiveness of our method, we performed a manual examination of top-1 mapping errors and compared them with baseline ToT. Errors were identified by contrasting LLM-generated mappings with ground-truth malicious behavior descriptions in the dataset. Our analysis revealed three primary error categories: (1) \textbf{Inference Errors} occur when the model misinterprets network behaviors or attacker intent. For instance, the technique \textit{T1071: Application Layer Protocol} (C2 communication) might be incorrectly mapped to \textit{T1046: Network Service Discovery} due to ambiguities in behavioral interpretation. Such errors often stem from the model's inability to discern subtle distinctions in attack semantics. (2) \textbf{Technique Confusion} arises when semantically similar techniques are conflated. Examples include \textit{T1595: Active Scanning} and \textit{T1046: Network Service Discovery}, which share overlapping behaviors but differ in tactical objectives. The LLM occasionally overlooks critical metadata or contextual cues, leading to error. (3) \textbf{Over Inference} manifests when the model overemphasizes localized malicious features while neglecting broader contextual signals. This is particularly prevalent in logs containing anomalous but benign patterns (\eg HTTP status codes 302/404 or SQL queries) that superficially resemble attack artifacts.

As summarized in Table~\ref{tab:error_analysis}, which presents the error rates of ChatGPT-4o across the DAPT2020, CICIDS2017, and IoT23 datasets, several key observations emerge.

\begin{itemize}
	\item \textbf{Inference Accuracy Improvement:} Our method reduces inference errors by 21.88\% (absolute) compared to ToT, demonstrating superior capability in disambiguating attack intent. This improvement is attributed to the structured reasoning module, which systematically decomposes behavioral patterns and tactical objectives.
	\item \textbf{Technique Confusion Mitigation:} While our approach reduces technique confusion errors by 31.99\% (relative), complete elimination remains challenging. Subtle distinctions between techniques (\eg scanning for reconnaissance vs. lateral movement) often require explicit guidance, which the refinement module partially addresses through standardized definitions. However, edge cases persist due to the inherent complexity of attack taxonomies.
	\item \textbf{Context-Detail Trade-off:} A marginal increase in over-inference errors (1.51\% absolute) was observed, likely due to the refinement module's emphasis on tactical scope narrowing. Although this focus improves precision, it occasionally leads to overfitting on local features. Despite this trade-off, our method achieves a net reduction in total errors.
\end{itemize}

Collectively, these results underscore our method's advantages over ToT in accuracy and error-mode reduction. Table~\ref{tab:error_overall} reports the per-class error rates of \tool across all four backbones. The overall trend is consistent with the analysis above.

\begin{table}[!t]
\caption{Error rates (\%) of \tool across four backbone LLMs.}
\label{tab:error_overall}
\centering
\scalebox{0.8}{
\begin{tabular}{lccc}
\hline
\textbf{Backbone} & \textbf{Inference Error} & \textbf{Technique Confusion} & \textbf{Over-inference} \\ \hline
ChatGPT-4o & 9.31 & 1.85 & 5.35 \\
Claude Sonnet 4 & 11.02 & 2.28 & 4.88 \\
Gemini 2.5 Flash & 9.65 & 1.64 & 4.94 \\
DeepSeek V3 & 10.45 & 1.65 & 6.09 \\ \hline
\end{tabular}}
\end{table}

\begin{table*}[]
	\centering
	\caption{Tactical consistency evaluation across three datasets (DAPT2020, CICIDS2017, IoT23). For each dataset, the table reports the number of techniques correctly identified (\# correct), the number of mismatched tactics (\# mismatched), and the mismatch rate (reported as a percentage). The evaluation is presented for top-1, top-3, and top-5 settings. The values for \# correct, \# mismatched, and rate are the averages from three independent experiments conducted on each dataset.}
	\label{tab:tactical_consistency}
    \scalebox{0.95}{
	\begin{tabular}{|c|c|ccc|ccc|ccc|}
\hline
 &  & \multicolumn{3}{c|}{\textbf{Top-1}} & \multicolumn{3}{c|}{\textbf{Top-3}} & \multicolumn{3}{c|}{\textbf{Top-5}} \\
\multirow{-2}{*}{\textbf{Model}} & \multirow{-2}{*}{\textbf{Dataset}} & \# Correct & \# Mismatched & Rate & \# Correct & \# Mismatched & Rate & \# Correct & \# Mismatched & Rate \\ \hline
 & DAPT2020 & 10,806.67 & 163.67 & 1.51\% & 16,505.00 & 367.33 & 2.23\% & 22,336.33 & 369.67 & 1.66\% \\
 & CICIDS2017 & 273,385.33 & 0.00 & 0.00\% & 376,053.67 & 4,970.33 & 1.32\% & 376,228.33 & 4,972.67 & 1.32\% \\
 & IoT23 & 72,994.33 & 0.00 & 0.00\% & 145,676.67 & 0.00 & 0.00\% & 145,680.00 & 1.33 & 0.00\% \\
\multirow{-4}{*}{\textbf{ChatGPT-4o}} & \cellcolor[HTML]{F2F2F2}\textbf{Total} & \cellcolor[HTML]{F2F2F2}357,186.33 & \cellcolor[HTML]{F2F2F2}163.67 & \cellcolor[HTML]{F2F2F2}0.05\% & \cellcolor[HTML]{F2F2F2}538,235.33 & \cellcolor[HTML]{F2F2F2}5,337.67 & \cellcolor[HTML]{F2F2F2}0.99\% & \cellcolor[HTML]{F2F2F2}544,244.67 & \cellcolor[HTML]{F2F2F2}5,343.67 & \cellcolor[HTML]{F2F2F2}0.98\% \\ \hline
 & DAPT2020 & 9,707.33 & 0.00 & 0.00\% & 16,652.33 & 0.00 & 0.00\% & 18,185.67 & 150.33 & 0.83\% \\
 & CICIDS2017 & 225,092.33 & 0.00 & 0.00\% & 347,457.33 & 0.00 & 0.00\% & 350,862.67 & 0.00 & 0.00\% \\
 & IoT23 & 71,696.67 & 0.00 & 0.00\% & 96,960.33 & 0.00 & 0.00\% & 97,134.67 & 0.00 & 0.00\% \\
\multirow{-4}{*}{\textbf{\begin{tabular}[c]{@{}c@{}}Claude\\ Sonnet 4\end{tabular}}} & \cellcolor[HTML]{F2F2F2}\textbf{Total} & \cellcolor[HTML]{F2F2F2}306,496.33 & \cellcolor[HTML]{F2F2F2}0.00 & \cellcolor[HTML]{F2F2F2}0.00\% & \cellcolor[HTML]{F2F2F2}461,070.00 & \cellcolor[HTML]{F2F2F2}0.00 & \cellcolor[HTML]{F2F2F2}0.00\% & \cellcolor[HTML]{F2F2F2}466,183.00 & \cellcolor[HTML]{F2F2F2}150.33 & \cellcolor[HTML]{F2F2F2}0.03\% \\ \hline
 & DAPT2020 & 8,926.67 & 460.33 & 5.16\% & 18,467.00 & 463.67 & 2.51\% & 18,755.00 & 463.67 & 2.47\% \\
 & CICIDS2017 & 267,187.33 & 1,324.00 & 0.50\% & 643,048.00 & 2,310.67 & 0.36\% & 652,190.00 & 3,634.67 & 0.56\% \\
 & IoT23 & 74,401.33 & 0.00 & 0.00\% & 191,186.00 & 0.00 & 0.00\% & 214,609.00 & 0.00 & 0.00\% \\
\multirow{-4}{*}{\textbf{\begin{tabular}[c]{@{}c@{}}Gemini 2.5\\ Flash\end{tabular}}} & \cellcolor[HTML]{F2F2F2}\textbf{Total} & \cellcolor[HTML]{F2F2F2}350,515.33 & \cellcolor[HTML]{F2F2F2}1,784.33 & \cellcolor[HTML]{F2F2F2}0.51\% & \cellcolor[HTML]{F2F2F2}852,701.00 & \cellcolor[HTML]{F2F2F2}2,774.33 & \cellcolor[HTML]{F2F2F2}0.33\% & \cellcolor[HTML]{F2F2F2}885,554.00 & \cellcolor[HTML]{F2F2F2}4,098.33 & \cellcolor[HTML]{F2F2F2}0.46\% \\ \hline
 & DAPT2020 & 7,756.00 & 0.00 & 0.00\% & 10,125.00 & 1.00 & 0.01\% & 10,214.33 & 1.67 & 0.02\% \\
 & CICIDS2017 & 260,432.00 & 0.00 & 0.00\% & 374,401.00 & 0.00 & 0.00\% & 381,858.67 & 1.33 & 0.00\% \\
 & IoT23 & 74,418.33 & 0.00 & 0.00\% & 98,075.67 & 0.00 & 0.00\% & 98,145.33 & 0.00 & 0.00\% \\
\multirow{-4}{*}{\textbf{DeepSeek V3}} & \cellcolor[HTML]{F2F2F2}\textbf{Total} & \cellcolor[HTML]{F2F2F2}342,606.33 & \cellcolor[HTML]{F2F2F2}0.00 & \cellcolor[HTML]{F2F2F2}0.00\% & \cellcolor[HTML]{F2F2F2}482,601.67 & \cellcolor[HTML]{F2F2F2}1.00 & \cellcolor[HTML]{F2F2F2}0.00\% & \cellcolor[HTML]{F2F2F2}490,218.33 & \cellcolor[HTML]{F2F2F2}3.00 & \cellcolor[HTML]{F2F2F2}0.00\% \\ \hline
\end{tabular}
}
\end{table*}

\subsubsection{Tactical Consistency Analysis}

We further evaluated the alignment between correctly identified techniques and their associated tactics. For each dataset, we computed the mismatch rate, which is the proportion of correct technique-to-tactic mappings that are incorrect. Table~\ref{tab:tactical_consistency} presents results for the top-1/3/5 settings across ChatGPT-4o, Claude Sonnet 4, Gemini 2.5 Flash, and DeepSeek V3:

\begin{itemize}
\item \textbf{ChatGPT-4o} shows a low total mismatch rate ranging from 0.05\% to 0.99\%, indicating occasional tactical misassignments. However, its high technique-level accuracy ensures minimal operational impact.
\item \textbf{Claude Sonnet 4} maintains nearly perfect consistency, with no mismatches for top-1/3, except for a minor mismatch (0.83\%) at top-5 on the DAPT2020 dataset, suggesting its dependable technique-to-tactic alignment.
\item \textbf{Gemini 2.5 Flash} demonstrates a total mismatch rate ranging from 0.33\% to 0.51\%, with a particularly high top-1 mismatch rate of 5.16\% on the DAPT2020 dataset. Despite this, Gemini achieves the highest top-1 accuracy of 88.45\%, suggesting that while it excels in overall accuracy, the model may still encounter occasional misclassifications on specific data points.
\item \textbf{DeepSeek V3} maintains near-perfect consistency, with no mismatches at top-1 across all datasets and only minor mismatches (range from 0.01\% to 0.02\%) at top-3/5, demonstrating its precise mapping approach that prioritizes tactical alignment over extensive coverage.
\end{itemize}

Claude Sonnet 4 and DeepSeek V3 maintain nearly perfect consistency, demonstrating the highest level of tactical alignment. Meanwhile, ChatGPT-4o shows strong consistency, slight mismatches are minimal in impact compared to its superior technique detection. On the other hand, Gemini 2.5 Flash achieves higher accuracy but sacrifices some precision, which occasionally leads to tactical misassignments. These results demonstrate the adaptability of our method across various backbone models, ensuring tactical coherence without compromising performance.

\subsection{RQ4: Practicality Study}
\label{sec:practicality}

To validate the real-world applicability of our approach, we integrate our system with an existing CNN+LSTM-based NIDS framework, evaluating \revise{how accurately it maps the traffic that the NIDS flags as malicious}. We conduct experiments on two benchmark datasets—CICIDS2017 (diverse attack types) and IoT23 (botnet-focused)—which reflect realistic attack scenarios. Our study assesses how effectively our system refines the mapping of attack techniques and tactics after the NIDS identifies malicious traffic.

\noindent \textbf{Experimental Setup.} We deploy the CNN+LSTM NIDS model from Bamber et al.~\cite{bamber2025hybrid}, pretrained on CICIDS2017 and IoT23. Each dataset is split into 80\% training and 20\% testing data, with the model trained for 5 epochs. \revise{The NIDS performs detection: it classifies traffic as benign or malicious. \tool then performs mapping on the detected subset only: for samples the NIDS labels malicious, \tool's abstraction, inference, and refinement modules map the behaviors to MITRE ATT\&CK techniques and tactics. Note that benign-flagged samples are not passed to \tool, and this study does not evaluate \tool's behavior on NIDS false positives or on malicious traffic the NIDS missed.}

We evaluate NIDS detection performance using standard metrics: True Positives (TP), True Negatives (TN), False Positives (FP), False Negatives (FN), Accuracy (AC), Precision (PR), Recall (RC), and F1 Score. Specifically, a TP occurs when the system correctly identifies a malicious network flow based on its feature representation, while a TN represents the correct classification of a benign flow. Conversely, an FP arises when the system misclassifies a benign flow as malicious, and an FN occurs when a malicious flow is erroneously classified as benign. For attack behavior mapping, we measure top-1, top-3, and top-5 accuracy for technique and tactic predictions.

\begin{table}[]
	\centering
	\caption{The performance metrics of the CNN+LSTM-based NIDS on the CICIDS2017 and IoT23 datasets, including True Positives (TP), True Negatives (TN), False Positives (FP), False Negatives (FN), and standard evaluation metrics—Accuracy (AC), Precision (PR), Recall (RC), and F1 Score.}
	\label{tab:nids_detection}
	\begin{tabular}{|c|cccc|}
\hline
\textbf{Dataset} & \multicolumn{4}{c|}{\textbf{Result Statistics}} \\ \hline
 & \cellcolor[HTML]{EFEFEF}\textbf{\# TPs} & \cellcolor[HTML]{EFEFEF}\textbf{\# TNs} & \cellcolor[HTML]{EFEFEF}\textbf{\# FPs} & \cellcolor[HTML]{EFEFEF}\textbf{\# FNs} \\
 & 103,150 & 242,404 & 49 & 68 \\
 & \cellcolor[HTML]{EFEFEF}\textbf{AC (\%)} & \cellcolor[HTML]{EFEFEF}\textbf{PR (\%)} & \cellcolor[HTML]{EFEFEF}\textbf{RC (\%)} & \cellcolor[HTML]{EFEFEF}\textbf{F1 (\%)} \\
\multirow{-4}{*}{\textbf{CICIDS2017}} & 99.97 & 99.95 & 99.93 & 99.94 \\ \hline
 & \cellcolor[HTML]{EFEFEF}\textbf{\# TPs} & \cellcolor[HTML]{EFEFEF}\textbf{\# TNs} & \cellcolor[HTML]{EFEFEF}\textbf{\# FPs} & \cellcolor[HTML]{EFEFEF}\textbf{\# FNs} \\
 & 1,398,062 & 1,575 & 82 & 42 \\
 & \cellcolor[HTML]{EFEFEF}\textbf{AC (\%)} & \cellcolor[HTML]{EFEFEF}\textbf{PR (\%)} & \cellcolor[HTML]{EFEFEF}\textbf{RC (\%)} & \cellcolor[HTML]{EFEFEF}\textbf{F1 (\%)} \\
\multirow{-4}{*}{\textbf{IoT23}} & 99.99 & 99.99 & 99.99 & 99.99 \\ \hline
\end{tabular}
\end{table}

\begin{figure}
	\centering
	\includegraphics[width=\linewidth]{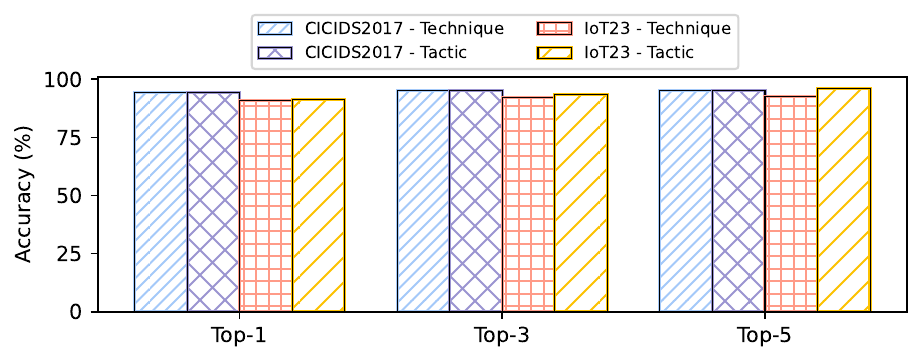}
	\caption{The top-1, top-3, and top-5 weighted accuracy rates of \tool in mapping detected attack behaviors to techniques and tactics for the CICIDS2017 and IoT23 datasets.}
	\label{fig:mapping_result}
\end{figure}

\noindent \textbf{Results and Analysis.} As shown in Table~\ref{tab:nids_detection}, the NIDS achieves near-perfect detection on both datasets (99.97\% AC for CICIDS2017; 99.99\% for IoT23), demonstrating the CNN+LSTM model's robustness. Post-detection, our system attains 94\% top-1 accuracy for technique/tactic mapping on CICIDS2017 (Figure~\ref{fig:mapping_result}). IoT23 exhibits slightly lower but still strong performance ($>$90\% top-1), with top-5 tactic accuracy exceeding 98\%, highlighting our method's adaptability to specialized attack patterns (\eg botnets).

These results underscore two key contributions:

\begin{itemize}
	\item \textbf{Integration feasibility:} \tool augments NIDS outputs with MITRE ATT\&CK technique mappings without compromising the underlying detection accuracy.
	\item \textbf{Scalability:} Strong performance across heterogeneous datasets (general-purpose CICIDS2017 and IoT-specific IoT23) suggests broader applicability.
\end{itemize}

This study confirms our approach's practicality \revise{as a post-detection mapping layer}, bridging the gap between detection and actionable threat intelligence.
\subsection{RQ5: Contamination Probe}

The three benchmarks used in Section~\ref{subsec:rq1_performance}, CICIDS2017, DAPT2020, and IoT23, predate the training cutoffs of every backbone we evaluate and have been widely analyzed in public literature. \tool's strong performance could therefore partly reflect prior exposure during pretraining rather than genuine flow interpretation. To address this concern, we evaluate \tool on a private dataset that no backbone could have observed during training, and ask whether the advantage observed in Section~\ref{subsec:rq1_performance} persists when this exposure is ruled out.

The dataset is a private collection assembled in 2026 that has never been publicly released. Nine attacks are executed in a multi-VM environment, spanning the major stages of the attack kill chain. These include HTTP vulnerability scanning, SSH brute-force, HTTP brute-force, vulnerability exploitation, reverse shell, SMB lateral movement, data exfiltration, denial-of-service, and command-and-control beaconing. The captured traffic is converted to flows via the same CICFlowMeter pipeline used on the public benchmarks, yielding 135,929 labeled flows across the nine techniques.

\begin{table}[ht!]
\centering
\caption{Top-1 technique-level micro- and macro-accuracy (\%) on the private contamination-probe dataset, reported as mean$\pm$std over three independent runs. Micro accuracy averages over all flows and is therefore dominated by high-volume techniques. Best result in each row is shown in bold.}
\label{tab:freshlab_top1}
\scalebox{0.85}{
\begin{tabular}{l|c|cccc}
\hline
Metric & Backbone & \tool & CoT & ToT & Vanilla \\ \hline
\multirow{4}{*}{Micro} & Claude & \textbf{99.58}$_{\pm0.07}$ & 2.33$_{\pm0.08}$ & 2.28$_{\pm0.02}$ & 2.30$_{\pm0.01}$ \\
 & ChatGPT & \textbf{95.59}$_{\pm7.03}$ & 1.11$_{\pm0.85}$ & 2.35$_{\pm0.11}$ & 1.91$_{\pm0.02}$ \\
 & Gemini & \textbf{95.10}$_{\pm6.64}$ & 2.27$_{\pm0.02}$ & 2.29$_{\pm0.02}$ & 0.65$_{\pm0.03}$ \\
 & DeepSeek & \textbf{99.61}$_{\pm0.13}$ & 2.36$_{\pm0.13}$ & 2.42$_{\pm0.11}$ & 1.12$_{\pm1.00}$ \\ \hline
\multirow{4}{*}{Macro} & Claude & \textbf{73.63}$_{\pm9.24}$ & 49.56$_{\pm12.45}$ & 46.29$_{\pm3.21}$ & 34.00$_{\pm4.60}$ \\
 & ChatGPT & \textbf{74.40}$_{\pm9.16}$ & 45.06$_{\pm7.01}$ & 42.99$_{\pm6.10}$ & 40.08$_{\pm1.85}$ \\
 & Gemini & \textbf{70.75}$_{\pm9.67}$ & 38.89$_{\pm0.07}$ & 46.30$_{\pm6.41}$ & 31.70$_{\pm3.96}$ \\
 & DeepSeek & \textbf{71.38}$_{\pm3.19}$ & 43.39$_{\pm6.78}$ & 47.09$_{\pm6.10}$ & 31.88$_{\pm6.25}$ \\ \hline
\end{tabular}}
\end{table}

Because high-volume techniques such as DoS and C2 produce disproportionately many flows and dominate aggregate accuracy, we report both micro and macro averages. As shown in Table~\ref{tab:freshlab_top1}, on micro accuracy, \tool exceeds 95\% on every backbone, while every baseline falls below 3\%. This gap is consistent with the over-inference behavior identified in Section~\ref{sec:error_analysis}. Baselines mismap DoS flows to probing-related techniques such as T1595 and T1083, because varying URL parameters latch the LLM onto a probing pattern and cause it to override the contradicting volumetric evidence. On macro accuracy, \tool reaches 70.75\% to 74.40\% across the four backbones, exceeding the strongest baseline by 24 to 29 points on every backbone. These results show that \tool's mapping capability does not depend on the backbone having encountered the dataset during pretraining. It is a substantive outcome of \tool's structured reasoning over flow content. By contrast, baselines cannot bridge the semantic gap between low-level traffic and high-level techniques, which leaves them unable to effectively leverage the malicious features in the input flows and drives their accuracy down.
\subsection{RQ6: Sensitivity to Input Perturbations}

To assess \tool's robustness under adversarial evasion attempts, we apply controlled input perturbations that mimic realistic attacker modifications while preserving the underlying malicious behavior. This experiment directly addresses two evasion categories: (i) behavioral modification (\eg slowing brute-force attempts to evade temporal descriptors such as ``rapid'' or ``high frequency'') and (ii) volumetric and protocol-level distortions. Prompt injection and cross-flow IP fragmentation are discussed in Section~VI.

We evaluate five perturbation operators: \textit{duration dilation} (stretching inter-arrival times and session durations), \textit{URL obfuscation} (percent-encoding and path mutation), \textit{user-agent spoofing} (identifier substitution), \textit{packet size increase} (payload padding), and \textit{packet count increase} (traffic splitting). The choice of operators follows the problem-space evasion framework of Pierazzi \textit{et al.}~\cite{pierazzi2020intriguing}, requiring that perturbations preserve attack semantics. Operators are applied at four increasing intensities across five representative samples (three HTTP-based DAPT2020 samples, one SSH brute-force from CICIDS2017, and one C2 beaconing from IoT23), producing 58 perturbation curves across four backbones with three trials each.

\begin{figure}[t]
	\centering
	\includegraphics[width=\columnwidth]{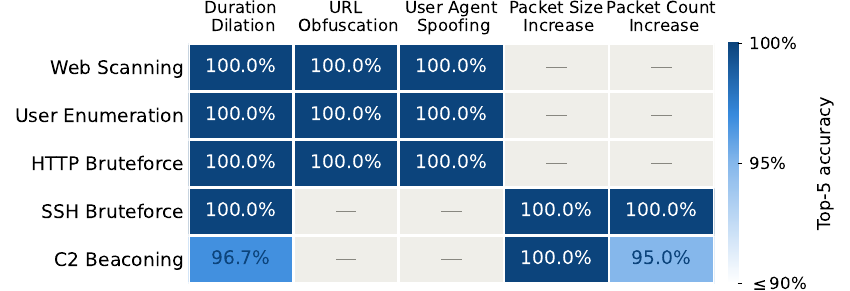}
	\caption{Top-5 accuracy of \tool under input perturbations. Each cell reports the fraction of evaluation runs whose top-5 candidates contain the ground-truth technique. Cells marked ``-'' indicate inapplicable operators.}
	\label{fig:p6_top5}
\end{figure}

As shown in Fig.~\ref{fig:p6_top5}, top-5 retrieval remains stable across the full perturbation range. The ground-truth technique appears in the top-5 candidates in 22 of 24 applicable sample-operator blocks; the remaining two blocks (C2 beaconing) retain 96.7\% and 95.0\% accuracy. Overall, 99.4\% of evaluation runs place the ground-truth technique within top-5. Critically, even under aggressive duration dilation that stretches brute-force attempts from seconds to minutes, removing any ``rapid'' temporal descriptor, T1110 remains in the top-5, indicating that \tool relies on multiple behavioral signals rather than any single, easily evaded feature.

\begin{figure}[t]
	\centering
	\includegraphics[width=\columnwidth]{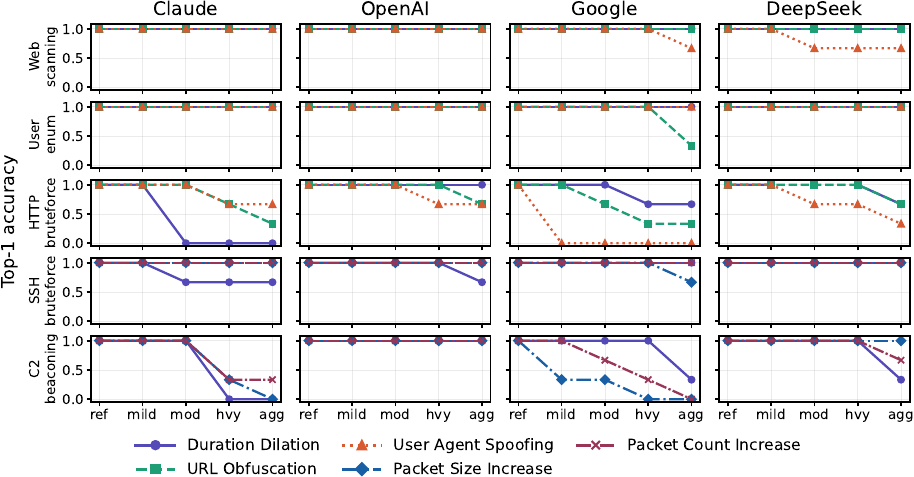}
	\caption{Top-1 accuracy of \tool under input perturbations, arranged as five samples (rows) and four backbones (columns). The horizontal axis spans five perturbation intensities (\ie reference, mild, moderate, heavy and aggressive); the vertical axis is top-1 accuracy.}
	\label{fig:p6_grid}
\end{figure}

Top-1 accuracy is more sensitive (Fig.~\ref{fig:p6_grid}). On HTTP brute-force under 20× duration dilation, T1110 remains in top-5 but is displaced from top-1 by T1595 (Active Scanning). On C2 beaconing under 10× dilation, T1071 stays in top-5 but is overtaken by T1041 (Exfiltration over C2 Channel). In both cases, the perturbation reorders candidates, but the ground-truth technique stays within the top-5, confirming that \tool's structured reasoning provides a fallback even when the top-1 prediction is evaded.

Together, these results demonstrate that while input perturbations can shift top-1 predictions, \tool aggregates evidence from multiple fields rather than relying on any single feature. As long as sufficient supporting evidence survives, even under 20× temporal stretching, the ground-truth technique remains within top-5, and the pipeline avoids outright misidentification. This resilience holds across evasion attempts that manipulate temporal, volumetric, and protocol-level features.

\section{Discussion}
\label{sec:discussion}

\begin{figure}[htbp]
	\centering
	\includegraphics[width=\linewidth]{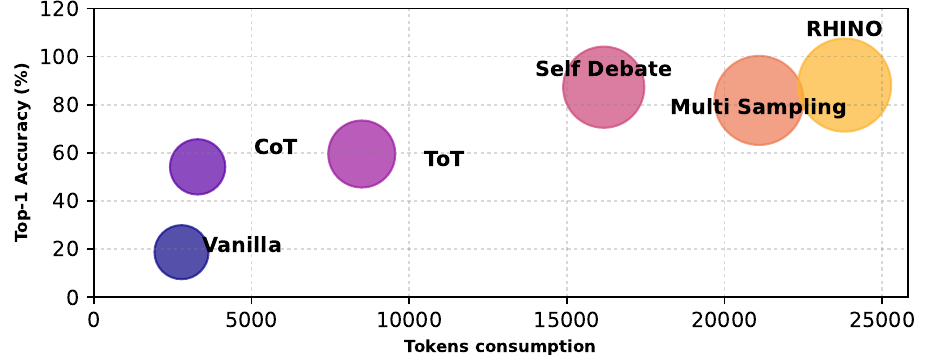}
	\caption{Cost-effectiveness analysis. Each dot's size represents the average token consumption of each method. The y-axis represents the average accuracy across three datasets.} 
	\label{fig:token_consumption}
\end{figure}

\noindent \textbf{Token Efficiency vs. Accuracy Trade-off.} Our method reveals a clear trade-off between token consumption and accuracy. As demonstrated in Figure~\ref{fig:token_consumption}, \tool achieves significantly higher accuracy at the expense of increased token usage. Simpler approaches such as Vanilla and Chain-of-Thought (CoT) consume fewer tokens but yield lower accuracy, whereas \tool improves accuracy by 65.22\% (from 22.93\% to 88.15\%) compared to Vanilla, despite its higher token cost. This indicates a fundamental balance between efficiency and performance: lightweight methods may suffice for latency-sensitive tasks, while \tool justifies its additional resource expenditure in scenarios demanding higher mapping accuracy.

\noindent \textbf{Computational Overhead Justified by Performance Gains.} Regarding computational overhead, \tool exhibits the longest inference time (12,768.3s) among the baseline methods. However, this represents only a 16.97\% increase over the baseline ToT (10,916.3s), while delivering a 29.90\% improvement in overall accuracy. This suggests that our method effectively balances runtime and performance, with the gains in decision-making quality outweighing the marginal increase in inference time. 
This absolute inference time is measured in a non-streaming context, but \tool is designed for offline analysis workflows such as forensic investigation, post-incident analysis, and proactive threat hunting, where the relevant time granularity is hours to days and inference latency is rarely the binding constraint. The higher token consumption does not directly translate to high deployment cost: running the full evaluation on CICIDS2017~\cite{sharafaldin2018toward} costs approximately \$11 with Claude Sonnet 4 and under \$1 with DeepSeek V3. For deployments where API costs or data residency are concerns, \tool can also run on a local open-weight model: as shown in Section~\ref{subsec:rq1_performance}, Llama-3.3-70B-Instruct achieves competitive mapping accuracy in this configuration.

\noindent \textbf{Exclusion of General-Purpose Reasoning LLMs.} We deliberately excluded general-purpose reasoning LLMs, such as ChatGPT-o1 or DeepSeek-R1, as backbone models in our evaluation for two key reasons. First, these models are not designed to enforce structured execution steps, which conflicts with the core design principles of our methodology. Second, their inherent latency and operational costs make them impractical for multi-step or modular tasks, further highlighting the need for specialized solutions like \tool.

\noindent \textbf{Privacy and Data Security Considerations.} Network logs processed by \tool may contain sensitive information (internal IPs, credentials, attack payloads), and sending such data to commercial LLM APIs creates data exposure risks and potential regulatory compliance issues (GDPR, HIPAA). To eliminate these risks, \tool can be deployed entirely on-premises using open-weight LLMs: as demonstrated in Section~\ref{subsec:rq1_performance}, Llama-3.3-70B-Instruct achieves competitive mapping accuracy (79.34\% micro-averaged top-1) without any API calls. When a commercial API must be used, \tool's modular architecture supports a hybrid deployment where only the abstracted behavioral narrative is sent to the API. Anonymization strategies such as IP pseudonymization, payload redaction, and aggregation-based summarization (replacing precise values with ranges) can be integrated as a preprocessing layer to further protect sensitive fields. Tuning these techniques to balance anonymization strength and mapping accuracy remains future work, but the availability of both a fully local open-weight deployment and feasible anonymization paths demonstrates that \tool can meet operational data-protection requirements without sacrificing accuracy.

\noindent \textbf{Deployment Considerations: Prompt Injection and Cross-Flow Fragmentation.} Two categories of adversarial threats, \textbf{prompt injection} via attacker-crafted payloads and \textbf{cross-flow fragmentation} across multiple source identifiers, lie outside \tool's core contribution. Prompt injection is a general vulnerability of LLM-based systems that ingest untrusted content~\cite{greshake2023not}; it is not specific to \tool's reasoning pipeline. Mitigating this risk requires input sanitization, instruction-data separation, or prompt isolation~\cite{hines2024defending,wallace2024instruction} applied at the \textit{abstraction-input boundary}, defenses that are orthogonal to \tool's design and can be integrated without modifying its multi-stage reasoning. Similarly, fragmentation attacks that distribute a logical session across many 5-tuple flows defeat any flow-level aggregator, not just \tool's preprocessing. Reconstructing such fragmented sessions is the responsibility of an upstream detection or correlation engine~\cite{NDSS23-HyperVision,gui2025principled}; \tool assumes that an aggregated behavioral unit has already been formed. Once that unit is provided, RHINO's mapping from aggregated behavior to MITRE ATT\&CK remains unchanged. Both threats are therefore \textbf{deployment-level concerns} addressable by existing techniques that operate \textit{before} \tool's pipeline, leaving the core contribution, bridging the semantic gap between raw logs and adversarial intent, unaffected. We leave this adversarial evaluation for future work.

\noindent \textbf{Limitations of Abstraction-Induced Anchoring.} Section~\ref{sec:error_analysis} identifies two related phenomena: over-inference errors, where the model overemphasizes localized malicious features (\eg a specific SQL keyword) while neglecting broader context, and potential anchoring bias from preliminary interpretations in the abstraction output (\eg ``suggests a DoS attack''). Our ablation study (Table~\ref{tab:ablation}) confirms the behavioral narrative as indispensable context (removing it causes a 68.46 percentage points accuracy drop), yet such reliance does not mean any single preliminary interpretation governs the final decision. In practice, over-inference is a \emph{ranking} error, not a \emph{mapping} one: in 97.6\% of cases the correct technique (Active Scanning, T1595) is recognized but ranked below top-1. Moreover, 99.16\% of these inputs contain a salient payload, and the model justifies its over-ranked choice (\eg T1190) by citing raw-traffic features such as SQL-injection keywords, rather than the abstraction's suggestion. Over-inference thus stems from overattention to salient local features rather than anchoring, consistent with known LLM susceptibility to salient distractors~\cite{shi2023large}. \tool's modular design reinforces this robustness: partitioned inference prevents premature commitment, knowledge-grounded refinement cross-checks candidates against canonical definitions, and the fusion prompt prioritizes structured attributes over interpretive claims, without retraining. Reducing this underlying susceptibility remains open, and \tool's modularity allows future advances to be incorporated.
\section{Related Work}
\label{sec:related_work}

Mapping security observables to MITRE ATT\&CK tactics and techniques (TTs) is critical for reconstructing adversarial behaviors. This task sits at the intersection of cyber threat intelligence automation and LLM-based structured reasoning.

\noindent \textbf{Rule and ontology-driven TT mapping.}
Early approaches rely on predefined mappings or static ontologies~\cite{izzuddin2022mapping,meng2024poster}. AttackDynamics~\cite{hankin2022attack}, for instance, correlates system topology with CAPEC, CWE, and CVE databases to infer plausible technique chains. Such methods produce deterministic and auditable outputs, but their dependence on manually curated patterns makes them ineffective against novel attacks.

\noindent \textbf{Learning-based approaches.}
Learning-based methods~\cite{milajerdi2019holmes,moskal2021translating} infer TTs from structured observations or system provenance. Representative examples include SAGE~\cite{nadeem2021alert} and ARKAVI~\cite{hakim2024predicting}, which model attack stages from IDS alerts and system logs, respectively. These methods either operate only at the coarse tactical level or rely on provenance graphs unavailable in network-only environments.

\noindent \textbf{LLM-based reasoning paradigms.}
The adoption of LLMs across cybersecurity tasks~\cite{xu2024large} has motivated parallel efforts to enhance LLM reasoning itself. Chain-of-Thought (CoT) prompting~\cite{wei2022chain} elicits step-by-step reasoning within a single generation pass. Tree-of-Thoughts (ToT)~\cite{yao2023tree} introduces deliberate search over multiple reasoning branches, enabling backtracking among competing hypotheses. Few-shot prompting~\cite{brown2020language} and retrieval-augmented generation (RAG)~\cite{lewis2020retrieval} inject task-specific knowledge through in-context exemplars or retrieved documents. These paradigms improve \emph{how} an LLM reasons but do not address the \emph{domain-specific} challenges of a particular task. For TT mapping, step-by-step reasoning or path sampling cannot on their own bridge the semantic gap between raw network observations and the high-level adversarial intent described by ATT\&CK techniques.

\noindent \textbf{LLM-based multi-agent systems.}
Beyond single-model prompting, multi-agent systems organize multiple LLM instances as autonomous agents. ReAct~\cite{yao2022react} interleaves reasoning traces with external tool calls, while frameworks such as AutoGen~\cite{wu2024autogen} and MetaGPT~\cite{hong2024metagpt} enable agents to assign roles, critique each other's outputs, and iteratively refine solutions~\cite{guo2024large}. These systems excel at open-ended tasks where the solution path is unknown in advance. TT mapping, by contrast, has a well-defined analytical scope and does not inherently require open-ended exploration or multi-party negotiation.

\noindent \textbf{Automated TT mapping with NLP and LLMs.}
Domain-specific pretrained models offer one path toward automation. SecureBERT~\cite{aghaei2022securebert} captures general security-domain semantics, but its pretraining objective optimizes for broad security-language understanding rather than for discriminating among individual ATT\&CK techniques, whose definitions often share substantial vocabulary. An alternative thread applies NLP and LLM techniques directly to TT mapping from \emph{natural-language CTI reports}. TTPDrill~\cite{husari2017ttpdrill} pioneers this direction by combining threat-action extraction with an attack-pattern ontology, and subsequent work extends coverage through knowledge-graph construction~\cite{li2022attackg} and fine-tuned classification~\cite{liu2022threat}. A parallel thread targets \emph{detection rules and SIEM alerts}: Daniel~\etal~\cite{daniel2023labeling} annotate Snort rule descriptions with MITRE labels, while RAM~\cite{wudali2025rule} maps SIEM queries to ATT\&CK techniques via prompt chaining.

A common property unifying these methods is that their inputs already reside in the \emph{semantic domain}: CTI reports are natural-language narratives, and detection rules are human-authored textual descriptions. The mapping task therefore reduces to matching natural-language text against ATT\&CK definitions. Raw network logs, by contrast, consist of numerical statistics and protocol-level metadata that carry no explicit natural-language semantics. Applying LLM-based mapping to this modality requires a cross-modal translation step that converts raw traffic observations into behavioral narratives before technique reasoning can begin, a challenge not addressed by existing TT-mapping methods. \tool is designed to bridge this gap.

\section{Conclusion}

RHINO reframes MITRE ATT\&CK mapping as a structured reasoning task, combining context-aware log abstraction, partitioned inference over the ATT\&CK tactic space, and knowledge-grounded refinement to outperform rule-based, learning-based, and monolithic LLM approaches. Experiments across APT, intrusion, and IoT botnet scenarios demonstrate $>$86\% top-1 accuracy \revise{over 11 techniques}, \revise{reducing the invalid-TT rate from 7.5\% to 2.6\%}, while maintaining compatibility with existing NIDS pipelines.

This work demonstrates that LLMs can \revise{recover the ATT\&CK techniques of documented attacks from network logs} when guided by a structured, multi-stage pipeline that decomposes the task along the structure of the ATT\&CK taxonomy. Future directions include extending RHINO to multi-modal threat intelligence (\eg logs + provenance graphs).

\bibliographystyle{IEEEtran}
\bibliography{reference}

\vfill

\end{document}